\documentclass{aastex}

\newcommand{\eg}{{\it e.g.}}

\newcommand{\ie}{{\it i.e.}}

\newcommand{\etc}{{\it etc.}}

\newcommand{\etal}{{\it et~al.}}

\newcommand{\sn}{$S/N$}

\newcommand{\um}{$\mu$m}

\newcommand{\gps}{\ensuremath{g_{\rm P1}}}
\newcommand{\rps}{\ensuremath{r_{\rm P1}}}
\newcommand{\ips}{\ensuremath{i_{\rm P1}}}
\newcommand{\zps}{\ensuremath{z_{\rm P1}}}
\newcommand{\yps}{\ensuremath{y_{\rm P1}}}
\newcommand{\wps}{\ensuremath{w_{\rm P1}}}

\newcommand{\PS}{\protect \hbox {Pan-STARRS}}
\newcommand{\PSone}{\protect \hbox {Pan-STARRS1}}
\newcommand{\PStwo}{\protect \hbox {Pan-STARRS2}}
\newcommand{\PSfour}{\protect \hbox {Pan-STARRS4}}
\newcommand{\knownserver}{{\tt known\_server}}

\begin{document}

\title{The \PS\ Moving Object Processing System}

\author{
Larry Denneau\altaffilmark{1} (denneau@ifa.hawaii.edu),
Robert Jedicke\altaffilmark{1},
Tommy Grav\altaffilmark{2},
Mikael Granvik\altaffilmark{3},
Jeremy Kubica\altaffilmark{4},
Andrea Milani\altaffilmark{5},
Peter Vere\v{s}\altaffilmark{1},
Richard Wainscoat\altaffilmark{1},
Daniel Chang\altaffilmark{1},
Francesco Pierfederici\altaffilmark{6},
N. Kaiser\altaffilmark{1},
K.~C. Chambers\altaffilmark{1},
J.~N. Heasley\altaffilmark{1},
Eugene.~A. Magnier\altaffilmark{1},
P.~A. Price\altaffilmark{7},
Jonathan Myers\altaffilmark{8},
Jan Kleyna\altaffilmark{1},
Henry Hsieh\altaffilmark{1},
Davide Farnocchia\altaffilmark{5,9},
Chris Waters\altaffilmark{1},
W.~H. Sweeney\altaffilmark{1},
Denver Green\altaffilmark{1},
Bryce Bolin\altaffilmark{1},
W.~S. Burgett\altaffilmark{1},
J.~S. Morgan\altaffilmark{1},
John~L. Tonry\altaffilmark{1},
K.~W. Hodapp\altaffilmark{1},
Serge Chastel\altaffilmark{1},
Steve Chesley\altaffilmark{9},
Alan Fitzsimmons\altaffilmark{10},
Matthew Holman\altaffilmark{11},
Tim Spahr\altaffilmark{12},
David Tholen\altaffilmark{1},
Gareth~V. Williams\altaffilmark{12},
Shinsuke Abe\altaffilmark{13},
J.D. Armstrong\altaffilmark{1}, 
Terry~H. Bressi\altaffilmark{14},
Robert Holmes\altaffilmark{15},
Tim Lister\altaffilmark{16},
Robert~S. McMillan\altaffilmark{14},
Marco Micheli\altaffilmark{1},
Eileen~V. Ryan\altaffilmark{17},
William~H. Ryan\altaffilmark{17},
James~V. Scotti\altaffilmark{14}
}

\slugcomment{57 Pages, 26 Figures, 13 Tables}

\altaffiltext{1}{Institute for Astronomy, University of Hawai`i, 2680 Woodlawn Dr., Honolulu, HI 96822}
\altaffiltext{2}{Johns Hopkins University, Baltimore, MD}
\altaffiltext{3}{Department of Physics, P.O. Box 64, 00014 University of Helsinki, Finland}
\altaffiltext{4}{Google, Inc.}
\altaffiltext{5}{University of Pisa, Pisa, Italy}
\altaffiltext{6}{Space Telescope Science Institute, 3700 San Martin Dr., Baltimore, MD 21218}
\altaffiltext{7}{Department of Astrophysical Sciences, Princeton University, Princeton, NJ 08544}
\altaffiltext{8}{University of Arizona, Tucson, AZ}
\altaffiltext{9}{Jet Propulsion Laboratory, Pasadena, CA}
\altaffiltext{10}{Astrophysics Research Centre, School of Mathematics and Physics, Queen's University Belfast, Belfast, BT7 1NN, UK}
\altaffiltext{11}{Harvard-Smithsonian Center for Astrophysics, 60 Garden St., Cambridge, MA 02138}
\altaffiltext{12}{Smithsonian Astrophysical Observatory, Cambridge, MA}
\altaffiltext{13}{Institute of Astronomy, National Central University, Taiwan}
\altaffiltext{14}{Lunar \& Planetary Laboratory, University of Arizona, Tucson, AZ}
\altaffiltext{15}{Astronomical Research Institute, 7644 NCR 1800E, Charleston, IL 61920}
\altaffiltext{16}{Las Cumbres Observatory Global Telescope Network, Inc., 6740 Cortona Dr. Suite 102, Santa Barbara, CA 93117}
\altaffiltext{17}{Magdalena Ridge Observatory, New Mexico Tech, 801 Leroy Pl., Socorro, NM 87801}

\shorttitle{\PS\ MOPS}
\shortauthors{Denneau \etal}

\begin{abstract}

We describe the \PS\ Moving Object Processing System (MOPS), a modern
software package that produces automatic asteroid discoveries and
identifications from catalogs of transient detections from
next-generation astronomical survey telescopes.  MOPS
achieves $>99.5\%$ efficiency in producing orbits from a synthetic but realistic
population of asteroids whose measurements were simulated for a \PSfour-class telescope. Additionally, using a non-physical grid population,
we demonstrate that MOPS can detect populations of currently unknown
objects such as interstellar asteroids. 

MOPS has been adapted successfully to the prototype \PSone\ telescope despite 
differences in expected false detection rates, fill-factor loss and relatively
sparse observing cadence compared to a hypothetical \PSfour\ telescope and survey.  MOPS remains
highly efficient at detecting objects but drops to 80\% efficiency at producing orbits. This loss is primarily due to configurable MOPS
processing limits that are not yet tuned for the \PSone\ mission.

The core MOPS software package is the product of more than 15 person-years of software
development and incorporates countless additional years of effort in third-party
software to perform lower-level functions such as spatial searching or orbit determination.
We describe the high-level design of MOPS and essential subcomponents, 
the suitability of MOPS for other survey programs, and suggest a road map for
future MOPS development.

\end{abstract}


\maketitle

\section{Introduction}
\label{s-Introduction}

As with most scientific endeavors, the history of asteroid and comet studies
depicts an exponential increase in the rate of discovery
since the identification of Ceres by Piazzi more than 200 years ago.
This work describes the next step in the evolution of asteroid surveys
--- an integrated, end-to-end moving object processing system (MOPS)
for the Panoramic Survey Telescope and Rapid Response System
\citep[Pan-STARRS][]{Kaiser2002,Kaiser2004,Hodapp2004}.  The system's prototype telescope
(PS-1) employs a 1.4 gigapixel camera able to detect asteroids and comets
faster than ever before.

The first asteroids were discovered serendipitously and laboriously by
eye until the dedicated photographic surveys of the 1950s and 1960s
like the Yerkes-McDonald \citep{Kuiper1958} and Palomar Leiden
Surveys \citep{vanHouten1970}.  The photographic surveys required a
major effort due to the need for human `blinking' of the images to
identify the moving objects.  The realization in the early 1980s by
\citet{Alvarez1980} that the extinction of the dinosaurs $\sim$65
million years ago was precipitated by the impact of a large asteroid
or comet with the Earth stimulated enhanced funding for wide field
asteroid surveys.  While those
surveys identified many asteroids and comets that may eventually
strike the Earth, only one object has been discovered that actually
hit the Earth --- 2008~TC$_3$
\citep[\eg][]{Jenniskens2009,Boattini2009}.

The photographic searches leveraged decades of experience in wide
field astronomical surveying but Spacewatch \citep{Gehrels1991,McMillan2007}
spearheaded the first use of CCDs in asteroid surveys.  In the
beginning, the small CCDs of the time limited the success of this
search program but the asteroid discovery rate improved dramatically
when they obtained a high quantum efficiency 2K$\times$2K CCD with an
$\sim$30~arcmin field of view and adopted a `drift scanning' survey
technique that eliminated the need for long readout times.  At the
same time, \citet{Rabinowitz1991} developed the first automated moving
object detection program to identify asteroids and comets in
Spacewatch's drift-scan images that launched the contemporary
generation of wide field surveys such as NEAT \citep{Helin1997},
LONEOS \citep{Bowell1995}, LINEAR \citep{Stokes2000} and CSS
\citep{Larson2007}.

These modern wide field asteroid surveys were prompted and subsequently funded by
the NASA Spaceguard Program \citep{Harris2008} with the goal of
identifying 90\% of Near Earth Objects\footnote{Near Earth Objects are
  asteroids or comets with perihelion distances of $\le$1.3~AU.  Most
  of these objects are in unstable orbits with dynamical lifetimes of
  $<$10~Myrs \citep[\eg][]{Gladman2000} and will quickly be ejected
  from the Solar System or impact the Sun or Jupiter but a small
  fraction will eventually strike Earth.} (NEOs) larger than 1~km
diameter before the end of 2008.  All the surveys broadly employ
similar techniques for asteroid and comet identification.  Their wide
field cameras have large pixel scales (typically $\ga$1 arcsec) and
image a field 3-5 times within about an hour.  Their moving
object detection software identifies `sources' in each image and then
spatially correlates the detections between the images to identify and
remove stationary objects.  All the `transient detections' are then
searched for consistency with a single object moving linearly
across the sky at a constant rate of motion between the exposures.
The constant motion requirement allows the software to achieve
per-detection signal-to-noise (\sn) levels of $\sim$1.5 to 3$\sigma$.  Sets of 3-5
linked detections (a `tracklet') representing candidate real moving
objects are reported to the Minor Planet Center (MPC), though some
groups review the detections by eye before submission to reduce the
false detection rate.  Overall, these groups were wildly successful,
identifying $\sim$79\% of the $\ge$1~km NEOs before 10 June 2008
\citep{Harris2008}.  \citet{Mainzer2011} report that the Spaceguard
goal of discovering 90\% of the $\ge$1~km diameter asteroids was
actually met some time later but before 2011.

The CCD surveys described above were designed to discover large,
extremely hazardous NEOs, not to characterize the size-frequency
distribution (SFD) of various solar system populations.  While
the process of searching for the NEOs, they discovered several
objects interesting in their own right (\eg\ Comet Shoemaker-Levy 9,
\citet{Shoemaker1995}; very fast moving objects, typically very small and nearby asteroids, \citet{Rabinowitz1993}; the Centaur object (5445) Pholus; 2008
TC$_3$, \citet{Jenniskens2009}); and {\it post facto} determinations of
their surveying efficiency allowed new estimations of the SFD of many
small body populations (\eg\ objects with orbits entirely interior to Earth's orbit, IEOs, \citet{Zavodny2008}; NEOs,
\citet{Rabinowitz1993b}; Main Belt, \citet{Jedicke1998}; Centaurs,
\citet{Jedicke1997}; Trans-Neptunian objects, TNOs, \citet{Larsen2001}; Distant objects,
\citet{Larsen2007}).  Despite the success of these programs there is room for
improvement in surveying techniques, moving object detection
software, and reporting and follow-up methods:

\begin{itemize}
\item Survey Technique and/or depth 

The surveys in operation in 2003 were not capable of meeting the
Spaceguard Goal due to their sky-plane coverage and/or limiting
magnitude \citep{Jedicke2003}.  These surveys concentrated on the
region of sky near opposition because asteroids are generally
brightest in that direction but objects on orbits that {\it will}
impact the Earth are under-represented towards opposition.  Their
sky-plane density increases at relatively small ecliptic latitudes and
solar elongations $\la 90\deg$ \citep[\eg][]{Chesley2004,Veres2009}.
However, asteroids in this region of the sky are notoriously difficult
to observe because of the limited time that they are above the
horizon, the high air mass when they are visible, their large phase
angle (reduced illumination of the visible surface) and faintness.
Furthermore, the 3-5 repeat visits/night to the same field is wasteful
of survey time.  More sky could be covered or the same sky could be
imaged to greater depth if there were fewer visits to the same field
each night.

\item Follow-up

To maximize survey coverage and their discovery rate most of the
surveys visit a field on only one night per lunation to obtain a set
of 3-5 linked detections.  The detections provide the sky plane
location and velocity vector for the object rather than an orbit.  To
identify NEOs, the surveys flag objects with unusual rates of motion
\citep[\eg][]{Rabinowitz1991,Jedicke1996} as NEO candidates and then
reacquire the objects in special follow-up efforts that reduce the time
available for discovery of new objects.  Alternatively, they report
the detections to the MPC who then perform a more sophisticated
probabilistic analysis to determine the likelihood that the object may
be a NEO.  If the object meets their likelihood cutoff, it is posted on
their NEO confirmation
webpage\footnote{http://www.minorplanetcenter.net/iau/NEO/ToConfirm.html}
and hopefully recovered by professional and amateur astronomers around
the world.  Given that telescope time is a valuable commodity and that
the next generation of professional asteroid survey telescopes will
discover more and even fainter objects, it is clear that the survey
telescopes must provide their own target follow-up and it should be
incorporated directly into their survey pattern.

\item Moving Object Detection and Orbit Determination Software

With limited follow-up resources and inefficiency in the ability to
determine if a candidate asteroid is likely a NEO based on its
magnitude and apparent rate of motion (its `digest' score), it would
be useful for a survey's moving object detection system to determine
orbits from observations over multiple nights instead of merely
reporting sets of detections (tracklets) to the MPC.  This allows for
improved discovery rates of NEOs that are `hidden in plain sight',
indistinguishable from main-belt asteroids by their brightness and
velocity vectors alone, and therefore scoring too low using the MPC's
NEO digest scoring to warrant follow-up.

\item Reporting

The system of reporting sets of detections to the MPC has been very
effective and the follow-up response of the international community of
amateur and professional observers has been fantastic as amply
demonstrated by the case of 2008~TC$_3$ \citep{Jenniskens2009}.
Still, the process of submission by email to the MPC, posting on the
NEO confirmation page, downloading and acquisition by observers is
unwieldy.  Since there is little coordination between observers there
is the potential for wasted time with multiple and unnecessary follow-up.
A modernized, automated reporting system tuned specifically to the
capabilities and interests of each follow-up observatory would be
useful to coordinate the follow-up effort.

\item Measured System Efficiency

Existing asteroid detection software packages do not monitor their own
detection efficiency and accuracy.  Instead, analysis of the survey
data must be performed {\it post facto}.  It would be convenient if
the packages incorporated an intrinsic near real-time measure of both
their efficiency and accuracy for the purpose of system monitoring and
subsequent data reduction.  

\end{itemize}

Some of the problems itemized above have been addressed by other
surveys, in particular the well-characterized main belt program of
\citet{Gladman2009} using software developed by \citet{Petit2004}.
They generate synthetic detections in the same images used to identify
their TNO candidates to measure the detection efficiency and accuracy.

The future of wide field moving object processing systems is embodied
in software such as the Panoramic Survey Telescope and Rapid Response
System \citep[\PS;][]{Kaiser2002,Kaiser2004} Moving Object Processing
System (MOPS), described in the remainder of this paper.  MOPS
incorporates state-of-the-art spatial searching, orbit computation and
database management into a cohesive package.  MOPS resolves many of
the issues described above and the system is capable of very high
efficiency and accuracy.

\section{\PS\ and \PSone}
\label{s-PS1Survey}

The University of Hawai`i's \PS\ project was formed in 2002 to design
and build a next-generation distributed-aperture survey system called
\PSfour\ \citep{Kaiser2002}.  The \PSfour\ design incorporates four
distinct telescopes on a common mount on the summit of Mauna Kea.
Shortly after inception, the project began development and
construction of a single-telescope prototype system called
\PSone\ \citep[][]{Hodapp2004,Morgan2006} on Haleakala, Maui to
validate essential components such as optics, camera and software
pipelines.  As of 2013, the \PS\ project is constructing a second
telescope called \PStwo\, essentially identical to \PSone, at the same
site on Haleakala.  The combination of \PSone\ and \PStwo\ is called
PS1+2, and it is expected that the two telescopes will be operated
together as part of a single science mission.

The \PSone\ telescope on Haleakala,
Maui, began surveying for asteroids in the spring of 2010.  This
1.8~m diameter telescope has a $\sim$7~deg$^2$ field of view and a
$\sim$1.4~gigapixel orthogonal transfer array (OTA) CCD camera
\citep{Tonry1997,Tonry2004} with 0.26\arcsec\ pixels.  \PSone's large aperture, field of
view and $\la$13~s between exposures allows imaging of the entire
night sky visible from Hawai`i to $r \sim 21.2$ in about 5 or 6 nights.
In practice, 2 to 8 images of the same field are acquired each night
and consequently it takes longer to cover the entire night sky.

The \PSone\ telescope is operated by the \PSone\ Science Consortium
\citep[PS1SC;][]{Chambers2006,Chambers2007} with a nominal mission of 3.5 surveying years.  The PS1SC
survey plan incorporates about a half dozen sub-surveys but only three
are suitable for moving object discovery (all the surveys are suitable
for moving object {\it detection} because they all obtain at least one
pair of images of each field each night) --- the 3$\pi$ all-sky survey,
the Medium Deep (MD) survey and the solar system survey.  

The \PSone\ detector system employs six passbands: \gps, \rps, \ips, \zps,
\yps\ and \wps.  \citet{Tonry2012} provide a detailed description of
the \PSone\ photometric system --- the first four passbands were designed to
have similar characteristics to the SDSS 
\citep[\eg][]{Karaali2005}, \yps\ was designed to take advantage of
the good sensitivity of the \PSone\ camera near 1\um\, and the
\wps\ has a wide bandpass with high and low wavelength cutoffs
designed to optimize the \sn\ for S and C class asteroids.
Transformations from each filter to the $V$ band are provided in
table~\ref{t.PS1FilterTransformations} for an object with solar colors
and also for an object with an average S+C class spectrum.  The difference between the solar and mean
asteroid class transformations are significant since the
\PSone\ photometric system currently provides better than 1\% photometry
\citep{Schlafly2012} in at least the 3$\pi$ survey's filters.

The 3$\pi$ survey images the entire sky north of $-30\arcdeg$ declination
multiple times in three passbands (\gps, \rps, \ips) in a single year.
About half the sky within $\pm30\arcdeg$ (two hours) of opposition in
R.A. is imaged in the different passbands each lunation.  The solar
system survey is performed in the wide filter (\wps) and most of that
time is spent near the ecliptic and near opposition to maximize the
NEO discovery rate.  A small percentage of solar system survey time is
used to image the `sweet spots' for Potentially Hazardous
Asteroids\footnote{PHAs are NEOs with orbits that approach to within
  0.05~AU of the Earth's orbit and have absolute magnitudes $H<22$.}
(PHAs) within about $\pm10\arcdeg$ of the ecliptic and as close to the Sun
as possible subject to altitude, sky brightness constraints, \etc\ The
MD survey obtains 8-exposure sequences at 10 different fixed footprints
on the sky, with an integration time of 1920 seconds per sequence.
MD sequences are observed using either 8 $\times$ 240-second exposures in a single filter 
(\ips, \zps, \yps), or back-to-back sequences of 8 $\times$ 120-second exposures
in \gps\ and \rps.

Visits to the same footprint within a night are usually separated by
about 15 minutes, a `transient time interval' (TTI), suitable for
asteroid detection.  The Image Processing Pipeline \citep[IPP;][]{Magnier2006} produces a
source list of transient detections, \ie\ new sources at some position
in the image and/or those that change brightness but are otherwise not
identifiable as false transients like cosmic rays or other image
artifacts.  The IPP then publishes the catalogs of transient
detections to MOPS which searches for moving objects.

\section{The \PS\ Moving Object Processing System (MOPS)}

\subsection{Overview}
\label{s.Overview}

MOPS software development began in 2003 under management by the
\PS\ Project, with a mission to discover hazardous NEOs and other
solar system objects while providing real-time characterization of its
performance.  Initial funds for \PS\ construction and engineering were
obtained by the University of Hawai`i via a grant from the United
States Air Force Research Laboratory (AFRL).  MOPS was the first and
only science client to be funded by the \PS\ Project --- systems
engineering, testing, performance tuning and creation of the Synthetic
Solar System Model (S3M, \S \ref{ss.SyntheticSolarSystemModel}) were
directly supported by project resources.  Additional in-kind software
contributions from external collaborators in the areas of orbit
determination \citep{Granvik2009, Milani2010}, spatial searching
\citep{Kubica2007} and algorithm design \citep{Tyson2001} were
incorporated via memoranda of understanding (MOUs).  In 2008, the NASA
Near Earth Objects Observations (NEOO) Program Office began its
support of continued MOPS development and \PSone\ operations.

\begin{figure}[btp]
\centering
\plotone{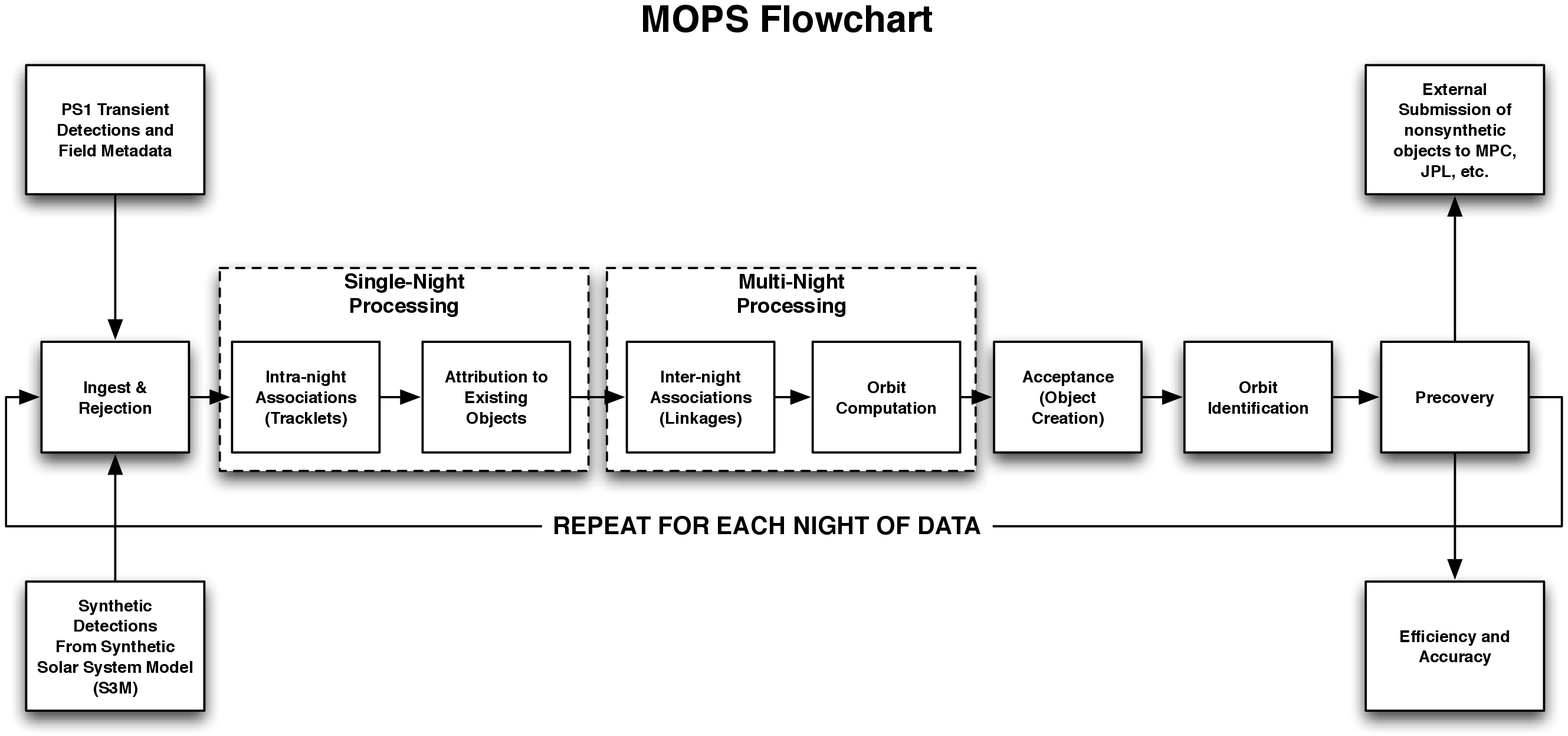}
\caption{High level flowchart for the \PSone\ Moving Object Processing
  System (MOPS).  The data processing proceeds from left to right in
  the figure, then repeats for every additional night of data ingested
  by MOPS.}
\label{fig.MOPS-Flowchart}
\end{figure}

MOPS is the first integrated detector system that
processes data from per-exposure transient detection source lists
through orbit determination, precovery and attribution. We sought
to create a system that could independently measure our system's
end-to-end throughput and efficiency for the purpose of correcting for
observational selection effects \citep{Jedicke2002}.  Our philosophy
was that it is more important to accurately know the system efficiency than
to be place all our effort in optimization.  
Figure~\ref{fig.MOPS-Flowchart} provides a high-level view of data
flow into, through and from MOPS.  Each of the MOPS data processing
sub-steps is described in the following subsections.

In the \PS\ system design there is a functional separation between
the image processing (IPP) and moving object processing (MOPS).
Unlike the object detection algorithm of \eg\ \citet{Petit2004}, the
MOPS design mandates no integration with the image data.  This was an
intentional design decision based on the scale and organization of the
\PS\ project.  The IPP is responsible for monitoring and reporting its
detection efficiency and accuracy as is MOPS.  In this way each
\PS\ subsystem can be developed and characterized 
independently.

MOPS data processing consists of assembling groups of transient
detections into progressively larger constructions until there are
enough detections to produce a high-quality orbit believed to
represent a real asteroid --- a `derived object'. Intra-night (same
night) detection groupings are called `tracklets' and inter-night
(multi-night) groupings are called `tracks'. All tracks created by
MOPS are evaluated by an orbit computation module after which a track
is either 1) deemed to represent a real asteroid, or 2) rejected and
its constituent tracklets returned to the pool of unassociated tracklets.
When a derived object is created MOPS uses the object's computed
(derived) orbit to search for additional detections of the asteroid
in MOPS data to further refine the orbit.

MOPS expects its incoming transient detection stream to be organized
by exposure, or \emph{field} in MOPS parlance. Each field within MOPS
is defined by its metadata \eg\ (right~ascension,~declination [RA,~dec]) boresight coordinates,
exposure date and duration, filter.  A MOPS detection by definition occurs in exactly
one field and is defined by its position \eg\ (RA,~dec) in the
field, observed magnitude, signal-to-noise ($S/N$), and associated
uncertainties.

\subsection{Pipeline Design}

The MOPS pipeline operates by using a linear nightly processing model
where data are ingested and processed in the order they are
observed. Nightly data are ingested from a live transient detection
stream and discrete processing stages are executed until all
processing is completed for the entire night. Many of the MOPS
processing stages build upon data structures created from prior
nights.

Out-of-order processing can be handled in a limited number of modes of
operation as necessitated by the current
\PSone\ system. Enhancements to the MOPS pipeline to perform full
processing on out-of-order data while preserving essential MOPS
efficiency computations is still under development.  The fundamental
difficulty in out-of-order processing lies with the amount of data
that needs to be recomputed when new observations are inserted 
in the middle of the temporal data stream.  The existing MOPS design
prefers nights to be added incrementally; insertion into the
middle of the existing dataset essentially forces all subsequent
nights to be reprocessed.

Pipeline resource scheduling and management are handled by the Condor
high-throughput-computing software \citep{Thain2005}.  Condor provides
effective, flexible management of hardware resources that need to run
the MOPS production pipeline simultaneously with test simulations or
experimental processing of MOPS data.

The MOPS pipeline is designed to be reliable in the event of cluster
resource limitations and hardware failure (\eg\ power outages, node
failures).  Working in tandem with Condor's process management
infrastructure, the MOPS pipeline can be restarted easily at the proper
point in the pipeline with all data structures intact. As an example,
in 2011 the MOPS production MySQL database suffered a complete
failure yet the MOPS pipeline was back online within hours after
restoring the database from a backup.

The discussion of each element of the MOPS pipeline is deferred to
\S\ref{s.ProcessingRealData} where we follow the processing of
detections from beginning to end of the pipeline and quantify the
performance of each step using \PSone\ data.

\subsection{Hardware}

\begin{figure}[btp]
\centering
\plotone{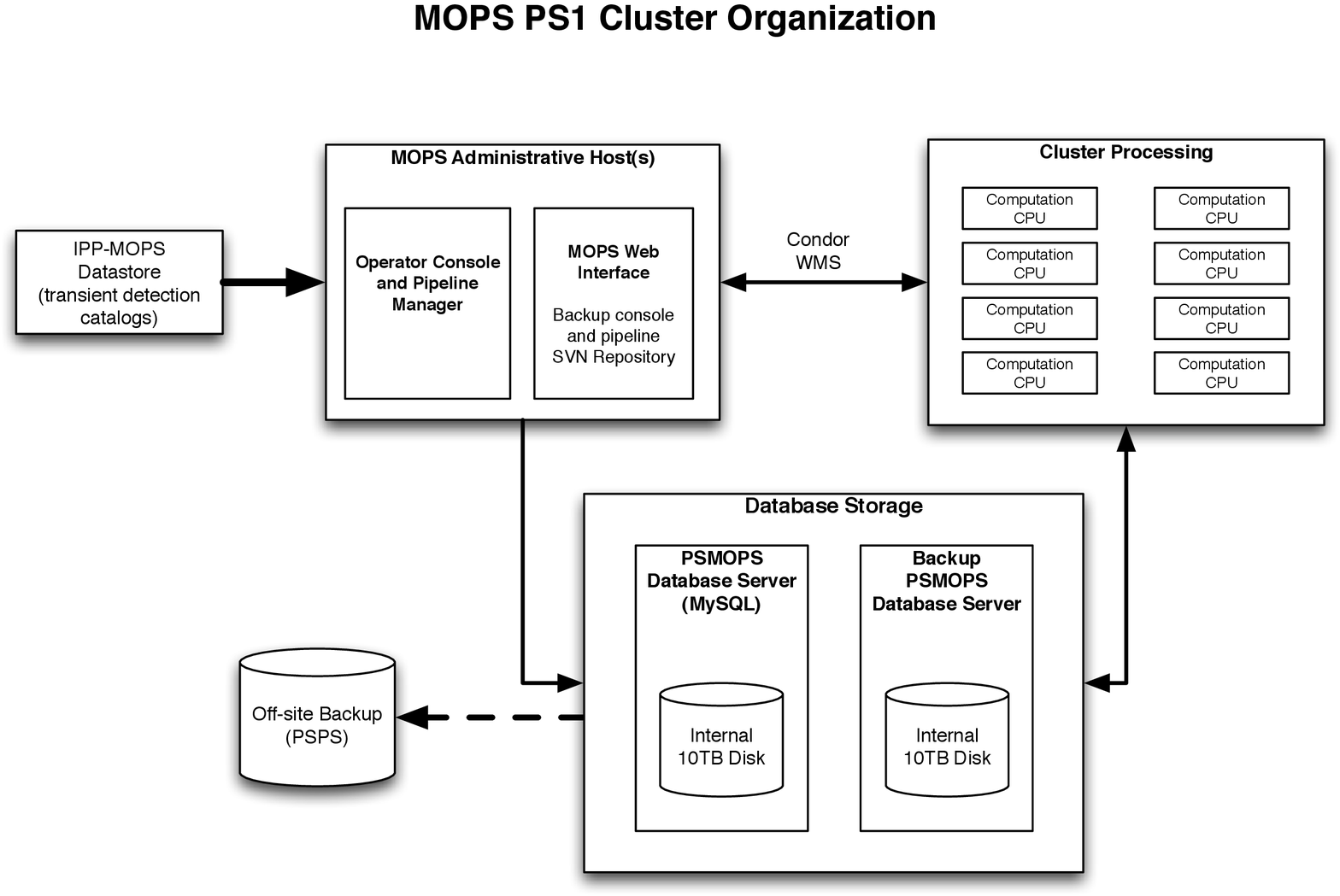}
\caption{MOPS \PSone\ cluster configuration.}
\label{f.cluster}
\end{figure}

The \PSone\ MOPS runs on a modest cluster of standard Linux
rack-mounted computers. MOPS makes no special demands on hardware so
long as the cluster can keep up with incoming data.  During early
stages of MOPS design we based hardware requirements on
estimates for transient detections stored and orbits computed per
night of observing for a \PSfour-like mission, and scaled down the
storage for \PSone\ volumes.  The current \PSone\ processing
hardware is capable of keeping up with a \PSfour-like
data stream except for the storage of detections in the database.  
Table~\ref{t.MOPSHardwareComponents} lists the
major hardware components employed by the \PSone\ MOPS production
processing cluster, and Fig.~\ref{f.cluster} shows the functional
relationships between the MOPS hardware components.

To ensure against most of the types of disk failures we are likely to
encounter, we use a multiple-parity redundant array of independent drives (RAID6) for our database storage, meaning multiple parity bits for every
byte stored on disk. This allows up to two drives to fail concurrently
and still keep the disk array operational.

Condor's workflow management automatically detects host-level failures
and redistributes jobs appropriately. An entire cluster computation
node can fail during a processing step and the MOPS pipeline will
adjust seamlessly.

MOPS pipeline administrative functions are handled by a `fat'
workstation-class computer that runs the main MOPS pipeline, Condor
manager and MOPS web interface (described later). We distribute these
functions between two identically configured hosts so that one can
fail completely, its responsibilities then assumed by the other node at
slightly degraded performance.
During \PSone\ commissioning and operations we have experienced
complete database host failures, accidental deletion of a production
MOPS database by a MOPS administrator, and numerous compute node
failures, without loss of data or significant interruption of MOPS
processing.

\subsection{Database}

\begin{figure}[btp]
\centering
\plotone{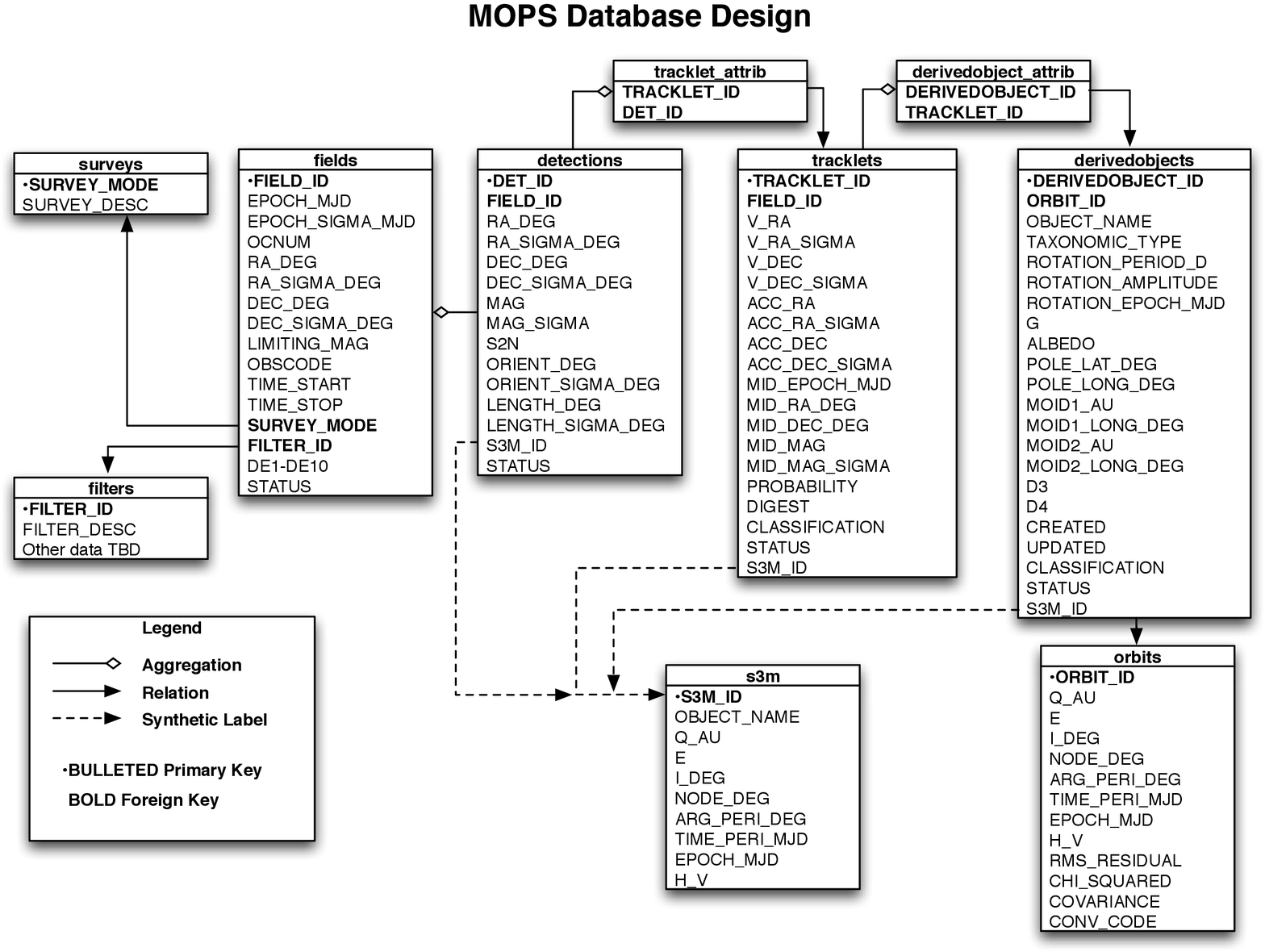}
\caption{MOPS database schema. S3M columns indicate synthetic data
for efficiency assessment. }
\label{f.database}
\end{figure}

Table \ref{t.MOPSDatabaseStorageRequirements} shows the storage
required for MOPS database components under \PSone\ and estimates for
\PSfour.  The catalog of high-significance detections and derived data
products are stored in an industry-standard relational database to
maximize interoperability with external analysis software and provide
data mining capability. Data are organized into multiple row-column
tables in which rows of one table may be related to rows in another
table (see fig.~\ref{f.database}). For example, the MOPS FIELDS table
contains exposure metadata describing the telescope pointing, exposure
time, filter, \etc\ for a single exposure. Each {\tt FIELD} record is
additionally assigned a unique identifier called {\tt FIELD\_ID}. A
second table called {\tt DETECTIONS} contains all transient detections
ingested by MOPS and, to relate detections to a particular exposure,
all detections from the exposure are assigned a {\tt FIELD\_ID} upon
ingest that matches their exposure's {\tt FIELD\_ID} . Subsequent
derived relations (\eg\ groupings of detections into an asteroid
`identification' and computation of its derived orbit) are similarly
maintained and are described in detail later.

A relational database allows MOPS data to be manipulated, analyzed and
queried by any external software that supports structured query
language (SQL). Requests for both a small amount of data (such as a
detection) or a large amount of data (\eg\ all derived objects with $q <
1.3$) are made through a SQL query and the results are returned in a
tabular representation. 

The MySQL relational database management system (RDBMS) was selected
for MOPS because of its combination of performance and cost but other
database vendors such as PostgreSQL and Oracle were considered and
evaluated early in MOPS development. Any modern database
system would work for MOPS as they all scale to billions of rows and
have additional features that promote data integrity and system
reliability.

\subsection{Low Significance Database}

MOPS was designed to perform searches for recovery observations using
a lower-significance data archive called the `low-significance
dataset' (LSD). In normal MOPS processing a nominal confidence level,
typically 5$\sigma$, is used as a baseline for all processing. 
The \PSfour\ camera and image processing are expected to deliver
$\sim 1500~5\sigma$ detections per exposure. Below
this confidence level the density of false associations becomes so
great that it overwhelms the data processing. However, the search for
recovery or precovery observations of an object with a known orbit may
constrain the predicted position and velocity enough to make it
possible to identify observations in the LSD.

Due to the much larger number of lower-significance sources in an
astronomical image (there are 1,000$\times$ more 3$\sigma$ than
5$\sigma$ detections) LSD detections are \emph{not} stored in the
MOPS relational database. Instead, they are stored in flat files on
the local or network file system.  The LSD archive is designed to be
compact and efficiently searchable by exposure epoch, right ascension
and declination. During only the attribution and precovery phases of
MOPS processing, candidate 3$\sigma$ recovery observations are
extracted for analysis from this dataset based on predicted positions,
and successful recoveries are than `promoted' into the high-confidence
database.

\subsection{Data Exchange Standard}

Significant work related to early MOPS development and simulations 
led to creation of the Data Exchange Standard (DES) for
Solar System Object Detections and Orbits\footnote{see 
  \PS\ document PSDC-530-004 by Milani \etal, {\it Data Exchange
    Standard (2.03) for solar system object detections and orbits: A
    tool for Input/Output definition and control.}}. The DES describes
a file format for dissemination of observations and orbits of solar
system objects and is used within MOPS and between MOPS and add-on
MOPS software developed by A. Milani and the OrbFit Consortium \citep{Milani2008,Milani2012}. It
provides mechanisms for reproducing detection$\Leftrightarrow$tracklet and
tracklet$\Leftrightarrow$derived object relations that exist in the MOPS database, and
allows for specification of statistical uncertainties for
observational measures.  The DES includes provisions for propagating
\begin{itemize}
\item orbit covariance and normal matrices,
\item detailed per-detection residuals,
\item ephemerides for real and synthetic asteroids,
\item detailed tracklet metrics beyond the scope of MOPS, and
\item radar and spacecraft observations.
\end{itemize}

\subsection{Web Interface}

Early in MOPS development we realized the utility of visual, web-based
interrogation of the MOPS database, leading to the creation of the
MOPS \emph{web interface}.  The MySQL RDBMS provides software `hooks'
into its internal code that allow web-friendly scripting languages to
perform queries so that web-based user interfaces and reporting tools
can be developed rapidly.

\begin{figure}[btp]
\centering
\plotone{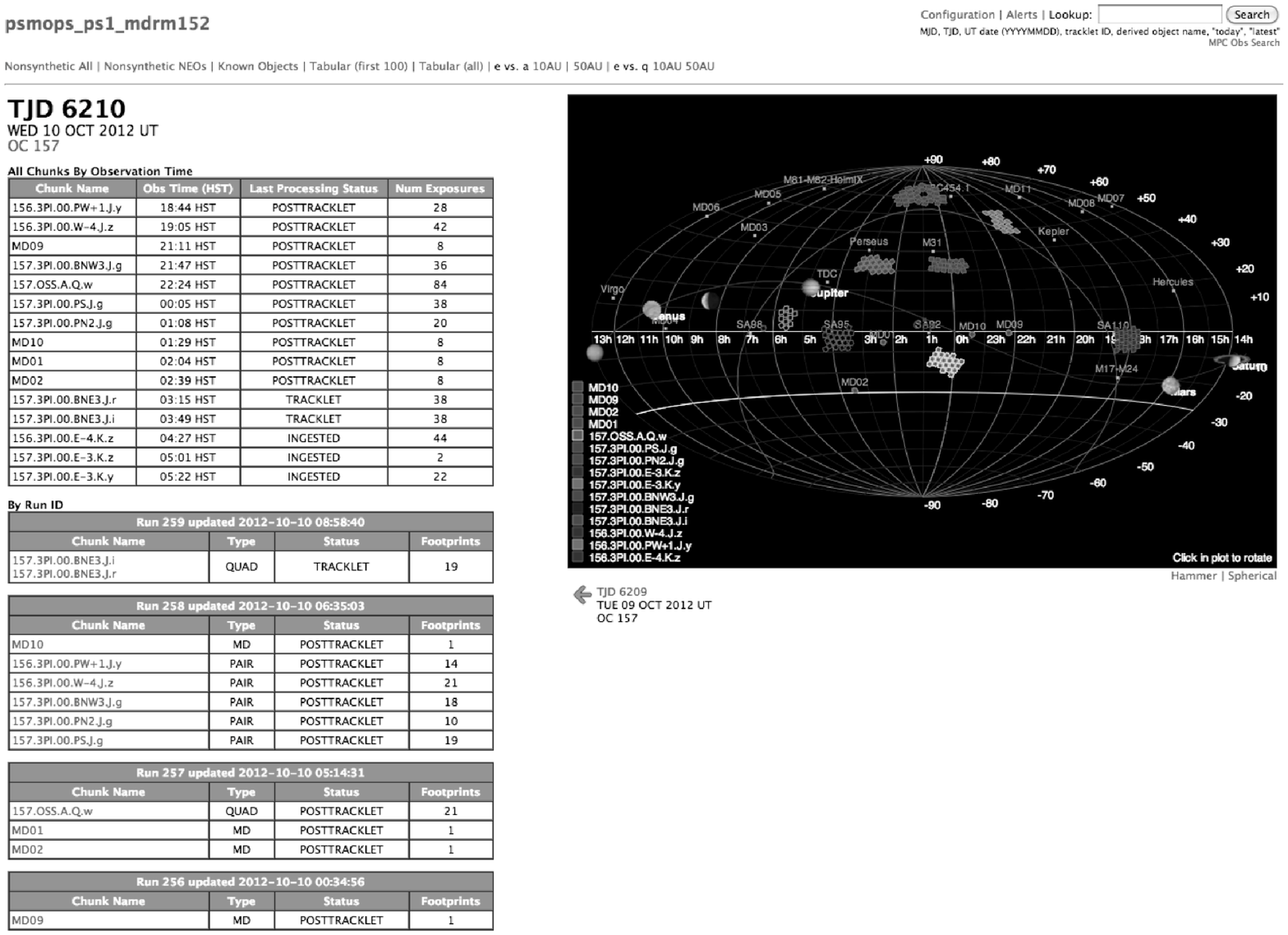}
\caption{\PSone\ MOPS web interface showing the sky map for
  the night of MJD 56210. On the left are lists of observing blocks
  executed by the telescope for the night.}
\label{f.webinterface1}
\end{figure}

The MOPS database user has ready visual representation of the MOPS processing stream, from a sky map of the nightly survey
pattern, detection maps of each observed field  indicating both
synthetic and real detections, to the orbit distributions of derived
objects, to the efficiency and accuracy of each of the processing
steps.  Pages for a single MOPS derived object show the object's linking
history and the evolution of its orbit at each step.
Fig. \ref{f.webinterface1} shows the nightly sky map of an active MOPS
database.

\subsection{Data Export}

The primary consumer of MOPS exported data products is the IAU Minor
Planet Center (MPC) that maintains the authoritative catalog of
observations and orbits for minor planets, comets and natural
satellites.  As of early 2013, the MPC prefers email data submission
because the \PSone\ survey is oriented toward reporting single-night
tracklets, not multiple-night derived objects.

Under the \PSfour\ processing model where MOPS produces derived
objects with secure orbits on a regular basis, MOPS is capable of
exporting monthly DES catalogs of detections, tracklets, orbits and
indentification records for known and unknown derived objects. Pilot
submissions to the MPC showed that this export and publication model
works as long as the false derived object rate is suitably low
(\eg\ $< 0.1$\%; Spahr, personal communication).

Early \PSone\ processing revealed a false derived object rate too high
for automatic submission to the MPC and an unsatisfactory NEO
discovery rate so, to improve the short-term science output and
satisfy funding agency requirements, the PS1SC abandoned the derived
object processing model in favor of the traditional tracklet reporting
method employed by other asteroid surveys.  Processing of derived
objects will be restored for the final post-survey MOPS processing but
the derived object throughput will be small due to the \PSone\ survey
strategy being optimized for single-night NEO discovery.

\begin{figure}[btp]
\centering
\plotone{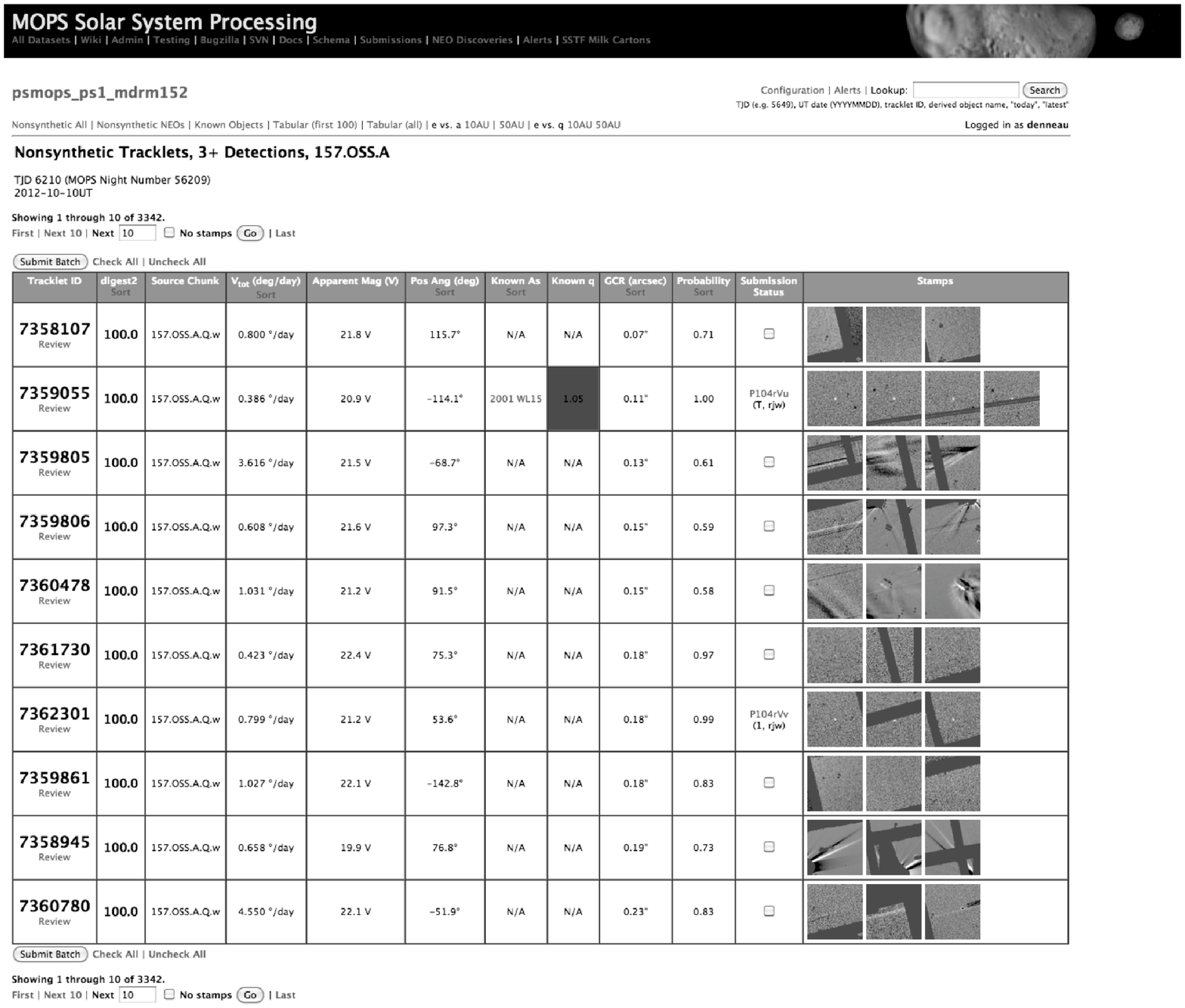}
\caption{\PSone\ MOPS web interface showing the NEO `czar'
  page. From this page, NEO candidates can be submitted to the IAU
  Minor Planet Center.  Rows 2 and 7 are 2001~WL$_{15}$ and 2012~TN$_{139}$ respectively.}
\label{f.webinterface2}
\end{figure}

Detections are distributed to the
MPC via email after a human `czar' visually verifies that candidates
are real using the MOPS web-based interface.  (NEO candidates consume
most of the czar's time because of the high rate of false tracklets at
high rates of motion.) In this display, a page lists nightly probable
real asteroid tracklets with highlighted NEO candidate tracklets (see
Fig. \ref{f.webinterface2}). Tracklets that belong to known numbered
and multi-opposition objects are identified by a MOPS add-on module
called \knownserver\ \citep{Milani2012} and automatically submitted to
the MPC. A MOPS administrative email account receives MPC confirmation
email to verify the MOPS submissions. MOPS maintains its own table of
submitted tracklets to the MPC for easy reproduction of submitted
detection details and to prevent multiple submissions of the same
tracklets.

\subsection{Alert System}

The MOPS Alert System is a software agent that runs independently of
standard MOPS processing and searches for events that require
immediate notification or follow-up. The specific type of event is
user-configurable using a small piece of Python code written by a MOPS
scientist or engineer to query for interesting events. A typical
example would be a fast-moving near-earth object (NEO) candidate
tracklet that is unlikely to be observed again by the survey telescope
and/or needs immediate follow-up so that it is not lost. In this case,
the alert system identifies the candidate tracklet by its sky-plane
rate of motion and the fact that the tracklet's positions cannot be
attributed to a known asteroid. Additional examples of alerts are
tracklets of unknown objects with comet-like motion or extended
morphology, or known objects with magnitude anomalies that indicate
activity.

The alert system allows alerts to be triggered from either a MOPS
derived object or an individual tracklet. The alert payload itself is
an extensible markup language (XML) VOEvent \citep[\eg][]{Seaman2011},
a standardized data structure that can be consumed by automatic data-processing
agents.  The alert deployment
infrastructure can be changed easily; the current MOPS
implementation uses both a restricted internet email distribution list and a private Twitter
feed to PS1 Science Consortium partners.

Alerts are organized using a publish-subscribe paradigm in which
multiple alert definitions can be published to various `channels'
that may be subscribed to by multiple users. For example, a `NEO
channel' might publish alerts for near-earth object alerts ($q < 1.3$
AU) and potentially hazardous objects (NEOs with $H < 22.5$ and minimum orbital intersection distance
$<0.05$~AU).  Such a channel could be subscribed to by NEO follow-up
organizations.  Alerts currently in use by the PS1SC include
unusual lightcurves within a single night and detections of objects
from asteroid families of interest (\eg\ Hildas).

\subsection{Synthetic Solar System Model}
\label{ss.SyntheticSolarSystemModel}

One major MOPS innovation is the incorporation of a synthetic solar
system model \citep[S3M;][]{Grav2011} that allows us to monitor
MOPS development and measure its performance in real-time
operations as quantified by the metrics defined in \S\ref{s.Overview}.
It is important to introduce realistic detections of solar system
objects into the processing system to ensure that it can handle real
objects and to measure a realistic detection efficiency.  The solar
system model has been described in detail by \citet{Grav2011} and we
only provide a brief overview here.

The S3M is a comprehensive flux-limited model of the major small body
populations in the solar system that consists of objects ranging from
those that orbit the Sun entirely interior to the Earth's orbit to
those in the Oort cloud.  It even includes interstellar objects
passing through the solar system on hyperbolic orbits.  The S3M
contains a total of over 13 million synthetic objects from 11 distinct
small body populations with the only requirement being that they reach
$V\le24.5$ during the time period from roughly 2005-2015 (the
anticipated operational lifetime of the \PS\ survey).  The time period
requirement only affects a sub-set of the populations in the distant
solar system.

The S3M has proven invaluable for developing and testing MOPS but
suffers from one major limitation --- it only tests the software
system for known types of objects.  Since we cannot anticipate the
properties of an unknown population we supplemented the S3M with an
artificial 575,000-object `grid' population.  Grid objects have a random and flat
distribution in eccentricity ($0\le e<1$), $\sin$(inclination) ($0^\circ\le i
< 180^\circ$) and in the angular elements ($0^\circ\le \Omega, \omega,
M <360^\circ$).  The semi-major axes of the objects were generated
using uniform random distributions over six different ranges as
provided in table~\ref{t.gridSemiMajorAxis}.  The ranges and number
of objects were selected so that the sky-plane density of objects in
each range is roughly equal and to provide denser coverage in the
semi-major axis ranges that have large numbers of known objects.
The absolute sky-plane density of the grid population is ~50 per
field, a factor of 20-30 less than the realistic S3M on the ecliptic
and much higher than the S3M off the ecliptic.

The absolute magnitudes of the objects in the grid population were
generated with uniform distributions in each semi-major axis range but
with lower limits that decrease (\ie\ the objects are made larger)
according to the square of the semi-major axis so that the objects are
visible from Earth.

A major benefit of testing MOPS with the grid population is that they
can appear at the poles where the sky-plane density of real or S3M
objects is small.  \S\ref{ss.GridSimulations} discusses the results
of simulations with the grid population.

\subsection{MOPS Efficiency Concepts}
\label{ss.MOPSEfficiencyConcepts}

During $\sim4$ calendar years of MOPS development we continuously benchmarked
its performance using 3 metrics for each MOPS subcomponent.  The metrics count the correct and incorrect associations of synthetic sources available at each MOPS processing step.  The metrics are:

\begin{itemize}

\item Efficiency\\
The fraction of available associations that were correctly
identified.

\item Accuracy\\
The fraction of associations that are correct, \eg\ consisting of detections from the same synthetic object definition.

\item Quality\\ 
The fraction of correct associations that meet or exceed a pre-defined quality metric.

\end{itemize}

To provide the detailed accounting required to measure system
efficiency and accuracy, data structures for each association created in a MOPS processing
phase are decorated with a label describing their disposition after
processing. These labels describe the detections incorporated in the
resulting data structure:

\begin{itemize}
\item {\tt AVAILABLE} - the data structure should have been created in the
  processing phase and is available to be assigned.
\item {\tt CLEAN} - the data structure contains only synthetic detections or
  tracklets that belong to the same synthetic object, \ie\ is a
  `correct' data structure. This label might describe a tracklet that
  is created from two detections of the same synthetic object, or a
  derived object consisting of three tracklets that are each {\tt CLEAN} and
  belong to the same synthetic object.
\item {\tt MIXED} - the data structure contains only synthetic objects but
  the detections or tracklets are from different synthetic objects.
\item {\tt BAD} - the data structure contains both synthetic and
  nonsynthetic detections. For tracklets this is a common and normal
  occurrence since we expect that synthetic detections will occasionally
  fall near real detections.
\item {\tt UNFOUND} - an expected data structure for a synthetic object was
  not created. {\tt UNFOUND} data structures are `dummy' structures that
  represent an operation that \emph{should have} happened. For
  example, if two sequential fields at the same boresight contain two
  detections of the same asteroid but a tracklet is not created, a
  dummy tracklet containing tracklet metadata is created so that the
  event can be counted and characterized.
\item {\tt NONSYNTHETIC} - the data structure contains `real' detections
  from actual telescope data.  {\tt NONSYNTHETIC} data consists of
  true detections of asteroids and false detections from image
  artifacts.
\end{itemize}

The pipeline must `peek' at the input and output data structures
before and after a MOPS processing step to assign synthetic labels but
this is the only time that the pipeline knows that a data structure is
synthetic --- the labels are never examined during any MOPS
computation or algorithm.  Data structures that are `contaminated' ---
contain mixtures of real and synthetic detections or different
synthetic objects --- continue to be propagated through the
pipeline. The only exception is for {\tt UNFOUND} objects for which dummy
structures are created and decorated with the {\tt UNFOUND} label so that
these occurrences can be counted by analysis tools.

MOPS uses the labels to calculate the incremental efficiency after
each processing stage.  For a derived object, the state at each processing stage is preserved so that a `paper trail' exists for each modification.
This allows a complete step-by-step reconstruction of a derived
object's history from its tracklets including the ability
to indentify the exact processing step where an object was lost or
incorrectly modified by MOPS.

\subsection{Simulations}
\label{ss.Simulations}

The MOPS pipeline is exercised through the creation and execution of a
`simulation'. Each simulation begins with a collection of telescope
pointings called `fields' and a population of synthetic solar system
objects whose positions and magnitudes are computed for every
field. Objects that `appear' in fields are converted into detections
and stored in the MOPS database as though they were reported from an
actual telescope. Detections can be `fuzzed' astrometrically and
photometrically using parameters that model the telescope
performance; these include sky background, detector noise, and the
point-spread function (PSF). The full
fuzzing model incorporates a baseline astrometric uncertainty from the
plate solutions added in quadrature with a flux-dependent
positional uncertainty.  Poisson-distributed (in \sn) false detections may be inserted in the
fields to simulate noise from a real detector.  After the injection of
synthetic detections for a night the MOPS pipeline is invoked as
though operating on real data.

MOPS simulations are used both to verify correct operation of the
pipeline and to interrogate performance of a hypothetical survey. A
simple verification test consists of creating a small
main belt object (MBO) simulation with several nights of synthetic
fields spanning two lunations, executing the pipeline, and verifying
that all expected MBOs were `discovered' through creation of a derived
object and `recovered' via attribution or recovery where possible.  In
this fashion all essential elements of the pipeline are exercised and
verified after software modifications.

For comprehensive testing, a larger, more realistic synthetic population of
objects is inserted into the simulation, the MOPS pipeline is
executed, and the output provides a quantitative assessment of the
system's efficiency and accuracy. The pipeline is designed to be
essentially 100\% efficient at creating tracklets and derived objects
for MBOs with the designed \PS\ performance.

When evaluating survey performance we configure MOPS to more
precisely mimic the astrometric and photometric
characteristics using filter- and field-dependent limiting magnitudes
and FWHMs. Given the calculated $V$-band apparent magnitude for a
synthetic asteroid, we calculate its expected signal-to-noise (\sn)
using per-field detection efficiency parameters, and then generate a
fuzzed magnitude and \sn. From the fuzzed \sn\ and the field's FWHM we
compute the asteroid's fuzzed right ascension and declination. If
there is variation in sensitivity due to sky brightness or airmass,
these effects can be modeled by supplying appropriate values for
photometric zeropoint or sky noise in the field's detection efficiency
parameters.

During MOPS development we ran thousands of small- to medium-scale
simulations for software unit testing and verification.  More
importantly, we ran several large simulations with synthetic solar
system models containing millions of asteroids spanning several years
of synthetic PS1 observations. These simulations were run on our
modest PS1 MOPS production cluster and took many weeks or months
to simulate the multi-year \PS\ mission.  For
instance, our {\tt full\_S1b} simulation used the full S3M
population in a 2~year simulation and required $\sim$180 calendar days
(see \S\ref{ss.2-yearS3M}) while our four year NEO-only simulation
({\tt neo\_4yr}) required 88 calendar days (see
\S\ref{sss.4-yearNEO}). At various times during the pipeline
processing the cluster was offline for maintenance or power outages
but MOPS is designed to be interrupted and restarted.

\begin{figure}[btp]
\centering
\epsscale{0.8}
\plotone{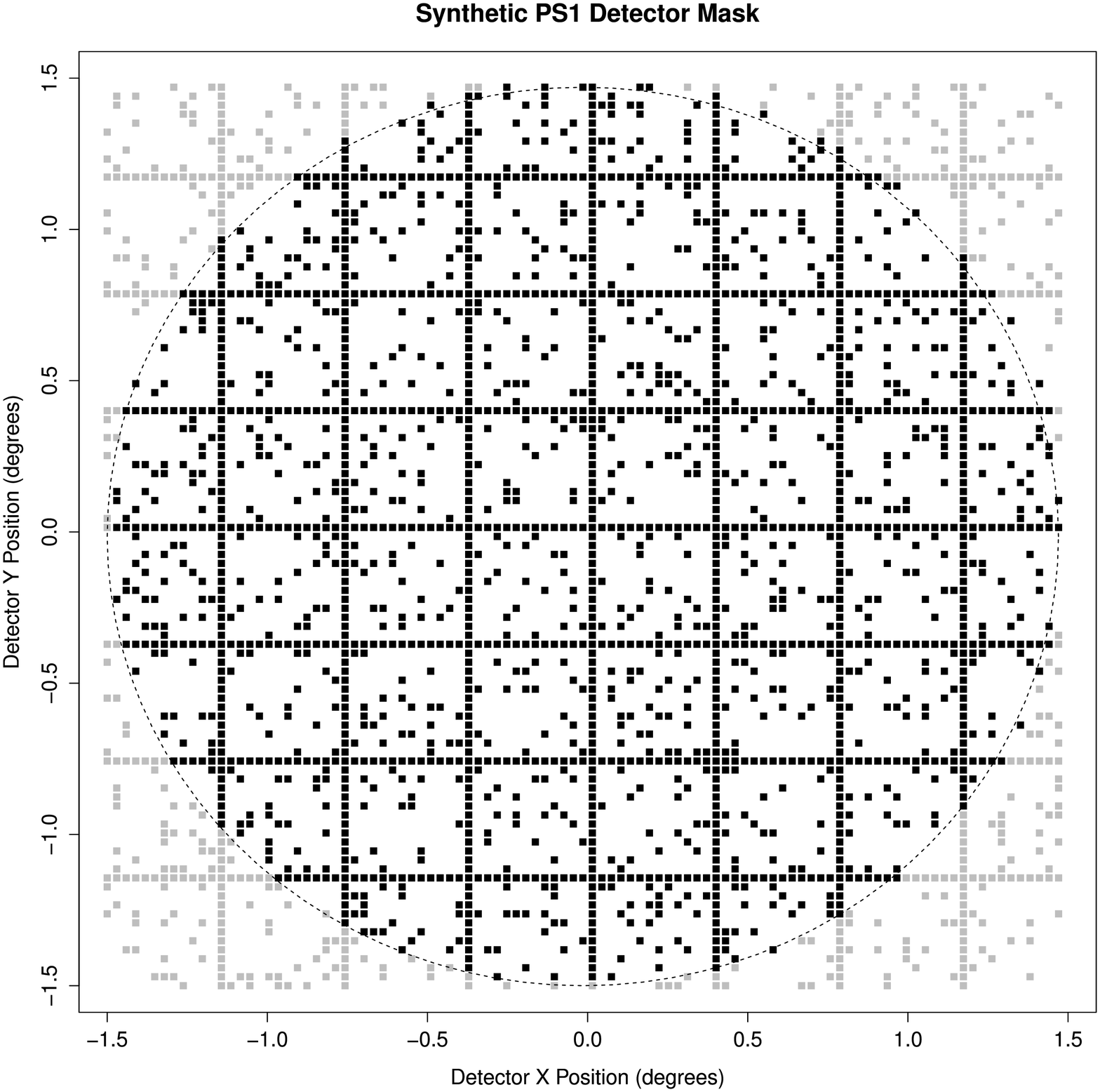}
\caption{Synthetic 7.0 deg$^2$ field-of-view static focal plane mask
  for \PSone\ MOPS simulations that simulates losses due to chip gaps
  and other masked area. The horizontal and vertical bands simulate
  the chip gaps while the remaining squares represent losses due to
  bad cells or those used for guide stars. The squares do not
  correspond to actual masked area on the real \PSone\ detector; they
  exist simply to simulate additional area loss.  The actual
  distribution of masked area on the real \PSone\ detector is very
  different --- there is more power on small scales than shown
  here. The overall simulated camera fill factor is fixed at 75\% in
  agreement with the measured values shown in
  Fig.~\ref{f.efficiency-vs-mjd}.}
\label{f.fillfactor75}
\end{figure}

The upstream image processing subsystem (\eg\ the \PS\ IPP) can
improve MOPS fidelity by providing a Live Pixel Server (LPS) that
reports whether a pixel at or near a specific (RA,dec) in an image
falls on `live' camera pixels.  This service improves MOPS simulations
in two ways: 1) by generating synthetic moving object data with
behavior that simulates real observations (\eg\ to account for the
loss of field due to the camera fill factor) and 2) accounting for
missing detections when attempting to recover known asteroids that are
in the field of view.  In a single \PSone\ exposure there are many
reasons an area of the detector can be inactive besides simple chip
gaps, such as saturation from a bright object in the field of view, or
a cell used for guide star tracking. Thus the LPS tells MOPS what
areas of the detector are active for any given exposure. If a LPS is
not available from the upstream image processing subsystem, MOPS has
provisions for allowing static specification of a detector mask that
defines areas on a detector where detections cannot occur (see
Fig.~\ref{f.fillfactor75}).  Synthetic detections that land on a dead
part of the detector (as determined by the LPS) are marked with a
special `unfound' decorator so that they can be accounted for but
otherwise omitted from all subsequent processing.

For increased simulation fidelity MOPS can accept a per-field detection efficiency for point sources and trailed
detections as a function of \sn\ and rate of motion so that MOPS can
generate more realistic synthetic photometry.

\section{MOPS Verification}
\label{s.MOPSVerification}

During MOPS development we undertook several large-scale simulations
to probe the performance limits of the MOPS design and implementation
and understand where further attention was needed in software
development.  These simulations consisted of a complete S3M or large
S3M subpopulation, a multi-year set of telescope pointings, and
detector performance that simulated a \PSfour\ system.  Our motivation
was to investigate MOPS performance for the \PSfour\ system as
specified in its design requirements, not the reduced, changing
performance of the single \PSone\ prototype telescope undergoing
development and commissioning.

\subsection{2-year S3M simulation ({\tt full\_S1b})}
\label{ss.2-yearS3M}

The MOPS 2-year S3M simulation consists of the full S3M and 100,424
fields spanning 245 distinct nights over nearly two years (29 Dec 2007
through 2009 Oct 22). The simulation took $\sim$3 months to execute
including various interruptions for cluster downtime and
maintenance. It assumes a nominal positional measurement uncertainty of
0.01$\arcsec$, a FWHM of 0.7$\arcsec$, a fill factor of 100\%, and a
constant limiting magnitude of $R=22.7$, representing a single
idealized \PSone\ telescope. Poisson-distributed false detections were
added to each field at a density of $\sim$200/deg$^2$. The simulated
field pointings were generated by the TAO survey
scheduler\footnote{\copyright 1999-2006 Paulo Holvorcem;
  http://sites.mpc.com.br/holvorcem/tao/readme.html} configured to
produce a \PS-like survey covering $\sim$3,600 deg$^2$ near opposition
and $\sim$600 deg$^2$ near each of the morning and evening
sweetspots. Each area is visited three times per lunation, with some
nights lost due to simulated poor weather. MOPS achieved
an overall 99.99997\% tracklet efficiency and 99.26\% derived object
efficiency. Tables~\ref{f.full_S1b_trk_eff} and \ref{f.full_S1b_eff} detail the tracklet and
derived object efficiencies per S3M sub-population.

For classes other than NEOs, Jupiter-family comets(JFCs) and long-period comets
(LPCs), MOPS has an efficiency better than 99\% for converting three or more tracklets to
a differentially corrected orbit when observed within a 14 day window.  For NEOs our performance is not as
good, but as stated elsewhere in this work, we have been more concerned
with quantifying MOPS performance for various populations than in
optimization. MOPS contains many runtime configuration parameters that
allow performance of individual modules to be tuned, often at the cost
of greater false results. \PSone\ operations have led
the project to a different mode of NEO discovery than originally
envisioned, so we have not focused on improving NEO derived object
performance. Similarly for LPCs, we can again improve efficiency by
tuning MOPS orbit determination modules to reject fewer parabolic or
hyperbolic orbits.

Our differential correction performance (\S \ref{ss.OrbitDetermination}) is probably overstated
because the OrbFit initial orbit determination (IOD) software 
\citep{Milani2005,Milani2010} employed by MOPS can perform its own differential
correction after IOD, and we elect to use it in this mode.  Therefore 
orbits handed to the JPL differential correction module are
almost always close to a minimum in the solution space where
convergence will occur.  MOPS supports other packages besides OrbFit
to perform IOD, and in early evaluation of these packages we saw
outstanding performance from the JPL differential corrector.

\subsection{4-year NEO simulation ({\tt neo\_4yr})}
\label{sss.4-yearNEO}

We ran the {\tt neo\_4yr} simulation to calculate upper limits on the
NEO detection and discovery rates with a \PSone-like system.  This
system differed from the {\tt full\_S1b} simulation by using a
0.1$\arcsec$ baseline astrometric uncertainty and by containing almost four years of TAO
simulated opposition and sweetspot observations over the expected \PSone\ mission from March
2009 through January 2013.  The simulation only used the 268,896 NEO
subset of the S3M. We were not able to simulate the \PSone\ focal
plane fill factor, as the code had not been incorporated into MOPS
yet, but we felt it would be reasonable to use a 100\% fill factor and
scale down our results based on true \PSone\ detector
performance. False detections were not added to this simulation to
speed up execution time, but the {\tt full\_S1b} results showed that
this would not impact our results.

The {\tt neo\_4yr} simulation found over 9,405 derived objects of
which nearly half had orbital arc lengths greater than 30 days.  As
with the 2-year {\tt full\_S1b} simulation the tracklet efficiency is
essentially 100\%, with 10 out of 105,439 tracklets lost.  MOPS
linking performance is similar to {\tt full\_S1b} (see
table~\ref{f.neo_4yr_eff}) but we find that orbit determination
performs less well, at 81.7\%, purely due to greater RMS residuals
across the orbit from the larger astrometric uncertainty. Again, this
acceptance threshold can be relaxed to improve efficiency (fraction of
correct linkages surviving orbit determination) at a cost of an
increase in false linkages.

\subsection{Grid Simulations}
\label{ss.GridSimulations}

We have run several grid simulations to ensure
that we were not tailoring the efficiency to localized regions in
$(a,e,i)$ phase space.  However, we do not inject the grid population
into the detection pipeline during normal MOPS operations as this
would needlessly increase the tracklet linking combinatorics in an
artificial and unrepresentative manner.  

\begin{figure}[btp]
\centering
\plotone{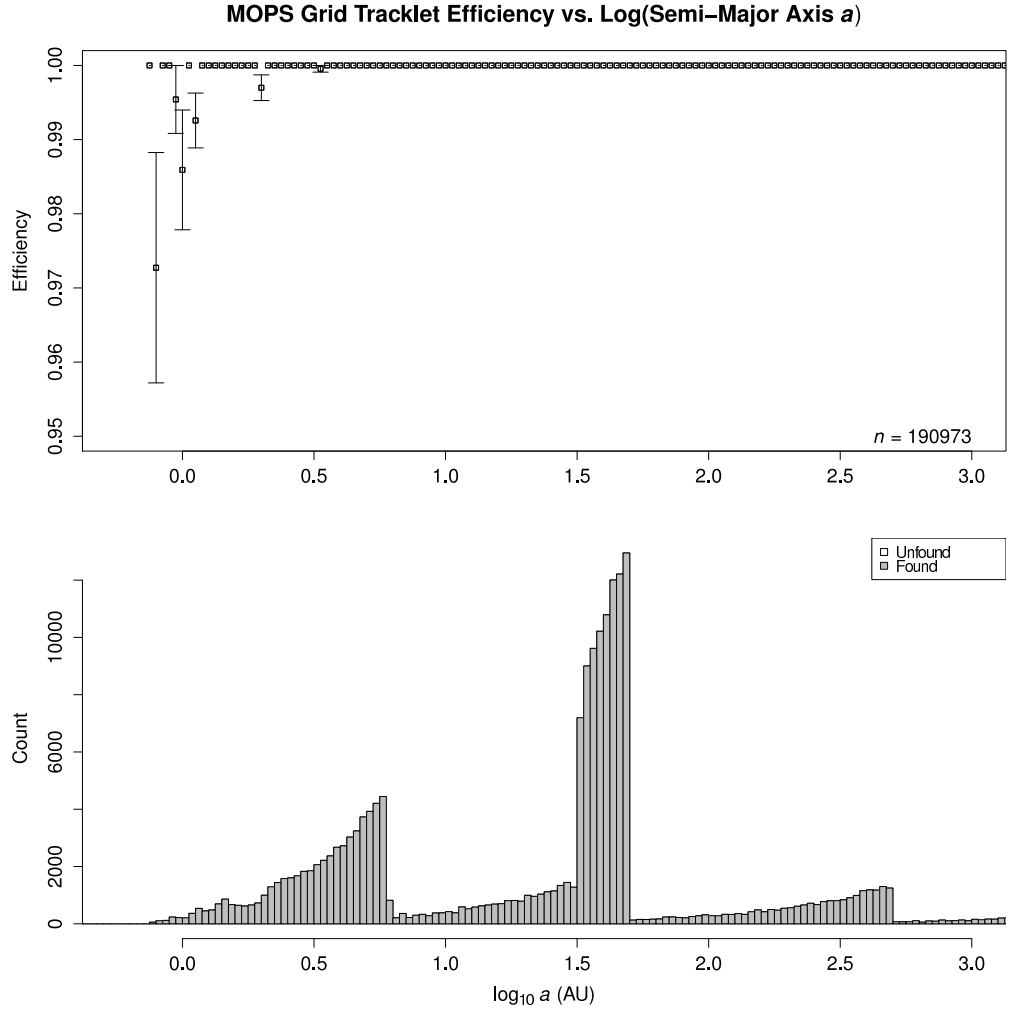}
\caption{Tracklet creation efficiency as a function of semi-major axis
  for a one lunation MOPS simulation using S3M grid objects. Note the
  truncated Y-axis range of 0.95 to 1.0. Fifteen fast-moving objects
  with sky-plane rates that exceed the configured limits for the
  production \PSone\ MOPS account for the reduced efficiency at small
  semi-major axis. These limits will be tuned upon reprocessing the
  full \PSone\ survey.}
\label{f.S3MGridTrackletBya}
\end{figure}

\begin{figure}[btp]
\centering
\plotone{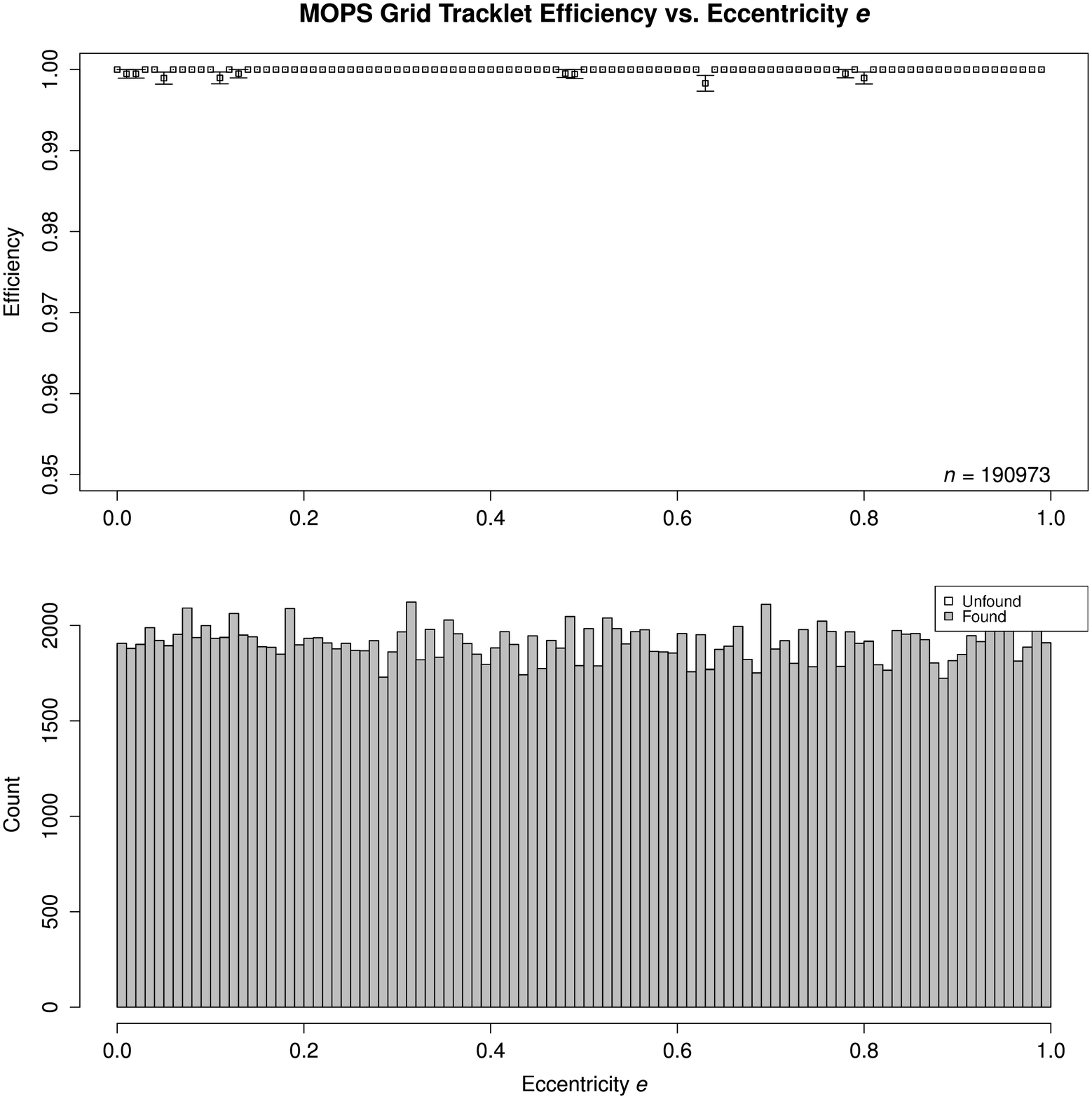}
\caption{Tracklet creation efficiency as a function of eccentricity
  for a one lunation MOPS simulation using S3M grid objects. Note the
  truncated Y-axis range of 0.95 to 1.0.}
\label{f.S3MGridTrackletBye}
\end{figure}

\begin{figure}[btp]
\centering
\plotone{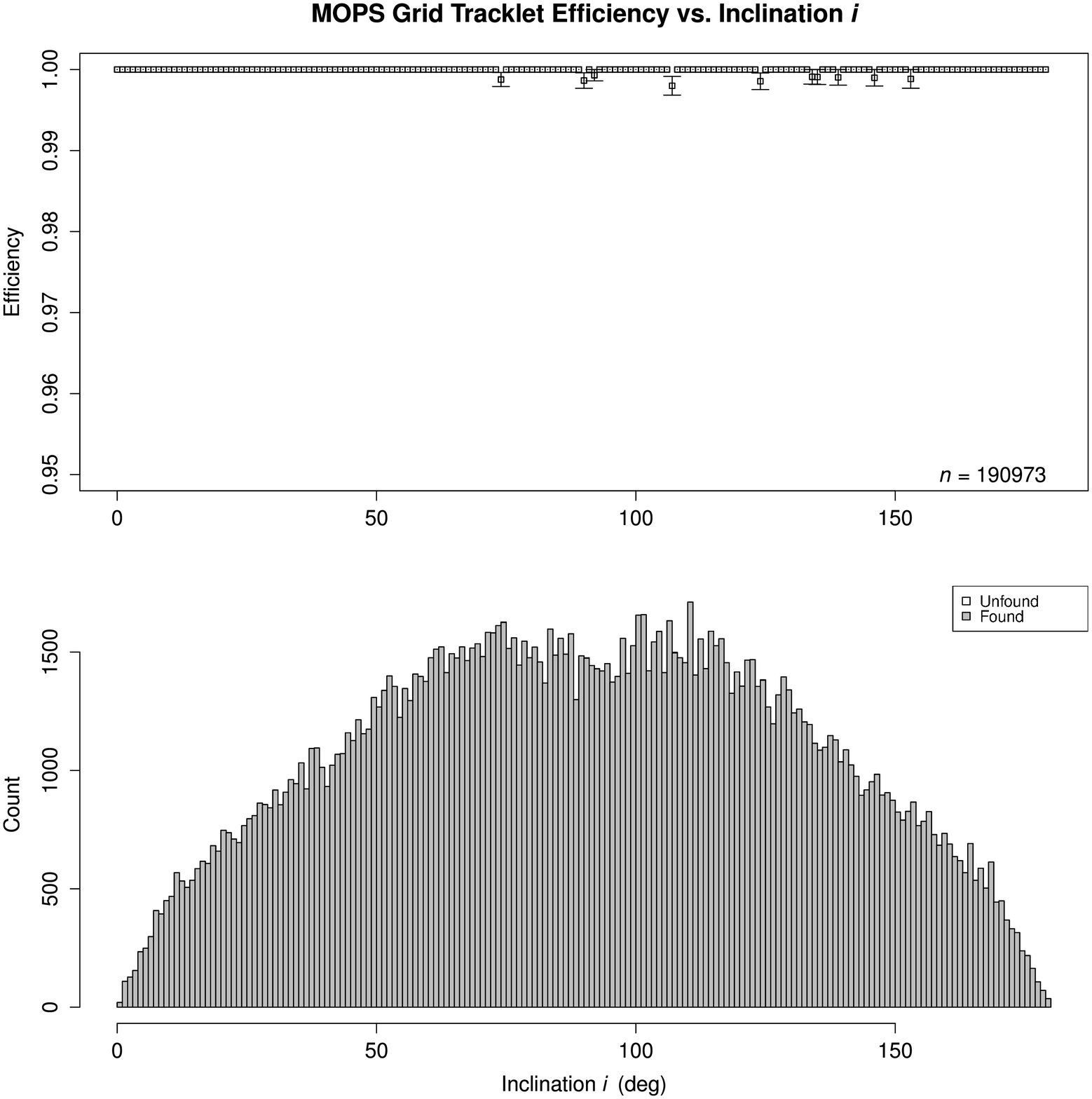}
\caption{Tracklet creation efficiency as a function of inclination for a one lunation MOPS simulation using S3M grid objects. Note the truncated Y-axis range of 0.95 to 1.0.}
\label{f.S3MGridTrackletByi}
\end{figure}

To illustrate MOPS operations over the entire phase space we ran the
real \PSone\ survey (\ie\ actual pointings and cadence) for two
consecutive lunations with the grid population and a realistic
($0.1\arcsec$) astrometric error model.
Figures~\ref{f.S3MGridTrackletBya} through \ref{f.S3MGridTrackletByi}
show that MOPS is $\sim100$\% efficient for semi-major axes in the
range from 0.75\,AU TO 5,0000\,AU, eccentricities in the range $[0,1]$
and all possible inclinations.  Fifteen fast-moving objects were
rejected (out of 190,973) by the rate limit of 5.0~deg/day in the
current \PSone\ production configuration to reduce false tracklets.

\begin{figure}[btp]
\centering
\plotone{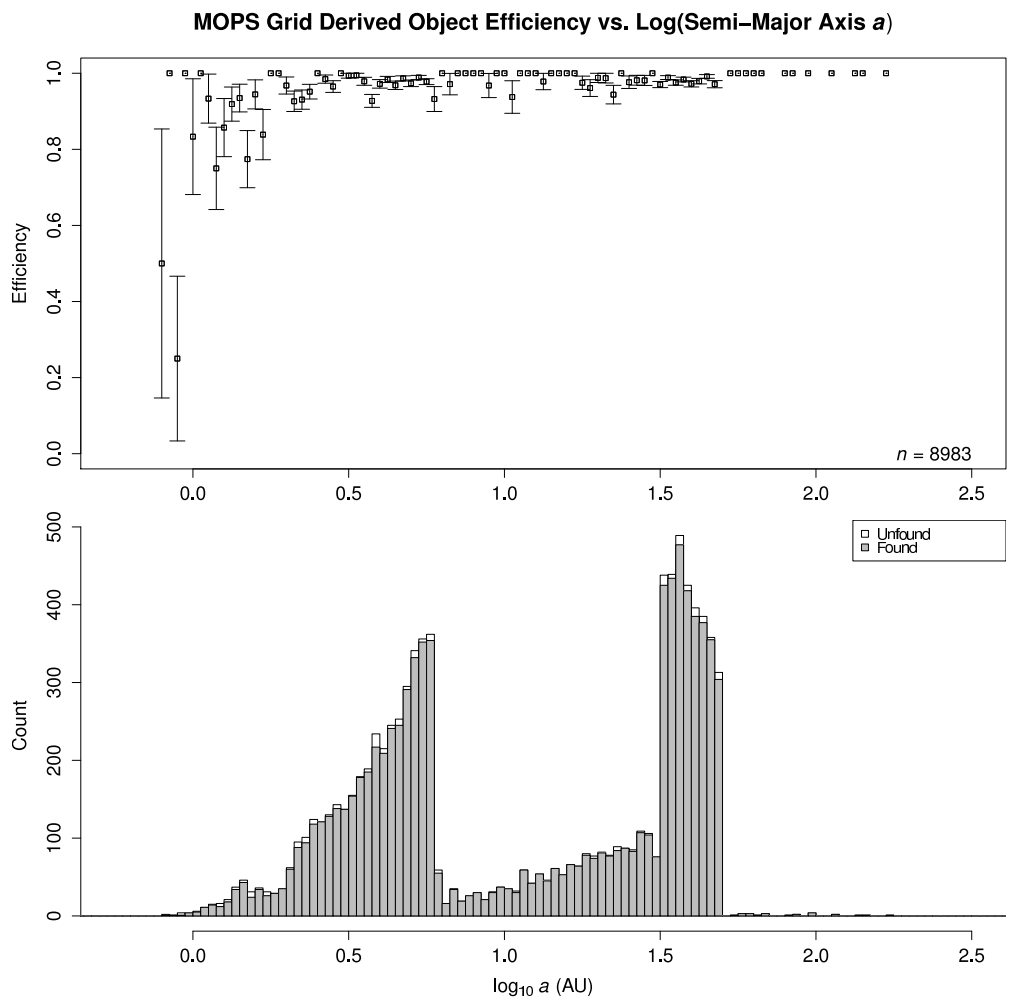}
\caption{Derived object creation efficiency as a function of
  semi-major axis for a one lunation MOPS simulation using S3M grid
  objects. The total efficiency is 97.5\%.  Losses at small semi-major
  axis are due to conservative acceptance limits in the current
  \PSone\ production MOPS that reject objects with large sky-plane
  accelerations and/or large RMS residuals. These limits
  will be tuned upon reprocessing the full \PSone\ survey.}
\label{f.S3MGridDOBya}
\end{figure}

\begin{figure}[btp]
\centering
\plotone{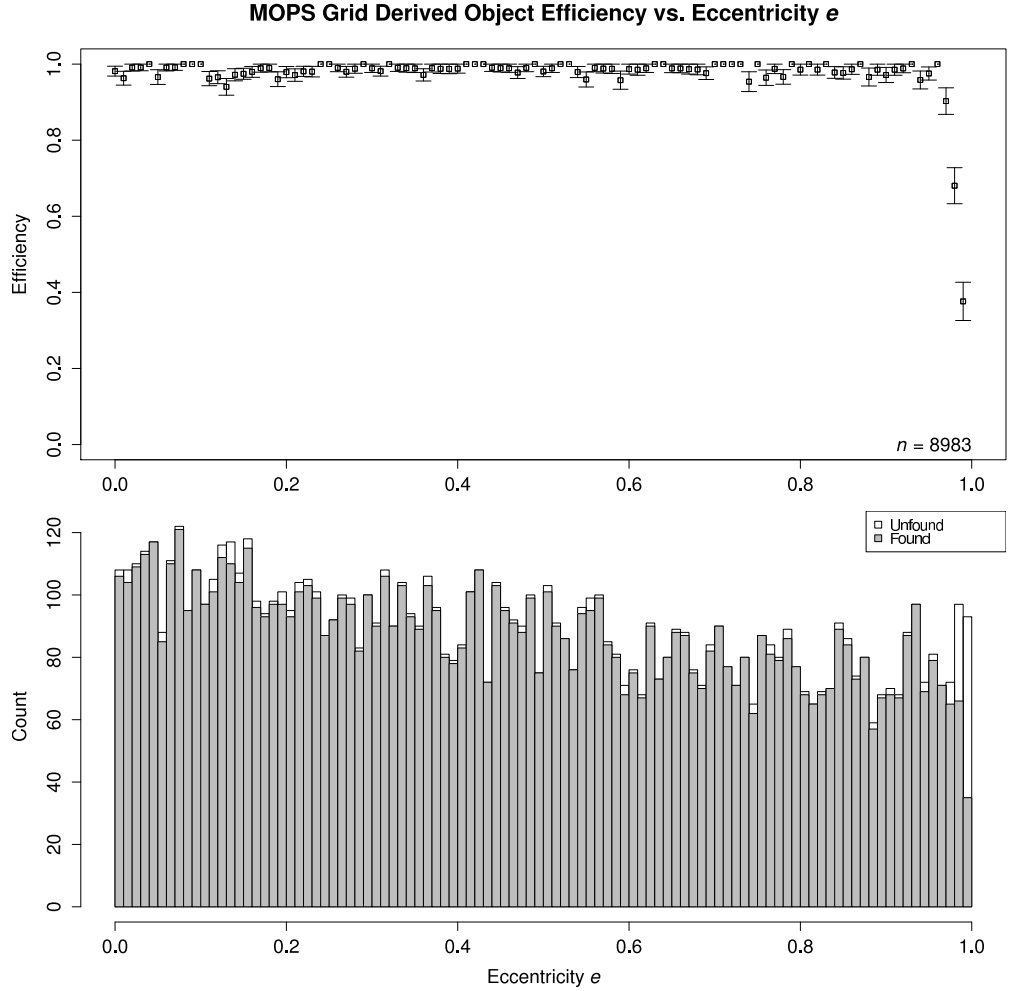}
\caption{Derived object creation efficiency as a function of
  eccentricity for a one lunation MOPS simulation using S3M grid
  objects. The failures at high eccentricity ($e > 0.99$) occur in
  initial orbit determination (IOD) for grid objects with semi-major
  axis $a \lesssim 5$. These failures can be corrected by modifying
  IOD acceptance parameters.}
\label{f.S3MGridDOBye}
\end{figure}

\begin{figure}[btp]
\centering
\plotone{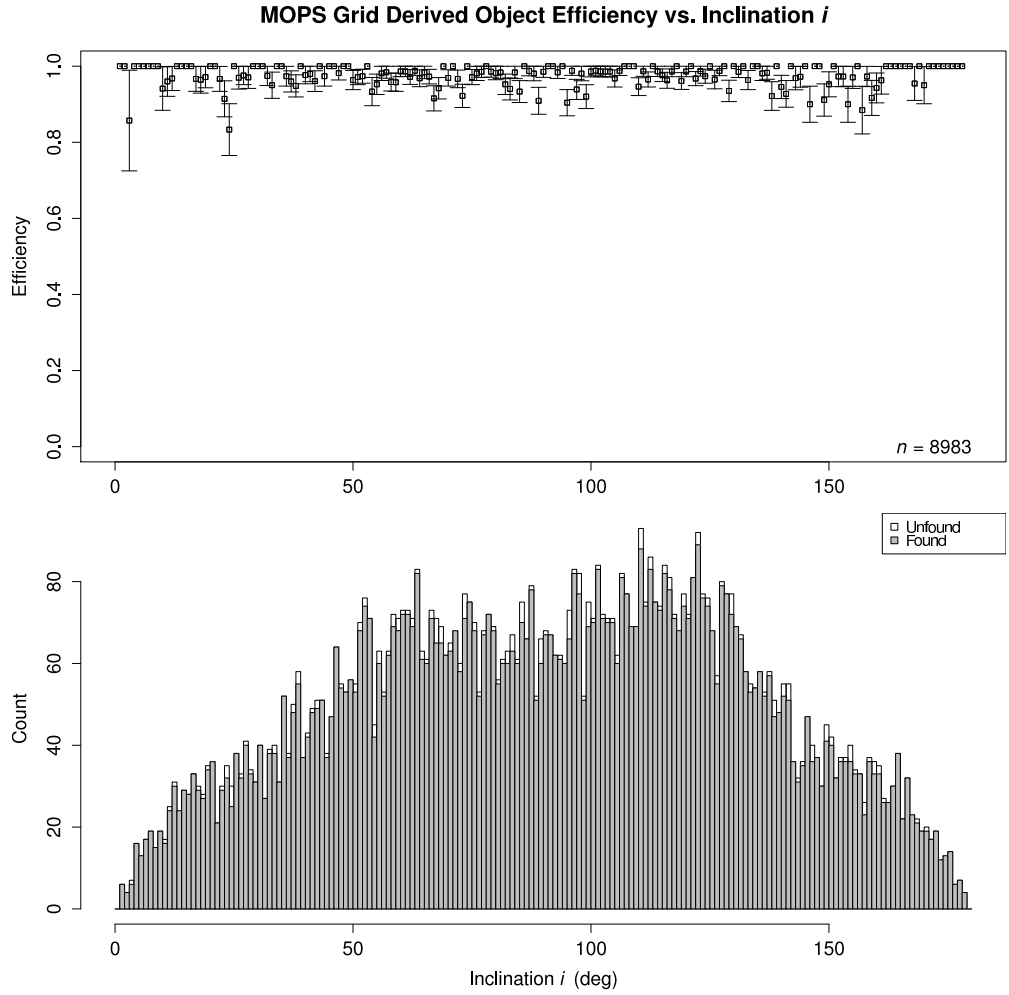}
\caption{Derived object creation efficiency as a function of
  inclination for a one lunation MOPS simulation using only grid
  objects.}
\label{f.S3MGridDOByi}
\end{figure}

Figures~\ref{f.S3MGridDOBya} through \ref{f.S3MGridDOByi} show that the 
derived object efficiency is consistently high for all synthetic grid objects on orbits ranging
from those with semi-major axes well within the Earth's orbit to
beyond 100~AU, over all inclinations with no deterioration in the
efficiency for retrograde orbits, and for nearly circular orbits to
those with $e \la 1$.  There is a degradation in the efficiency for $e
\sim 1$ due to the orbit determination failing at the highest
eccentricities.

\begin{figure}[btp]
\centering
\plotone{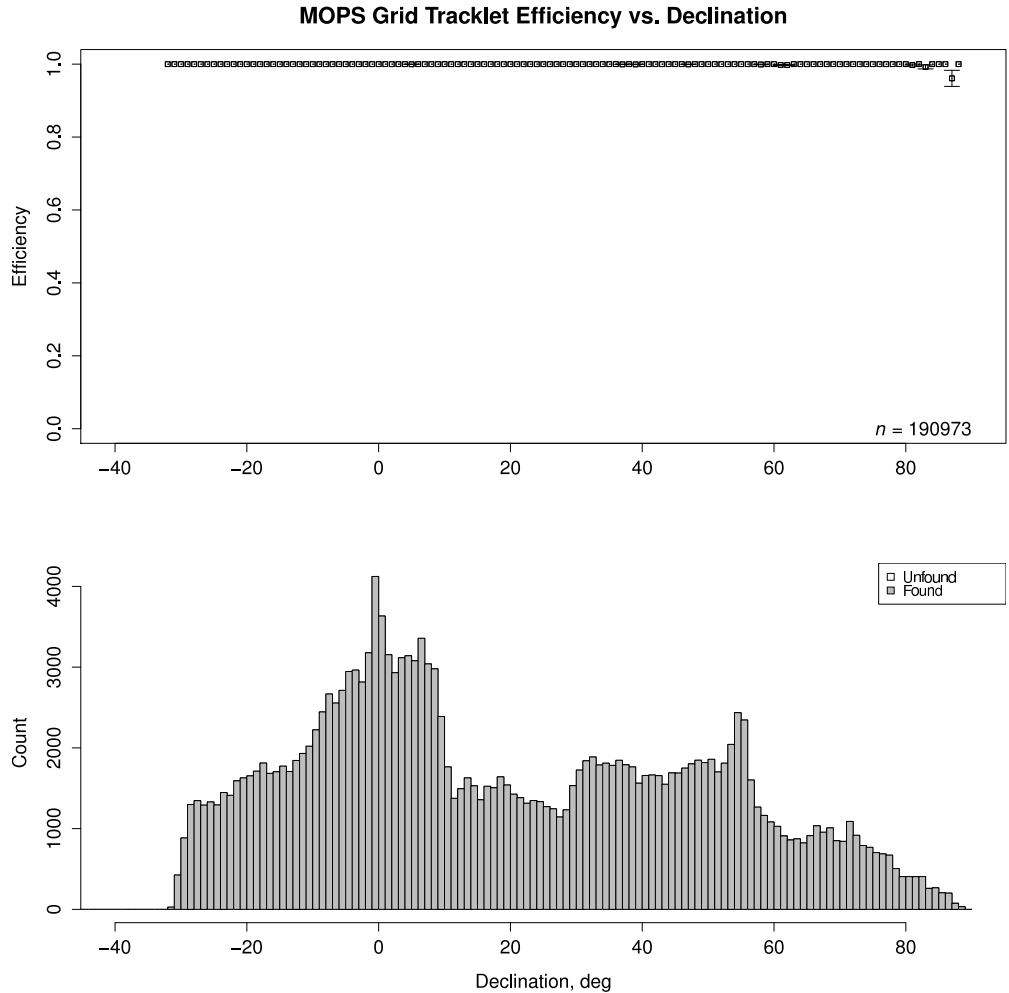}
\caption{Tracklet creation efficiency as a function of declination for
  a one lunation MOPS simulation using PS1 telescope pointings and S3M
  grid objects with \PSone\ realistic (0.1$\arcsec$) astrometric
  uncertainty. The uneven declination coverage is due to uneven
  coverage of the \PSone\ survey during the lunation (14 Aug 2011
  through 12 Sep 2011). Several fast-moving objects in the
  declination=89$\arcdeg$ bin were lost due to overly conservative tracklet acceptance
  parameters.}
\label{f.S3MGridTrackletDeclinationEfficiency}
\end{figure}

\begin{figure}[btp]
\centering
\plotone{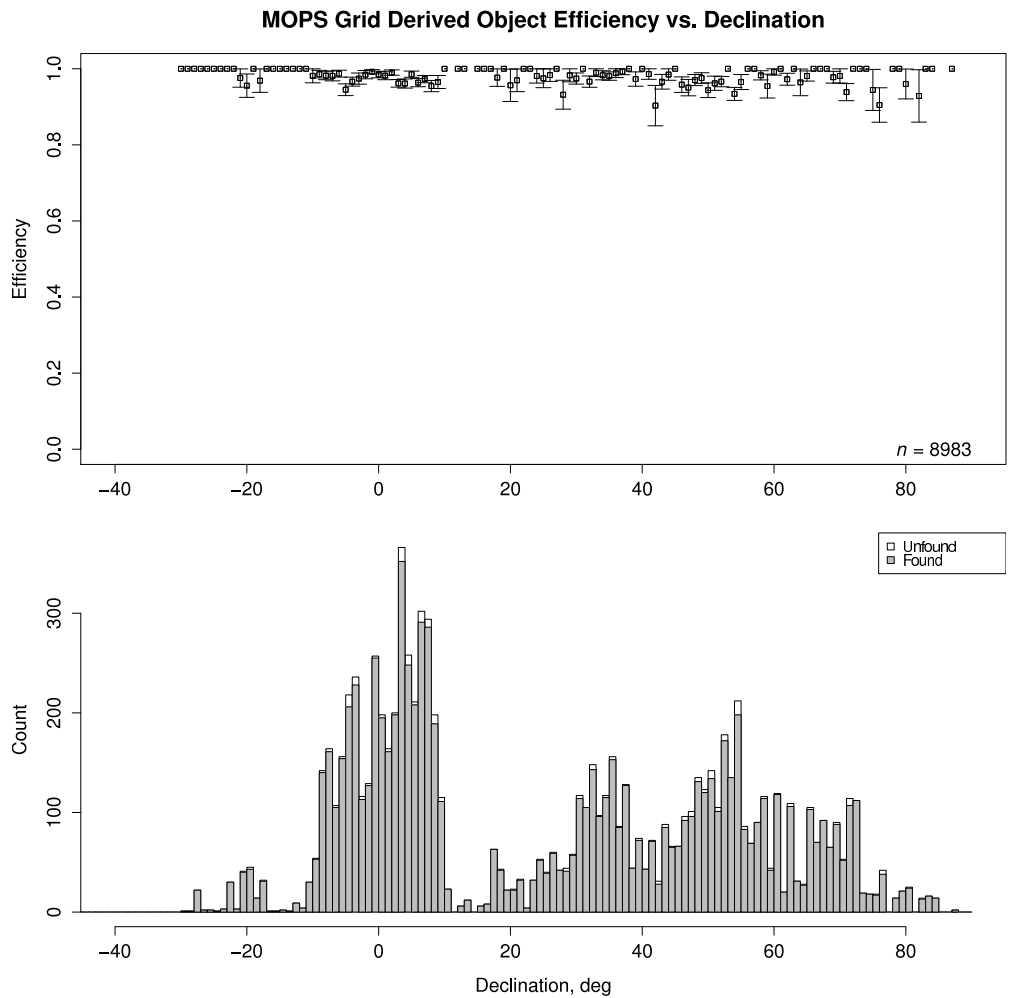}
\caption{Derived object creation efficiency as a function of
  declination for a one lunation MOPS simulation using S3M grid
  objects with \PSone\ realistic (0.1$\arcsec$) astrometric
  uncertainty. The uneven declination coverage is due to uneven
  multi-night coverage of the \PSone\ survey during this particular
  lunation (14 Aug 2011 through 12 Sep 2011). While the total grid
  derived object efficiency is 97.5\%, we believe this number can be
  improved further by tuning of MOPS linking parameters.}
\label{f.S3MGridDODeclinationEfficiency}
\end{figure}

One important feature of the grid population is that it includes
objects that can appear near the poles where the sky-plane density of
real solar system objects is negligible, \ie\ it allows us to check
that MOPS works correctly even for objects at high declinations as
shown in Figs.~\ref{f.S3MGridTrackletDeclinationEfficiency} and
\ref{f.S3MGridDODeclinationEfficiency}.  While tracklet creation
efficiency is  nearly 100\% for all declinations $\la88\arcdeg$, within 2$\arcdeg$ of the pole the grid simulation revealed that slightly
conservative tracklet acceptance parameters rejected several
fast-moving objects.  The derived object efficiency is typically
$\ga$95\% for declinations ranging from $-30\arcdeg$ to the north
celestial pole indicating that the tracklet linking algorithm works
across the entire sky.  The drops in efficiency to $\sim$90\% are
largely due to an inability to compute an initial orbit (see
\S\ref{ss.InitialOrbitDetermination}) which is in turn usually due to
a sparse observation cadence for the track.

\section{MOPS Validation and Real Data}
\label{s.ProcessingRealData}

Processing of data from the current generation of surveys has been
addressed by software pipelines already employed by
Spacewatch \citep[\eg][]{Rabinowitz1991,Larsen2001}, the Catalina Sky
Survey, LINEAR, \etc.  We did not want to reproduce their work.  MOPS
was designed for use with a high-quality transient detection stream
delivered by a next-generation \PSfour\ or LSST-class survey system
with nearly 100\% fill-factor and a relatively modest number of
systematic false detections.  In many respects, despite the much
higher data volumes expected from these telescopes, moving object
algorithm design is simpler due to the high quality astrometry and
photometry of the input data.  MOPS performs less effectively with
real \PSone\ data because it is the {\it prototype} for the next-generation
\PSfour\ survey --- not a next-generation survey itself.

That said, the \PSone\ telescope represents a significant step forward
in wide field survey astronomy, and has become a major contributor in
NEO detection and discovery and the largest single-telescope source of
asteroid observations while satisfying multiple, disparate survey
programs. We have adapted MOPS to \PSone\ to maximize asteroid
detection and NEO discovery from the \PSone\ data stream through the
creation of tools to reject false detections and tracklets and allow
for human review of data before they are submitted to the Minor Planet
Center.  Additionally, due to a realized survey cadence that is
inefficient at producing enough repeat observations to compute
reliable orbits, we have de-emphasized MOPS's derived object
functionality in normal \PSone\ processing.  In practice, the MOPS
pipeline is executed through the tracklet-creation stage, with some
``add-on'' tools that are described below.  MOPS is still used to its
full capability as a simulation and research tool.

In this section we describe in detail
all the \PSone\ MOPS processing steps and quantify its performance on the
transients provided by the prototype telescope. 

\subsection{\PSone\ Surveying}

The \PSone\ `effective' camera fill factor ($f$) as measured by MOPS
is $\sim$75\%.  The loss of image plane coverage results from gaps
between the OTAs, dead cells and gaps on the CCDs themselves,
problematic pixels (\eg\ anomalous dark current or non-linearity), and
the allocation of sub-sections on the OTAs (known as cells) to high
speed acquisition of bright guide stars.  The original MOPS design
assumed $f\sim100$\% (\PSfour) and three 2-detection tracklets/lunation for new object discovery
so that with $f<100$\% the maximum achievable system efficiency
($\epsilon_{max}$) for moving objects must be somewhere in the range of $f^6
\le \epsilon_{max} \le f^1$, depending on the object's rate of motion,
the distribution of `dead' image plane pixels/area and the telescope
pointing repeatability.  This implies that the maximum \PSone\ MOPS end-to-end
derived object efficiency for objects in the field of view must lie in the
range 26-75\% even for objects that would be imaged at high \sn.

In part to compensate for the fill-factor, in October 2010 the
\PSone\ survey was modified to obtain a four-exposure `quad' in a
manner similar to other NEO surveys instead of a single pair of
exposures.  The 4 images of a quad are typically acquired at the same
boresight and rotator angle with a transient time interval (TTI) of
about 15~minutes between sequential detections.  Thus, a tracklet
containing four detections would have a typical arc length of about
45~minutes.  We then search for tracklets with three or four
detections in the quad.  The false tracklet rate decreases as the
number of detections in the tracklet increases but even with 4
detections per tracklet there are still false tracklets.  To cope with
the false tracklet rate we implemented a NEO `czaring' procedure and
an associated infrastructure for extracting `postage stamp'
$100\times100$ pixel images centered on every detection incorporated
into a tracklet.  A human observer, the `czar', vets every tracklet
before submission to the MPC.

The tracklet processing for pair observations by other \PSone\ surveys
is specifically directed towards NEO detection and is divided into two
rate of motion regimes --- slow and fast, corresponding to 0.3 to
0.7~deg/day and 1.2 to 5~deg/day respectively.  \ie\ Tracklets
containing just two detections are not detected if they have rates of
motion between 0.7 and 1.2~deg/day.  The reason for the two-regime
processing is that the confusion limit from false detections becomes
unmanageable around 0.7~deg/day, but beyond $\sim$1.2 deg/day we can
use trailing information to reduce the false tracklet count.  The slow
regime's lower limit corresponds roughly to a rate of motion that
easily distinguishes between NEOs and main belt asteroid motions at
opposition.  The upper limit in the slow regime is set empirically at
the rate at which the confusion limit becomes too high for the czar.
The gap between 0.7~deg/day and 1.2~deg/day leaves a `donut hole' in
the velocity space where pairwise tracklets will not be found. Recent
improvements in our ability to screen false detections will allow this
hole to be removed when \PSone\ is observing away from the Galactic Plane.

In the fast pairwise tracklet regime MOPS makes use of morphological
information (\eg\ moments) provided by the IPP as a proxy for trailing
information. Quality cuts require similar morphology between the two detections and that the position angles (PA) calculated from the detection
moments be aligned with the PA of the tracklet.

\subsection{Detections}
\label{ss.Detections}

\begin{figure}[btp]
\centering
\plotone{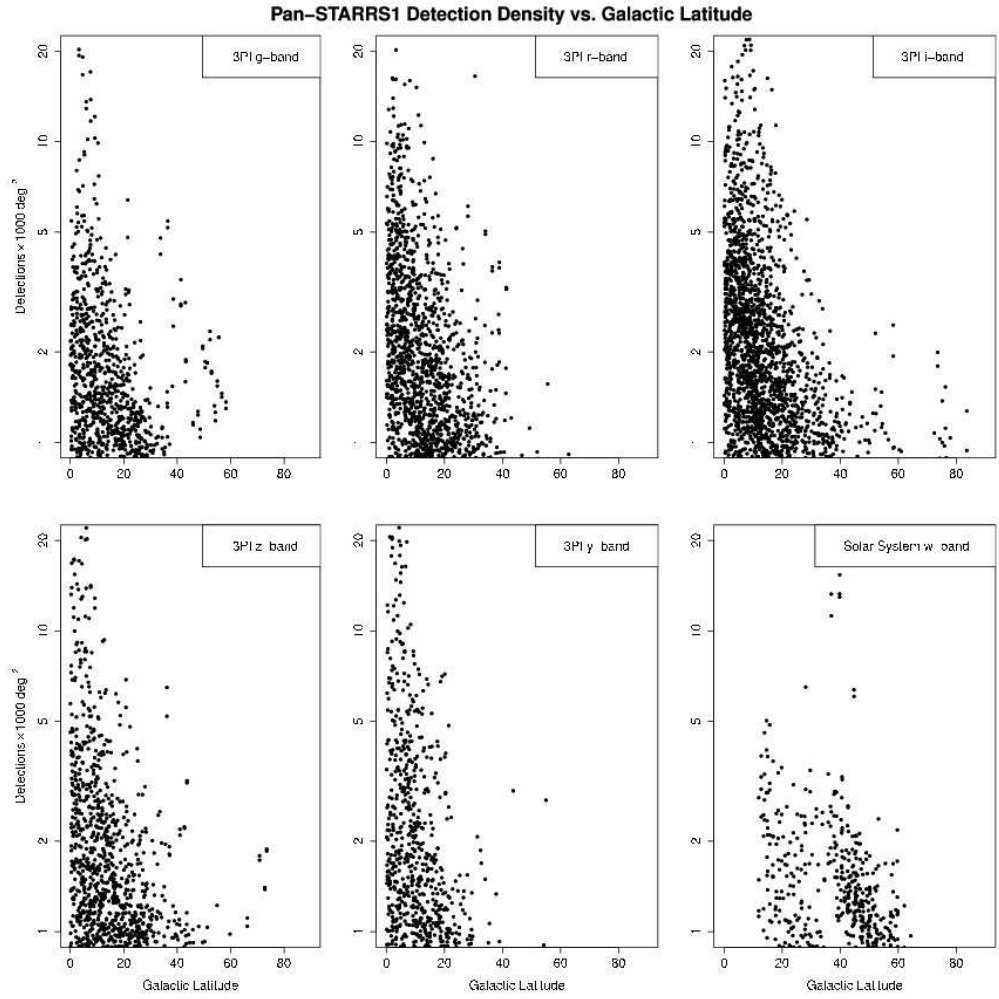}
\caption{Number of transient detections reported to the MOPS by the
  IPP as a function of galactic latitude in all six \PSone\ filters.
  Almost all the transient detections are false.}
\label{f.FalseDetections-vs-GalacticLatitude}
\end{figure}

The false detection rate delivered by the \PSone\ IPP to MOPS is shown
in Fig.~\ref{f.FalseDetections-vs-GalacticLatitude}.  The rate is dominated by systematic image artifacts rather than
statistical noise.  At high galactic latitudes the transient detection
rate is about the expected value from statistical fluctuations in the
background but the rate increases dramatically as the field center
approaches the galactic plane where the false detection rate is
$10-50\times$ higher than expected.  The increase in false detections
with proximity to the galactic plane is due to image subtraction
issues in regions with high stellar sky plane density and residual
charge in the CCDs from bright stars.  The transient rate in the
\wps\ filter does not have data for galactic latitudes $\la30\deg$
because this filter was solely used for solar system surveying and
detection of moving objects.  The solar system team does not survey
into the galactic plane because we learned early-on that the false
detection and tracklet rates are too high to efficiently detect solar
system objects.

\begin{figure}[btp]
\centering
\plotone{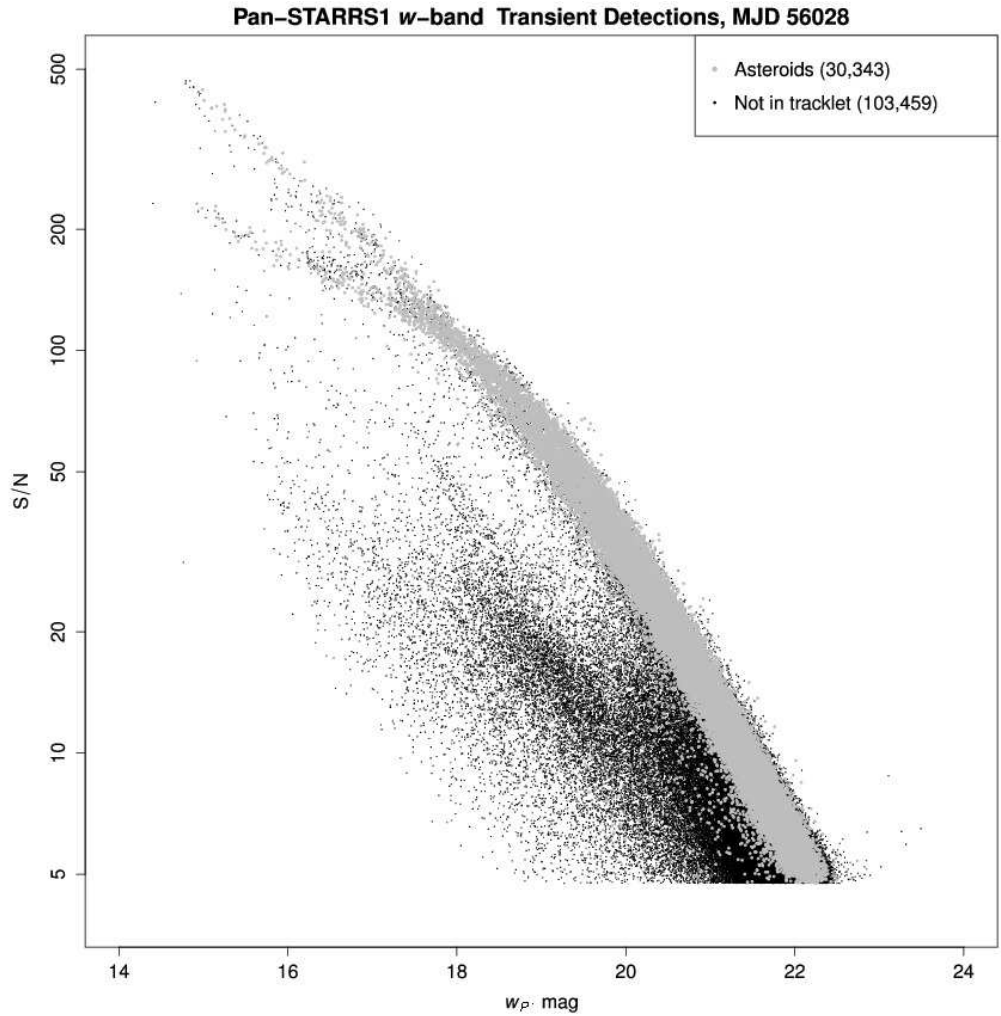}
\caption{\PSone\ transient detections reported to MOPS by the IPP as a
  function of \sn\ and \wps\ magnitude on MJD~56028.  Grey dots are detections in tracklets that
  are real asteroids. Black dots are all other detections.  The `fork' in
  the distribution at $\wps \sim 18$ is due to a change in observing
  conditions that caused distortion of PSFs for bright objects.}
\label{f.detections-s2n-vs-w-mag}
\end{figure}

Overall, the rate of transient detections at $\ge$5$\sigma$ is $\sim
8200$/deg$^2$ --- orders of magnitude higher than expected with Poisson statistics and with far more systematic structure, \eg\ from
internal reflections and other artifacts.  This structure is
especially problematic because, unlike Poisson-distributed false
sources that are evenly distributed across the detector, systematic
false sources tend to be clumped spatially and form many false
tracklets that can clog and contaminate subsequent data
processing.  A large number of reported false detections for which the \sn\ is
much lower than expected at a given magnitude could
easily be removed with a field-dependent
cut on \sn\ vs. magnitude (see Fig.~\ref{f.detections-s2n-vs-w-mag}).
  
\begin{figure}[btp]
\centering
\plotone{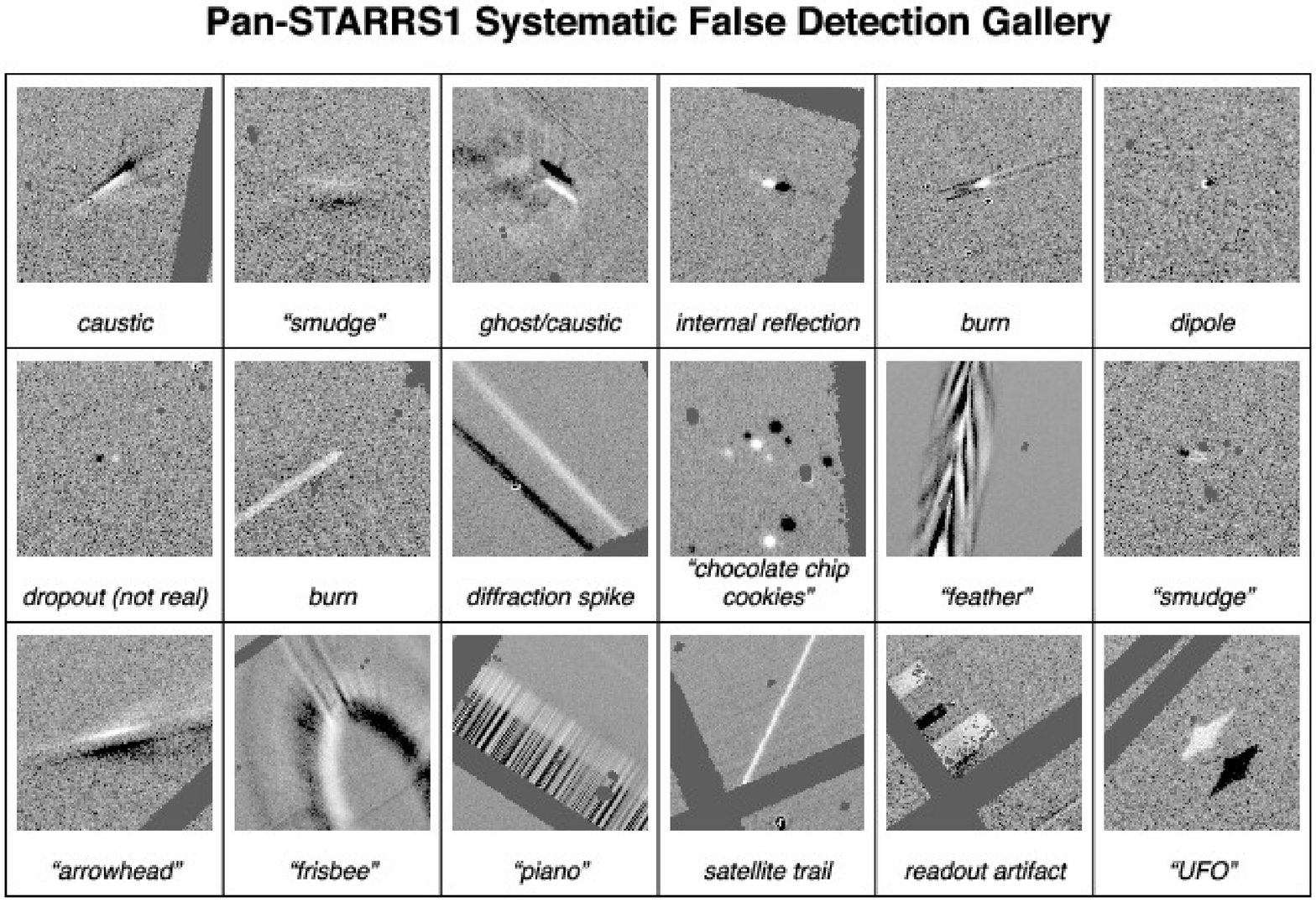}
\caption{A sample of false source detections delivered to MOPS.  The
  cause of the detection is provided under the image when known.  The
  source detections are in the positive image (white pixels).  Each
  image is a $200\times200$ pixel difference image with the original
  source detection in the center.  \ie\ the difference between two
  successive images acquired at the same boresight.  Solid dark grey
  regions represent gaps between CCDs or cells or masked regions.}
\label{fig.FalseDetectionMenagerie}
\end{figure}

Figure~\ref{fig.FalseDetectionMenagerie} shows a representative sample
of the types of false detections provided to MOPS by the IPP.
Accurate machine classification of these artifacts using methods such
as those described by \citep{Donalek2008} can greatly reduce
contamination and is under exploration by the PS1SC.  We use
imaginative monikers like `arrowheads', `chocolate chip cookies',
`feathers', `frisbees', `pianos' and `UFOs' to classify them when
iterating with the IPP on improving the detection stream.  Many of the
false detections are easily explained as internal reflections, ghosts,
or other well-understood image artifacts but some are as yet
unexplained.  MOPS screens false detections using their morphological
parameters but many survive the cuts into tracklet formation because
they must be conservative so that diffuse comet detections are not
removed.

Finally, when the density of false detections spikes to rates that are
difficult to manage, \eg\ near a very bright star, we invoke a last-ditch spatial
density filter to reduce the local sky-plane density to no
more than 10,000 detections/deg$^2$ in a circle of 0.01$\arcdeg$ radius.
The filter eliminates false detections by discarding those with the highest PSF-weighted masked fraction until the detection density is manageable.

\begin{figure}[btp]
\centering
\plotone{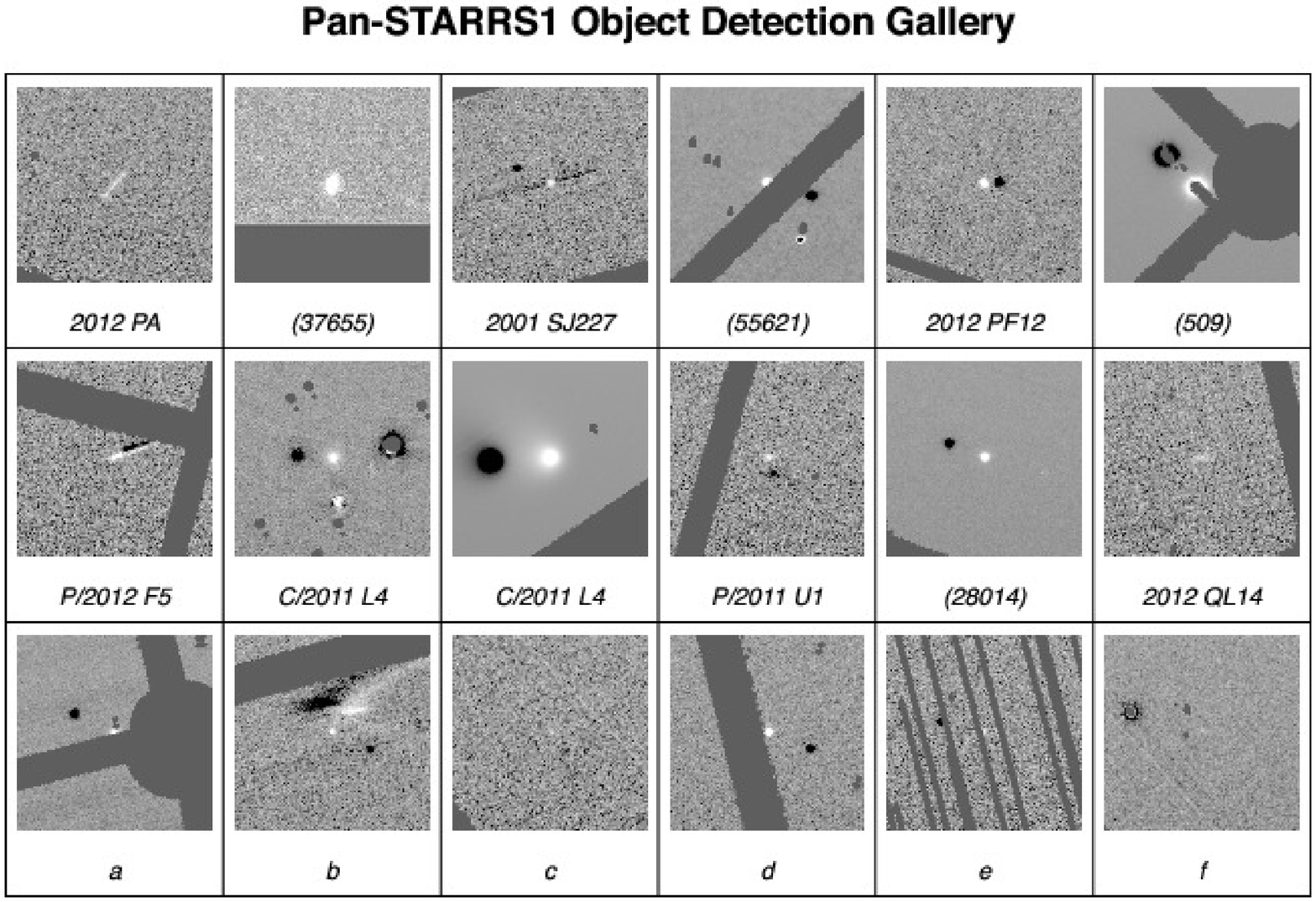}
\caption{A sample of real object detections delivered to MOPS via
  difference imaging. Each image is a $200\times200$ pixel difference
  image with the positive source detection in the center. Many
  detections lie near artifacts or in heavily masked regions or are
  difficult to distinguish from systematic false detections. Objects
  \emph{a} through \emph{f} are unknown asteroids.  Solid dark grey
  regions represent gaps between CCDs or cells or masked regions.}
\label{fig.RealDetectionMenagerie}
\end{figure}

Even real detections are often not well formed PSFs as illustrated in
Fig.~\ref{fig.RealDetectionMenagerie}.  Trailed detections often
intersect with cell and chip gaps (a), PSF-like detections of
slower moving objects have overlapping PSFs (2012~PF$_{12}$), or sit on the
boundary of a masked region (\eg\ e).  Whenever a detection is
poorly formed it creates difficulties in astrometry and
photometry and therefore in linking the detections into tracklets and
tracks.

Despite these problems \PSone\ astrometry and photometry reported to
the MPC for slow moving objects is currently excellent by contemporary
standards.  The average RMS astrometric uncertainty is about
$0.13\arcsec$ \citep{Milani2012}.  The intrinsic photometric uncertainty in the
calibrated\footnote{\PSone\ began reporting calibrated magnitudes in
  May 2012.} \PSone\ data is $<10$~mmag in the \gps, \rps, and
\ips\ filters and $\sim10$~mmag in the \zps\ and \yps\ filters
\citep{Schlafly2012}.  The asteroid photometry is not yet as good as
in the 5 primary \PSone\ filters because 1) much of the data is in the
still to be fully calibrated \wps\ filter and 2) the detections are
often trailed enough to cause problems with the photometric fit.  We
expect to solve these problems soon and begin reporting \wps\ band
photometry for moving objects with a calibration accuracy of a few tens of mmag.

\subsection{Tracklets}
\label{ss.Tracklets}

\citet{Kubica2007} provide detailed information on our algorithms and
their performance for linking detections of moving objects on a single
night into `tracklets' or linking tracklets between nights into
`tracks'.  Their techniques rely on the use of
`kd-tree' structures to provide fast and scalable performance.

Their Table~1 shows that the algorithms yield nearly 100\% tracklet
creation efficiency for the S3M in the presence of the expected but
random false detection rate of 250~deg$^{-2}$.  About 10-15\% of the
generated tracklets are {\tt MIXED} or {\tt BAD} but this tracklet accuracy is perfectly
acceptable because false tracklets should not link together across
nights to form a set of detections with a good orbit.  The mixed
tracklets cause no loss of objects because the tracklet creation is
non-exclusive \ie\ the same detection can appear in the correct {\tt CLEAN}
tracklet and in multiple other tracklets. \citet{Kubica2007} note that
the technique still worked with high efficiency even without using the
detections' orientations and lengths but at the expense of a much
higher false tracklet rate.  Furthermore, they showed that the system
maintains its integrity even at $>10\times$ higher random false
detection rates.

The algorithm for linking tracklets between nights uses a kd-tree to
quickly identify candidate tracks. Each track is then fit to a
quadratic in RA and Dec with respect to time to eliminate tracks based
on their fit residuals.  The number of candidate tracks increases
exponentially with the number of detections as shown in Table~3 in
\citet{Kubica2007} --- while 98.4\% of the correct tracks are properly
identified only 0.3\% of the tracks that pass all the cuts are real.
The false tracks are eliminated in the next step through initial orbit
determination (IOD) described in the following sub-section.

As discussed in the previous section, tracklet formation must take
place in the presence of false detections due to statistical
fluctuations on the sky background but also the usually much more
numerous and problematic systematic false detections.  We note that
\citet{Kubica2007} performed their tests using 2-detection tracklets
and with assumptions on the astrometric and photometric performance
expected for the cosmetically clean images of an idealized \PS4
survey.  Their simulations are thus not directly applicable to the
real \PSone\ system that includes more noise and degraded astrometry.
The \PSone\ false detection rate is $10-50\times$ greater than
expected at $\ge5\sigma$ and almost all of them are systematic.  They
pose problems for MOPS in that the detections ``line up'' well enough
to pass the tracklet formation cuts and the tracklets represent
objects with asteroid-like rates of motion.  The high false tracklet
rate significantly slows down derived object processing because of the
resulting tremendous false track rate.

The realized tracklet creation efficiency and accuracy was measured
with synthetic detections injected into the real \PSone\ data.  In
this data stream typically 4 images are acquired at the same boresight
with roughly a TTI separating each image in the sequence.  We form
tracklets of 2, 3, or 4 detections with some limits on the effective
rate of motion depending on the number of detections in the tracklet.
Table~\ref{t.MOPSTrackletEfficiencyAndAccuracy} shows that the
tracklet creation efficiency is 99.98\% --- the algorithm correctly
identifies almost every possible set of detections in a tracklet even
in the presence of real detections and noise.

The algorithm returns
additional {\tt MIXED} and {\tt BAD} tracklets though at a small fraction of
the {\tt CLEAN} rate (see \S\ref{s.Overview} and \S\ref{ss.MOPSEfficiencyConcepts}
 for definitions).  Tracklets are also
identified for the non-synthetic detections \ie\ detections of real
objects or false detections.  For the purpose of the efficiency and
accuracy calculation both types of non-synthetic detections are
`noise' as viewed from the perspective of the synthetic solar system
objects.  Thus, non-synthetic tracklets may be sets of detections of
real solar system objects --- but the numbers in Table~\ref{t.MOPSTrackletEfficiencyAndAccuracy} indicate that there 
is a problem.  There are
roughly $7\times$ more non-synthetic detections than synthetic but, if
our simulation and synthetic population are an accurate representation
of reality, and all the non-synthetic detections are real, the two
values should be roughly the same.

Tracklets can only be formed from detections in images acquired within
a reasonably short time frame with sequential detections of the same
object separated by about a TTI.  Linking
detections into tracklets across longer time intervals creates a
combinatoric explosion that can not be addressed by the current
linking algorithms.  On the other hand, it often happens that multiple
tracklets for the same object are created on the same night separated
by time intervals large compared to a TTI.  The most common cause is
overlap between adjacent fields acquired at times separated $\gg$TTI.
Even more problematic are `deep-drilling' situations where $\gg 4$
images are acquired at the same boresight in rapid succession.  In this
situation a real object moving between the images is not necessarily
detected in every image due to \eg\ chip and cell gaps, passing over a
star, masking, photometric fluctuations, \etc, and detections from the
same object may appear in different tracklets.  Furthermore, the
detections within the tracklets may overlap in time due to astrometric
fluctuations.  (\eg\ detections from images 1, 3 and 6 appear in
tracklet A and detections from images 2, 5, and 8 appear in tracklet B
with no detections in images 4 and 7.)  Thus, when there are $>4$
exposures in a sequence at the same boresight we post-process the
tracklet list with {\tt collapseTracklets}.

{\tt collapseTracklets} merges co-linear tracklets within a single
night or data set using a method similar to a Hough transform.  The
approximate sky plane location of each tracklet is determined at the
mid-time of all tracklets in the set assuming that their motion is
linear in RA and Dec during the time frame ($\dot\alpha$ and
$\dot\delta$ respectively).  Co-linear tracklets corresponding to the
same object should have similar positions and motion vectors making
them straightforward to identify as `clumps' in
$(\alpha,\delta,\dot\alpha,\dot\delta)$-space using a series of range
searches implemented with a 4-dimensional kd-tree. Given candidate groupings
of tracklets in the transform space, {\tt collapseTracklets} attemps to merge
them using straightforward RMS residual acceptance criteria, 
where there is a basic tradeoff between creating false merges vs. allowing
too many duplicates. The cost of duplicate
tracklets in the deep-drilling sequences is that if there are duplicate tracklets
for the same object, a potential derived object will rightly
use the distinct  tracklets to create two distinct derived objects for the same object,
and derived object discordance rejection (see \S\ref{ss.OrbitDetermination}) will disallow the derived objects on
the grounds that multiple tracklets for an object cannot exist in the same
set of fields.

We adopted {\tt collapseTracklets} essentially unmodified upon delivery
from the development team at LSST, and beyond simple tests that
verify basic capability, we have not attempted to maximize its efficiency
with \PSone\ deep-drilling cadences or data quality.  Informal evaluations
against the \PSone\ 8-exposure Medium Deep (MD) sequences show a duplicate
ratio of $\sim$25\% with essentially zero lost tracklets. 

In the time span from February 2011 through May 2012 MOPS created
$1,513\times 10^3$ tracklets of which $534\times 10^3$ were real
(35\%) and $345\times 10^3$ were automatically attributed to numbered
or multi-opposition objects (see
\S\ref{ss.AttributionOfKnownObjects}).  The situation is much worse
for tracklets containing just 2~detections: $11,691\times 10^3$
created with just $780\times 10^3$ attributed (6.7\%).  Assuming that the
attributed:real ratio is the same as for the 3- and 4-detection
tracklets it implies that only $\sim$10\% of 2~detection tracklets are
real.

In theory, the \PSone+MOPS system is capable of detecting the highest
proper motion stars within a single lunation.  Consider the unusual
case of Barnard's star with an annual proper motion\footnote{{\tt http://simbad.u-strasbg.fr/simbad/sim-id?Ident=V*+V2500+Oph}} of
$\sim10.3\arcsec$/year or about $0.0282\arcsec$/day.  \PSone's average
RMS astrometric uncertainty is about $0.13\arcsec$ \citep{Milani2012}
so that Barnard's star shows noticeable astrometric motion in about 5
days.  Barnard's star would be provided as a transient detection to
MOPS if the \PSone\ IPP created difference images using a static sky
created from earlier images \citep{Magnier2008}.  On each night MOPS would create a
`stationary' transient and if the survey provided more detections 5
and 10 days later MOPS would link the tracklets together.  Thus, even
in normal operations it is theoretically possible that MOPS could
provide detections of high proper motion stars light years from the
Sun.  It takes little imagination to realize that MOPS could process
`stationary' tracklets identified in three successive {\it months} to
discover proper motion stars to even larger distances.  The false detection
rates for image differencing using a static sky should be no 
worse than pairwise image differences at the same \sn.

\subsection{Tracks}
\label{ss.Tracks}

Tracklets are linked into `tracks' across
previous nights over a time window of typically 7-14 days using another kd-tree
implementation called {\tt linkTracklets} (\citet{Kubica2007}).  The MOPS
runtime configuration specifies how many days prior to the current
night to search and how many tracklets are required to form a linkage. The
tracklets search locates combinations of tracklets that collectively
fit quadratic sky-plane motion to within some configured error.  Of
course actual asteroid motions are not truly quadratic, but within the
time interval of track creation this approximation holds for most
objects.

Column~4 in table~\ref{t.MOPSDerivedObjectEfficiency} shows that the tracklet
linking efficiency is $\sim80$\% for most classes of solar system
objects ranging from the inner solar system to beyond Neptune.  The
linking efficiency is reduced from our grid simulations for several
reasons: a) linkages are contaminated with false tracklets and fail
orbit determination; b) Medium Deep sequences generate multiple
tracklets for the same object, causing rejections in the discordance
checks (see \ref{ss.OrbitDetermination}); and c) we simply have not yet tuned {\tt linkTracklets}' 
operational parameters to \PSone's current survey
mode and astrometric performance.  Early MOPS tests with synthetic and real data
obtained from Spacewatch \citep[\eg][]{Larsen2001} showed that the
derived object efficiency can be increased to nearly 100\% with a
suitable survey strategy, false detection rate, and set of {\tt
  linkTracklets} configuration parameters

Table~\ref{t.MOPSDerivedObjectEfficiency} does not show {\tt
  linkTracklets}'s accuracy but it is similar to the $\ll1$\% accuracy
reported by \citet{Kubica2007}.  \ie\ $\ga$99\% of all the tracks are
{\it not} real.  However, very few of the tracks survive the
subsequent motion, acceleration, astrometric and photometric residual
cuts, \etc\  Those that do pass all the cuts are subject to initial and
differential orbit determination to eliminate essentially all the bad
tracks as described in the following two subsections.

\subsection{Orbit Determination}
\label{ss.OrbitDetermination}

MOPS identifies tracks that represent a real or synthetic asteroid
through `orbit determination'. A six-parameter orbit is calculated for
each track and only those that do not exceed a RMS residual
requirement are accepted and passed on to the next processing
phase. Orbit determination is a two-stage process within MOPS
beginning with initial orbit determination (IOD) and followed by a
least-squares differentially corrected orbit determination. Tracks
that do not satisfy the residual requirement are discarded and their
tracklets made available for use in other linkages.

Occasionally there may be cases where a tracklet is included in
multiple distinct linkages whose fitted orbits produce RMS residuals that meet
MOPS acceptance thresholds. When this occurs the tracklet cannot
logically be included in both derived objects, so MOPS includes code
`discordance identification' software that identifies this situation
and either

\begin{itemize}
\item chooses one linkage as `correct' if its RMS residual is significantly smaller than all others, or
\item rejects all linkages as bad since no correct linkage can be determined.
\end{itemize}

Similarly, two distinct tracklets may share a common detection and be
included in otherwise separate linkages with acceptable RMS
residuals. In this case it is logically impossible for the detection
to belong to two different objects, so the discordance identification
routines select one linkage as described above or rejects them all if
none stands out as correct.  Tracklets in rejected linkages can be recovered
later in processing through precovery (see \S\ref{ss.PANDA}).

\subsection{Initial Orbit Determination (IOD)}
\label{ss.InitialOrbitDetermination}

Orbit determination using a small number of observations over a short
time interval has been studied for centuries, since Kepler first
described the laws of planetary motion. Many techniques have been
devised, each with its own advantages, and modern computers allow many different
methods to be evaluated for a set of input detections.

For MOPS's purposes, IOD is a module that produces six-parameter
orbits given a set of detections. We implemented the `OrbFit' package
developed by the OrbFit Consortium \citep{Milani2010}. Our selection
criteria for an orbit determination package were 1) efficiency at
producing an orbit given a `true' linkage; 2) orbit accuracy for
observations using \PS\ cadences and astrometric uncertainties; 3)
speed of computation; and 4) availability and support. We tested the orbit
determination software against millions of synthetic tracks to measure
performance. As with other MOPS software, while we wanted high
efficiency and accuracy in our orbit determination software, it was
equally important to precisely measure the orbit determination
efficiency.

Table~\ref{t.MOPSDerivedObjectEfficiency} shows that the IOD
efficiency is essentially 100\% for all classes of solar system
objects. \ie\ OrbFit successfully provides an orbit when provided a
correctly linked set of tracklets (a track).

\subsection{Differential Correction}
\label{ss.DifferentialCorrection}

MOPS attempts to improve the IOD by navigating the parameter space of
orbit solutions to minimize the least-square RMS residual of the
fitted detections. As with initial orbit determination, differential
correction is a long-studied problem, spanning many decades of research,
 made routine by today's computer hardware. Modern differential
correction techniques use precise models for the motion of the Earth
and other large solar system bodies and are able to predict positions
on the sky to within hundreths of an arcsecond many years into the
future given enough input observations.

Through a memorandum of understanding (MOU) with the Jet Propulsion
Laboratory (JPL) \PS\ obtained permission to use a subset of
JPL's Solar System Dynamics (SSD) software to compute asteroid
ephemerides and differentially corrected orbits. SSD is a workhorse of solar system analysis,
used for guiding spacecraft missions to their asteroid and comet
destinations and for assessing impact probabilities for high profile
NEOs.

Table~\ref{t.MOPSDerivedObjectEfficiency} shows that the differential
correction efficiency for objects that have passed IOD is essentially 100\% for all classes
of solar system objects.

Despite the overall effectiveness of the OrbFit IOD+JPL differential corrector combination, certain geometric configurations can lead to convergence to an incorrect orbit --- the motion of objects in the small-solar-elongation
sweetspots leads to dual solutions for orbits computed with short arcs. For short windows around
the orbit computation, typically up to 30 days, the ephemeris uncertainty using the `wrong' orbit
is small enough that an attribution or precovery tracklet can still be found. With the additional
arc length from the attribution or precovery the new orbit will collapse to a single solution.

\subsection{Derived Objects}

The tracklets and orbital parameters of tracks that survive the RMS residual cuts for
orbit determination are stored in the MOPS database as a `derived
object'. A derived object represents a moving object `discovered' by MOPS; in other words, the detections are
believed to belong to the same body with an orbit that allows the
body's motion to be predicted well enough to recover the body
in an adjacent lunation to the discovery lunation. 

The derived object efficiency and accuracy is provided in
table~\ref{t.MOPSDerivedObjectEfficiency}.  The realized efficiency
for the \PSone\ survey is currently in the 70-90\% range for most
classes of solar system objects.  Due to the high false
detection rates the survey has concentrated on identifying candidate
NEO tracklets by their anomalous rates of motion
\citep[\eg][]{Rabinowitz1991,Jedicke1996} and has temporarily abandoned the idea of
creating derived objects.  Despite a survey pattern, cadence and un-optimized MOPS
configuration that are not particularly well-suited to creating derived
objects, the system still achieves 70-90\% efficiency for objects
that appear in multiple tracklets within a lunation.

\subsection{Attribution of known objects}
\label{ss.AttributionOfKnownObjects}

MOPS was designed to operate agnostically on all tracklets regardless
of whether they correspond to known or unknown objects.  \ie\ no {\it
  a priori} information about known objects is used when creating MOPS
tracklets, tracks or derived objects.  The motivation was to create a
final set of `derived objects' from \PSone-only data with good
controls on observational selection effects.

On the other hand, for the purpose of extracting more science from the
data, and as an alternate means of measuring the system's detection
efficiency and accuracy, we also attribute tracklets to known numbered
and multi-opposition asteroids using \citet{Milani2012}'s
\knownserver\ module.  They show that it has essentially
100\% efficiency and 100\% accuracy using \PSone\ data for those
classes of objects and subsection \S\ref{sss.DetectionEfficiency}
discusses the realized \PSone+MOPS detection efficiency
characteristics using the known objects.

\begin{figure}[t]
\centering
\plotone{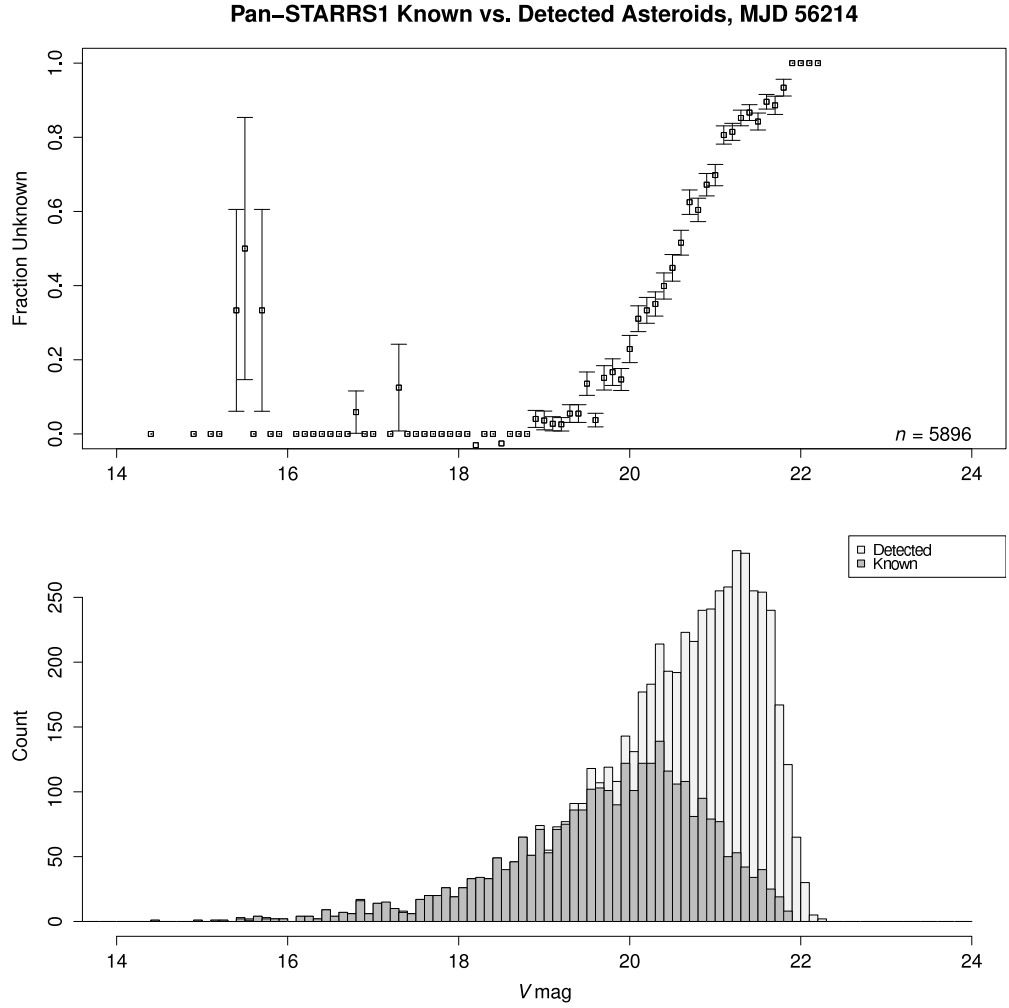}
\caption{Overlapping detected and known asteroid
  distributions for a single night of \PSone\ solar system
  observing.}
\label{f.knownvsdetected}
\end{figure}

As of October 2012 we have reported detections for approximately 240,000 numbered
and 84,000 multi-opposition objects identified by
\knownserver\ to the MPC.  These detections represent about 73\% of the total 3.4
million detections, the remainder
being mostly previously unknown asteroids.  The fraction of known
objects decreases with increasing $V$-magnitude such that
50\% of the reported objects are unknown for $V>21.6$ (see Fig.~\ref{f.knownvsdetected}).

\subsection{Precovery \& Attribution (`PANDA') of derived objects}
\label{ss.PANDA}

In keeping with the principle of MOPS agnosticism with respect to
previously known objects we implemented the capability within MOPS to
`attribute' new tracklets each night to MOPS's derived objects.
Recall that the time window for creating a MOPS derived object is
typically 7-14 days, but we want to associate individual tracklets
from new data with existing derived objects if possible (attribution), and search the MOPS database for individual tracklets
observed previously that did not form a derived object because not
enough tracklets were observed at the time (precovery). Early
\PS\ work showed that with
sufficiently high-quality astrometry, an orbit with arc length of
typically 10-14 days would have a prediction uncertainty small enough
to locate the object in an adjacent lunation. Then, after a successful
precovery or attribution extends the arc to beyond 30 days, the orbit
is secure.

The PANDA algorithm is a simplified version of that described by
\citet{Milani2012} --- in essence we integrate every derived object's
motion to the time of observation of each image and compare the
predicted location and velocity to all nearby tracklets to see if
there is a match.  If there is a match we attempt a differential orbit
computation and if the resultant fit and residuals are within
acceptable bounds we attribute the tracklet to the derived object.
In practice, we pre-determine the locations and velocities of all
derived objects at the beginning, middle and end of the night and then
use {\tt fieldProximity} \citep{Kubica2005} which interpolates their
locations and velocities to each image time and uses a kd-tree
implementation to make candidate associations between derived objects
and tracklets.

Table \ref{t.MOPSAttributionEfficiency} shows that the attribution
efficiency is $\sim$93\% but it is important to keep two points in
mind: 1) this is the cumulative efficiency for {\it all} possible
attributions on that night and 2) the statistics are dominated by the
main belt asteroids.  In regard to the first point, our efficiency
determination software knows when a new synthetic tracklet is detected
for an existing synthetic derived object but it does not account for
the accuracy of the derived orbit's ephemerides at the time of
observation \eg\ the derived object's arc-length.  It may be that the
available attribution was for a derived object detected on each of the
last 3 nights or it may be a derived object whose last observation was
3 years beforehand.  Thus, we expect that attribution efficiency will
decrease as a function of time.  Even though the results in
table~\ref{t.MOPSAttributionEfficiency} are dominated by the main belt
the attribution efficiency mostly increases with the semi-major axis
of the object --- it is relatively easy to attribute distant, slow
moving objects because their sky plane density is low and their motion
is mostly along a great circle, and it is difficult to attribute
nearby objects for the opposite reasons.

The mixed and bad attributions in
table~\ref{t.MOPSAttributionEfficiency} are of particular concern (see
\S\ref{ss.MOPSEfficiencyConcepts} for definitions of the terms).  In
these types of attributions {\it unassociated} detections are added to
the derived object.  \ie\ the derived object is being contaminated by
noise or detections of other objects.  Since the synthetic MOPS
objects are intended to mimic the behavior of the real objects we
assume that the real objects suffer the same `contamination' levels
(from real and synthetic detections and from false detections).  Some
level of contamination is always unavoidable but it will degrade the
quality of the derived orbits so that future attributions are less
likely to be real.

There are several ways to mitigate the contamination
including 1) tighter controls on the residuals of candidate detections
added to derived objects or 2) {\it post-facto} `scrubbing' of all
derived orbits to identify outlying detections associated with derived
objects.  We never implemented these techniques and have essentially
abandoned improvement of the attribution algorithm because it was designed for a survey pattern that would
provide $\ge 3$ tracklets for most observed asteroids in each
lunation.

The complement to attribution is `precovery' in which historical
unattributed tracklets are linked to new derived objects.  Our
precovery algorithm is essentially identical to attribution but
applied backwards in time.  The only enhancement is that when a
successful precovery enhances the derived orbital element accuracy
with a consequent reduction in the ephemeris errors we iterate on the
precovery attempts until all possible precoverable tracklets are
associated with the new derived object.\footnote{The precovery
  algorithm is invoked any time any orbit changes.  \ie\ the
  attribution of a new tracklet to an existing derived objects also
  triggers the precovery algorithm.}  We also allow for the precovery
of tracklets intermediate in time between existing tracklets in the
derived object.  

When the observed arc length of a derived object exceeds a
configurable time span such that the prediction uncertainty over the
entire survey is below some threshold, the object has reached a stage
where the orbit solution is stable, all possible detections have
already been associated with the object, and it no longer needs to be
precovered. This optimization reduces the processing time required for
precovery (see Fig.~\ref{f.timing}) because over a long enough survey
many objects will be `retired' out of precovery when their orbits
become sufficiently accurate. For synthetic objects, MOPS records all
successful and failed precoveries so the pipeline can produce
statistics on objects that have reached orbit stability but still have
recoverable tracklets (see table~\ref{t.MOPSPrecoveryEfficiency}).

\begin{figure}[btp]
\centering
\plotone{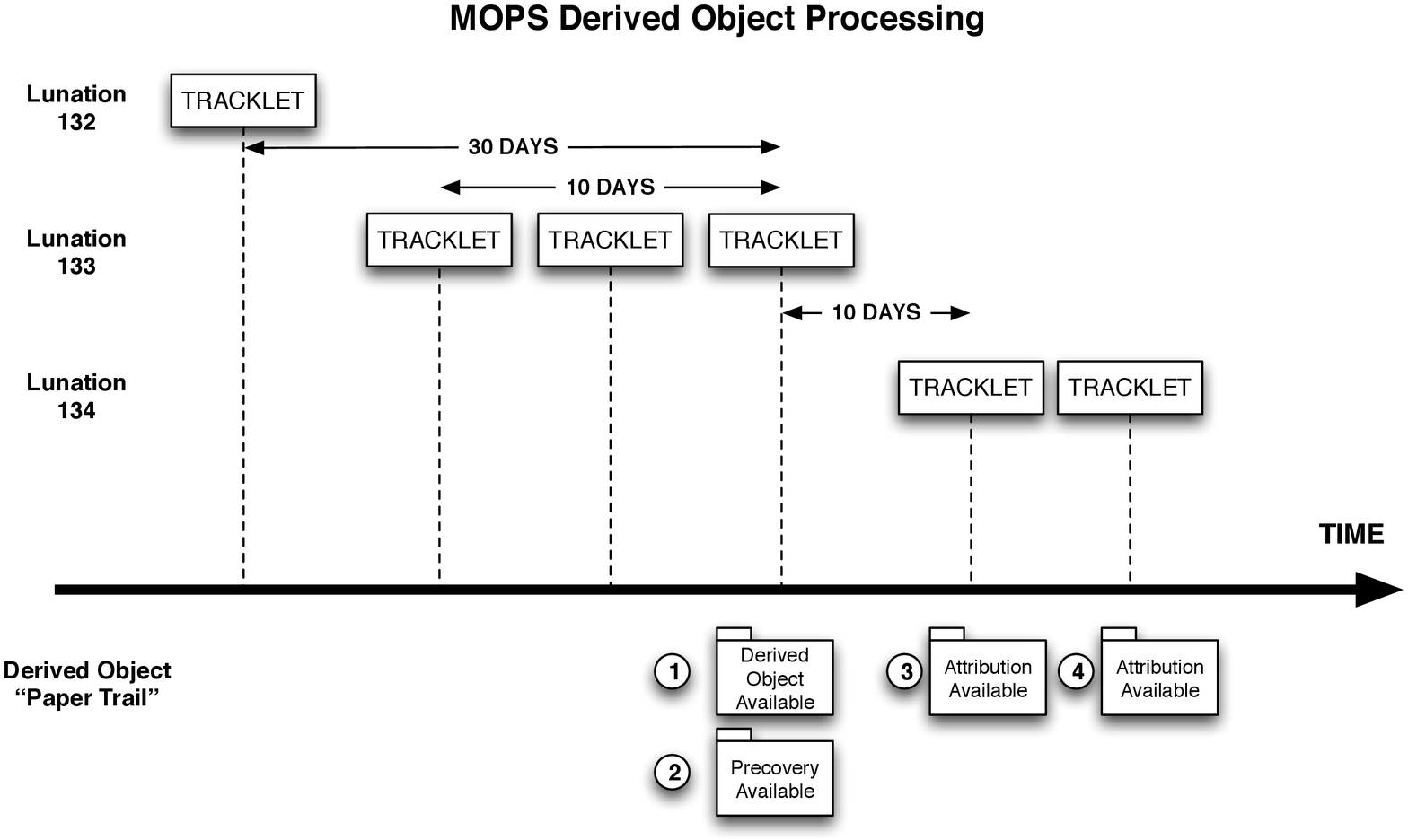}
\caption{MOPS derived object processing overview for a single hypothetical asteroid, showing observed tracklets
and the created derived object. In lunation 132, no derivation is available because there is only a single
tracklet observed. In lunation 133, three tracklets are observed in a 10 day window, so a derived object
is created and the precovery in lunation 132 becomes available. In lunation 134, two attributions are available
using predictions from the object's derived orbit. Numbers in circles show the sequence of `paper trail' records
inserted into the MOPS database.
}
\label{f.derivedobject}
\end{figure}

Fig. \ref{f.derivedobject} illustrates the sequence of events within
MOPS for a hypothetical asteroid that is observed over three successive 
lunations. For synthetic objects, the `paper trail' allows us to interrogate
derived object performance and understand how and what kinds
of objects get lost in the system. For each derived object `event' (derivation,
attirubtion, or precovery), the object's orbit is modified (hopefully improved) and
recorded, allowing easy inspection of the orbit history for the derived object
within MOPS.

\subsection{Orbit Identification}
\label{s.orbitidentification}

When a new derived object ($A'$) is created it is automatically
checked for possible precoveries as described above.  If the arc
length is still short even after precovery it is possible that much
earlier detections of the object $A'$ appear in the database that can
not be precovered because the ephemeris uncertainty becomes too large
when the current derived orbit is integrated backwards in time.  On
the other hand, those much earlier detections may also have been
incorporated into a derived object ($A$) that similarly can not be
attributed to the current detections because integrating derived
object $A$ forward in time generates too large an ephemeris
uncertainty.  Thus, the same object may exist in two separate derived
orbits that are unlinkable by our precovery and attribution
techniques.

Rigorous techniques exist for `orbit identification' --- associating
short arcs of detections within an apparition to the same object
observed with a short arc in another apparition
\citep[\eg][]{Granvik2008,Milani2005} --- but given the high astrometric 
quality we could assume for \PSfour, we adopted a simplified strategy
that searches for neighboring orbits in orbit parameter space.

If each of the two derived orbits are themselves reasonably accurate
and precise then it is possible to identify the two derived objects as
being identical simply by checking that the derived orbit elements are
similar.  Given that we expected to find $>10^7$ objects with \PS\ and
thousands of new derived objects each night we implemented the search
for similar derived orbits in a kd-tree ({\tt orbitProximity}).  For
any new derived object $A'$ {\tt orbitProximity} identifies a set of
candidate derived orbits ($X=\{A,B,C,D\}$) for which all six orbit
elements match $A'$ to within the tolerances shown in
table~\ref{t.OrbitIdentificationMatchingTolerances}.  Then we attempt
a differential correction (\S\ref{ss.DifferentialCorrection}) to an
orbit including all detections from pairwise combinations of $A'$ with
each of the candidate orbits.  \eg\ $\{A,A'\}$, $\{B,A'\}$.  If the
differentially corrected orbit meets our $RMS$ residual requirements
then the two derived objects are merged into one derived object
(\eg\ $A$ and $A'$).

Our MOPS simulations for a \PSfour\ system showed that the orbit
identification efficiency and accuracy were close to 100\% using the
tolerances in table~\ref{t.OrbitIdentificationMatchingTolerances}.
The realized orbit identification efficiency for \PSone\ is $\sim$26\%
as show in table~\ref{t.MOPSOrbitIdentificationEfficiency}.  The
reduced efficiency is because the \PSone\ cadence and astrometry
yields orbits with uncertainties larger
than our configured orbit identification tolerances.

\subsection{Data Rates \& MOPS Timing}
\label{ss.DataRatesMOPSTiming}

\begin{figure}[t]
\centering
\plotone{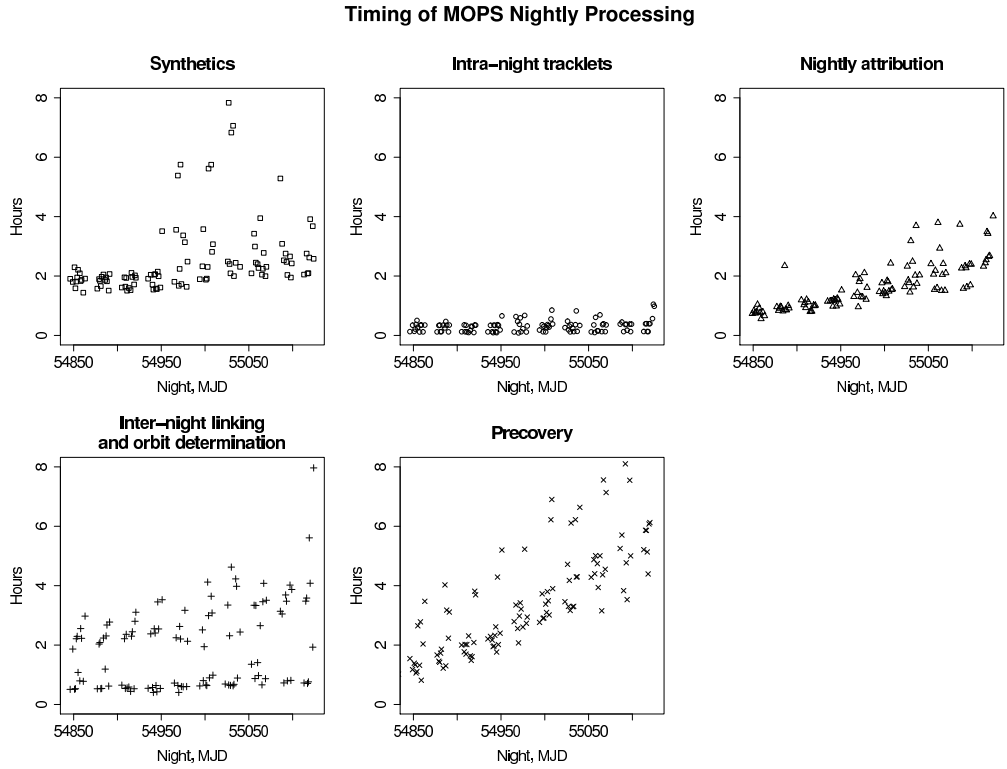}
\caption{Processing times for different MOPS pipeline stages over one
  simulated year of a two-year simulation using \PSfour\ data
  volumes. Some data prior to MJD 54850 was discarded because MOPS
  pipeline software was not separating processing times for synthetic
  and tracklet stages. }
\label{f.timing}
\end{figure}

Figure~\ref{f.timing} shows execution times for various MOPS
processing stages over one year of a 2-year MOPS simulation with the S3M. Processing times for
per-night stages within the pipeline (generation of synthetics and
tracklets) are generally constant, while stages that operate on the
simulation's internal catalog of derived objects grows linearly or
worse, depending on gross algorithmic considerations. Naively, the
precovery stage exhibits quadratic growth, as the number of derived
objects seen by MOPS and the number of precovery images to search both
grow linearly as a function of simulation time, but optimizations in
the precovery algorithm that limit the search windows reduces this
growth below quadratic ($\sim O(n \cdot log\;n)$).

The bimodality of timing results in Fig.~\ref{f.timing} is due to the
survey containing nights with either sweetspots (168 exposures) or
opposition regions (660 exposures) or both. The slow growth of the
synthetic generation stage is due to the increased integration time
required to produce a position for an asteroid at the observation
epoch from the survey start date. Recent MOPS versions eliminate this
overhead by periodically propagating the epoch of S3M orbits to the
current end-of-survey. Inter-night linking and orbit determination 
computation time grow slowly as the size of the MOPS database increases.

\subsection{Detection efficiency}
\label{sss.DetectionEfficiency}

\begin{figure}
\epsscale{1.0}
\plotone{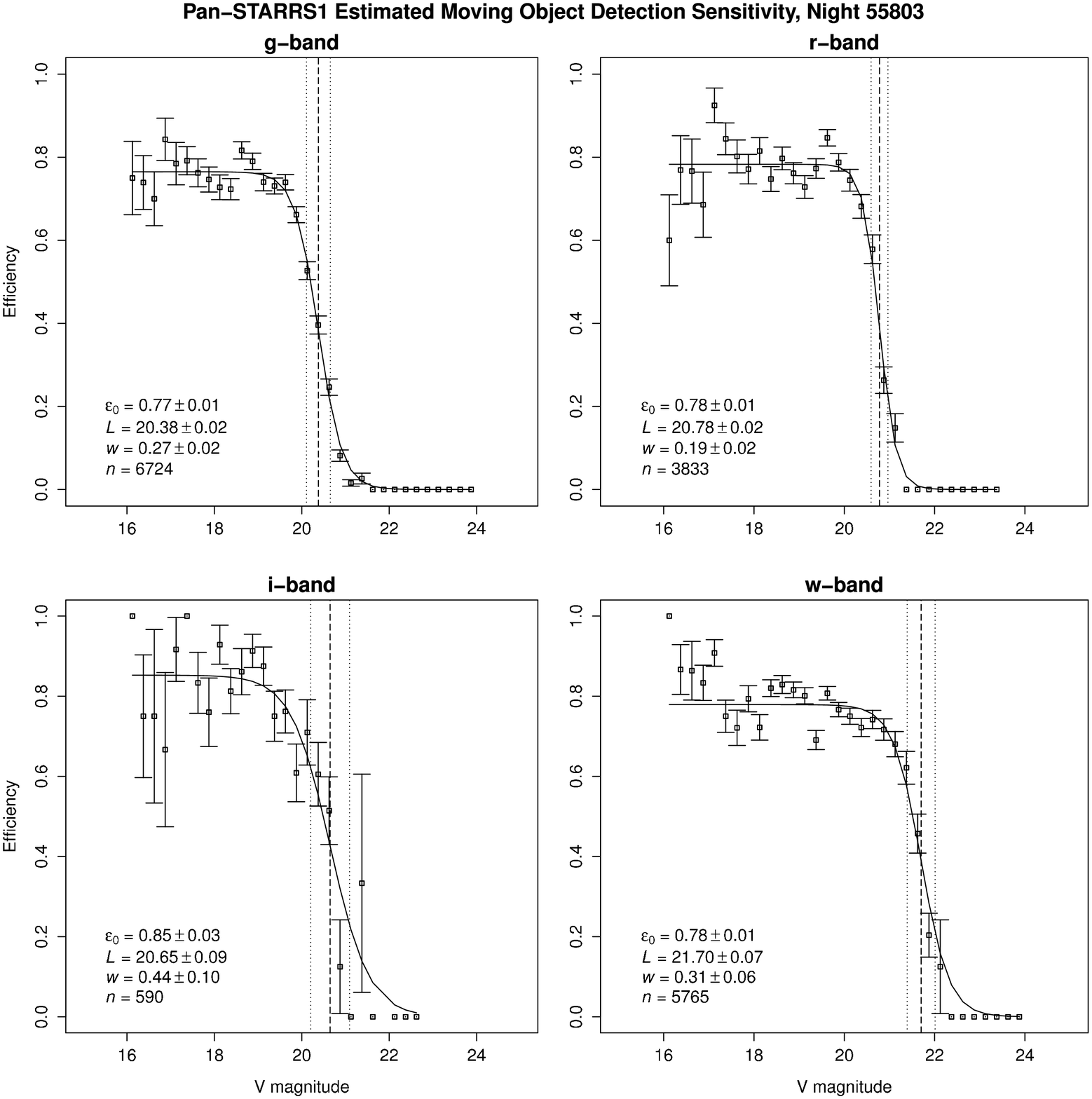}
\caption{\PSone\ moving object detection efficiency on a single night (MJD
  55803) for
  detections of known numbered and multi-opposition asteroids in each
  filter as a function of $V$~magnitude.  The data in each filter were
fit to the function $\epsilon=\epsilon_0 \; \bigg[ 1 +
  \exp\bigg([V-L]/w\bigg) \bigg]^{-1}$ as described in \S\ref{sss.DetectionEfficiency}.  The vertical dashed line is at the $V$~magnitude where the efficiency drops to 50\% of the maximum ($L$).  The vertical dotted lines provide the range $[L-w,L+w]$ over which the efficiency drops quickly. }
\label{f.efficiency-vs-V}
\end{figure}

Figure~\ref{f.efficiency-vs-V} illustrates the realized tracklet
detection efficiency of the combined \PSone+MOPS systems as measured
using asteroids attributed (or not) by \knownserver\ (see
\S\ref{ss.AttributionOfKnownObjects}).  The data in each filter were
fit to the function $\epsilon=\epsilon_0 \; \bigg[ 1 +
  \exp\bigg([V-L]/w\bigg) \bigg]^{-1}$ where $\epsilon_0$ represents
the maximum efficiency for bright but unsaturated detections, $V$ and
$L$ are the apparent and limiting $V$-band magnitudes respectively,
and $w$ is the `width' of the transition from maximum to zero
efficiency.  The limiting magnitude $L$ is the magnitude at which the
efficiency is 50\% of the maximum value.  The magnitudes in all bands
were converted to $V$ using a mean C+S class asteroid type as shown in
table~\ref{t.PS1FilterTransformations}.  The function is a good representation of the
efficiency near the limiting magnitudes and the scatter at
bright values is simply due to low statistics.

The maximum efficiency in all 4 filters in
Fig.~\ref{f.efficiency-vs-V} is $\sim$78\% which is almost entirely
driven by the effective \PSone\ camera fill factor of $\sim$80\%.
This implies that when a asteroid is imaged on a live and unmasked
camera pixel it is detected with $\sim$97\% efficiency.  The
limiting magnitudes in \gps, \rps, and \ips\ are roughly equal at
$V\sim20.5$ and any difference between the bands is due to unbalanced
exposure times.  The wide-band filter \wps\ goes $\sim1$ $V$-mag
deeper through a combination of the 45~second exposure time and its
$\sim3\times$ higher bandwidth.

\begin{figure}[btp]
\centering
\plotone{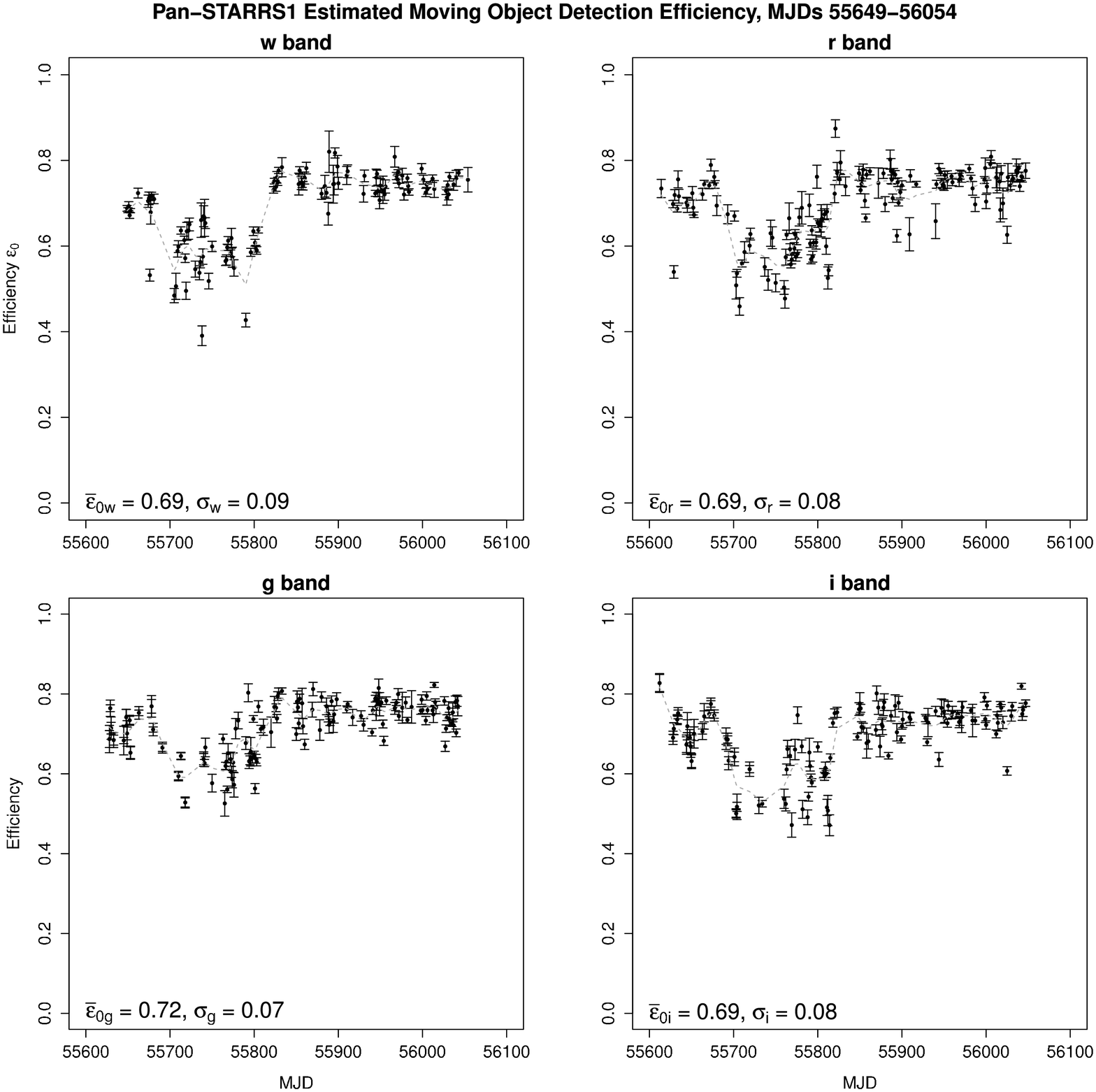}
\caption{\PSone\ moving object detection efficiency for bright
  non-saturated detections as a function of MJD corresponding to the
  time period from approximately  Feb 2011 through Jun
  2012. Around MJD 55820, MOPS began employing
  less-aggressive false detection filtering of IPP transient
  detections, boosting per-exposure detection efficiency to 75\%
  consistently. The dashed line is a spline-fit to the data.}
\label{f.efficiency-vs-mjd}
\end{figure}

\begin{figure}[btp]
\centering
\plotone{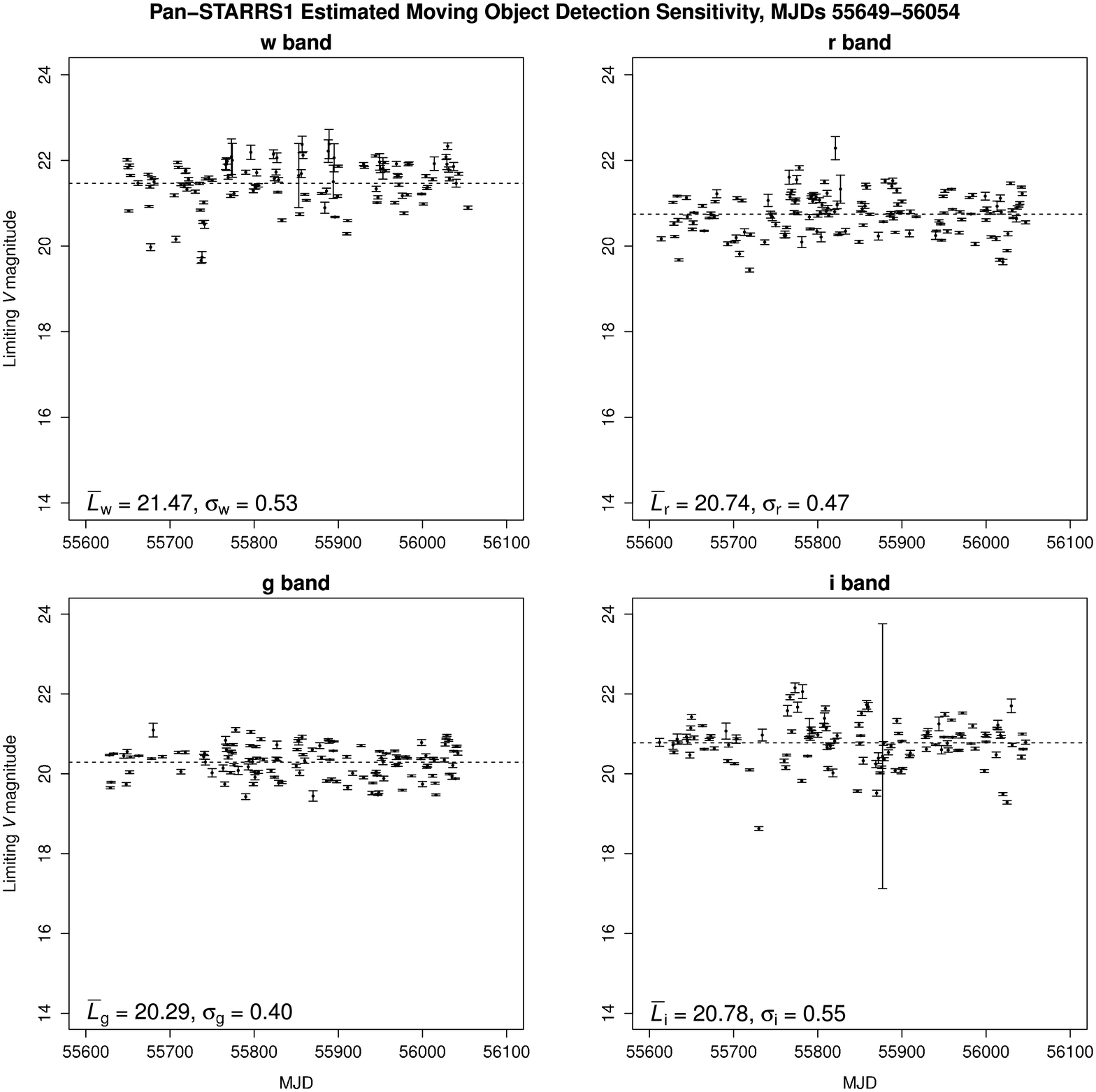}
\caption{\PSone\ nightly limiting $V$ magnitude in each of the four main
  filters used to detect moving objects as a function of MJD
  corresponding to the time period from approximately Feb 2011
  through Jun 2012.  The dashed lines represent fits to the passband
  data as a function of time.}
\label{f.limitingmag-vs-mjd}
\end{figure}

Figures~\ref{f.efficiency-vs-mjd} and \ref{f.limitingmag-vs-mjd} show
that \PSone+MOPS performance was roughly constant over the first year
of operations.  The dip in efficiency in the range
$55700\la$MJD$\la$55800 is due to overly aggressive filtering of false
detections during that time period.  The fit to the limiting magnitude
in each passband as a function of time is consistent with there being no change in any filter.  

\begin{figure}[btp]
\centering
\plotone{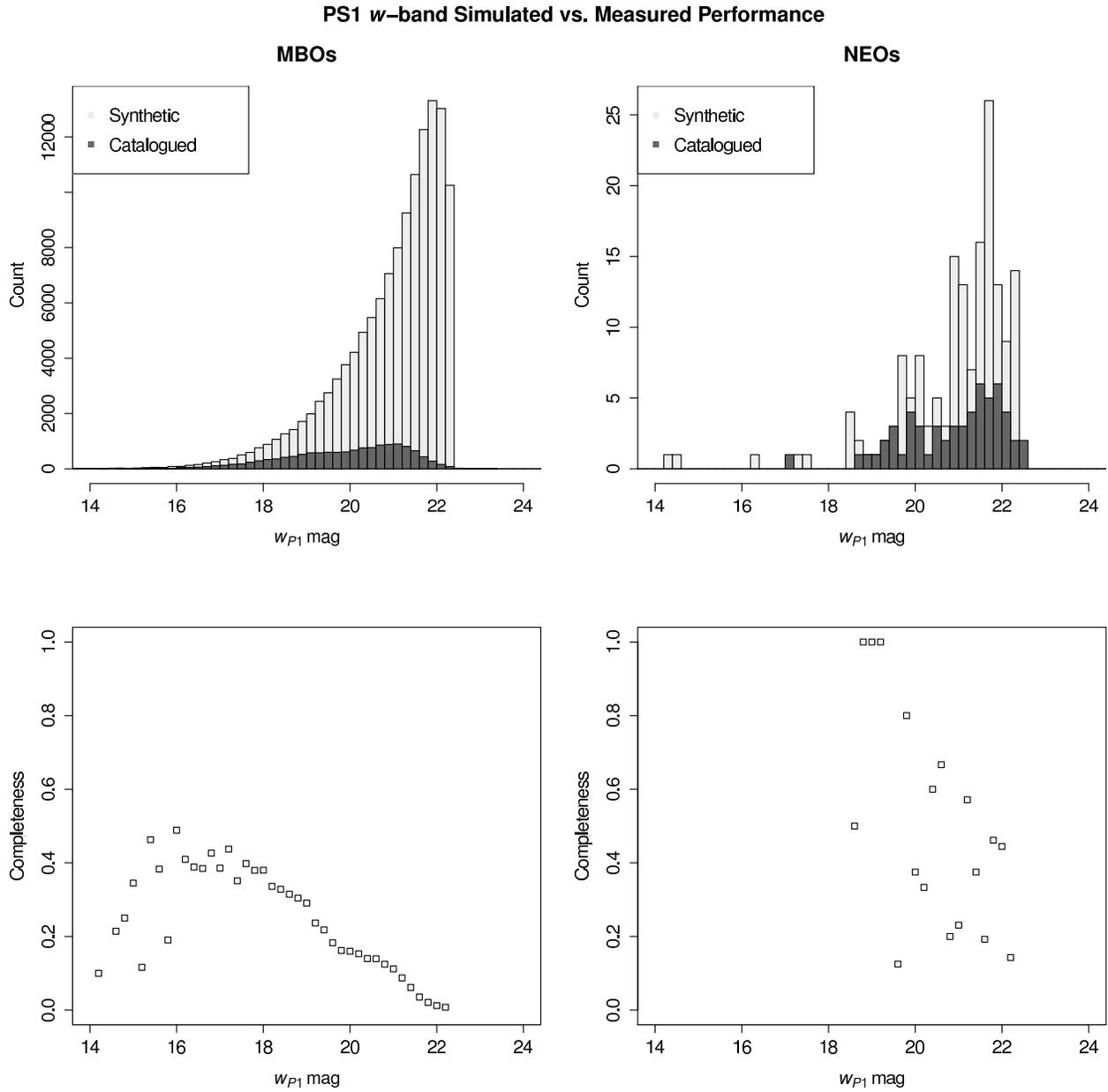}
\caption{Comparison of MOPS synthetic tracklet counts to \PSone\ submitted tracklets during the period 13 Aug 2011 through 11 Oct 2011 for (left) main belt objects and (right) near-Earth objects.}
\label{f.real-vs-sim}
\end{figure}

In Fig.
\ref{f.real-vs-sim} we compare the predicted number of tracklets using the S3M to \PSone\ 
tracklets reported to the MPC during the period 13 Aug 2011 through 11 Oct 2011.
For bright main-belt asteroids with $\wps < 19$, \PSone\ reported about 1/3 the predicted number.  Fainter
than $\wps=19$ the reported rate drops further until the sensitivity
limit is reached.  Many factors account for the discrepancy between the predicted and realized rates:

\begin{enumerate}
\item objects can be lost (primarily MBOs) in image differencing when observing away from opposition because of stationary points (\ie\ the differencing subtracts all or part of the detection)
\item tracklets submitted to the MPC might not yet be assigned a designation (a `one night stand' tracklet)
\item per-exposure live detector fraction may fall below the nominal 75\% due to \eg\ aggressive masking in regions with high stellar sky-plane density
\item true night-to-night sensitivity in \wps\ is poorer than the constant model used for synthetics
\item the synthetic \sn\ model overestimates faint-end performance
\item the S3M may overstate the number of objects in some sub-populations.
\end{enumerate}

The situation is better for the NEOs that show better agreement between
the model and \PSone\ down to the sensitivity limit.  The effect of
(1) almost disappears because NEOs will rarely have stationary
sky-plane motion, and (2) is reduced because of the NEO confirmation
process for one-night tracklets coordinated by the MPC.

\section{Current \PSone\ Performance}

As of late 2012, great strides have been made image processing and
telescope scheduling to increase the yield of asteroids from the \PSone\ 
survey.  Better measurement and modeling of detector readout noise has
resulted in greater detection sensitivity in image processing.  Finer characterization
of detection morphology by the IPP allows MOPS to reject many classes of
false detections upon ingest so that they do not contaminate the 
NEO tracklet review process. The \PSone\ Modified
Design Reference Mission survey \citep{Chambers2012} has increased the fraction of
3$\pi$ survey time spent observing in quads.  With the recent completion of the
3$\pi$ static sky processing, there will be additional improvement in the system
limiting magnitude of at least 0.4 mag as IPP moves from pairwise difference imaging to static sky
differencing.

\begin{figure}[btp]
\centering
\plotone{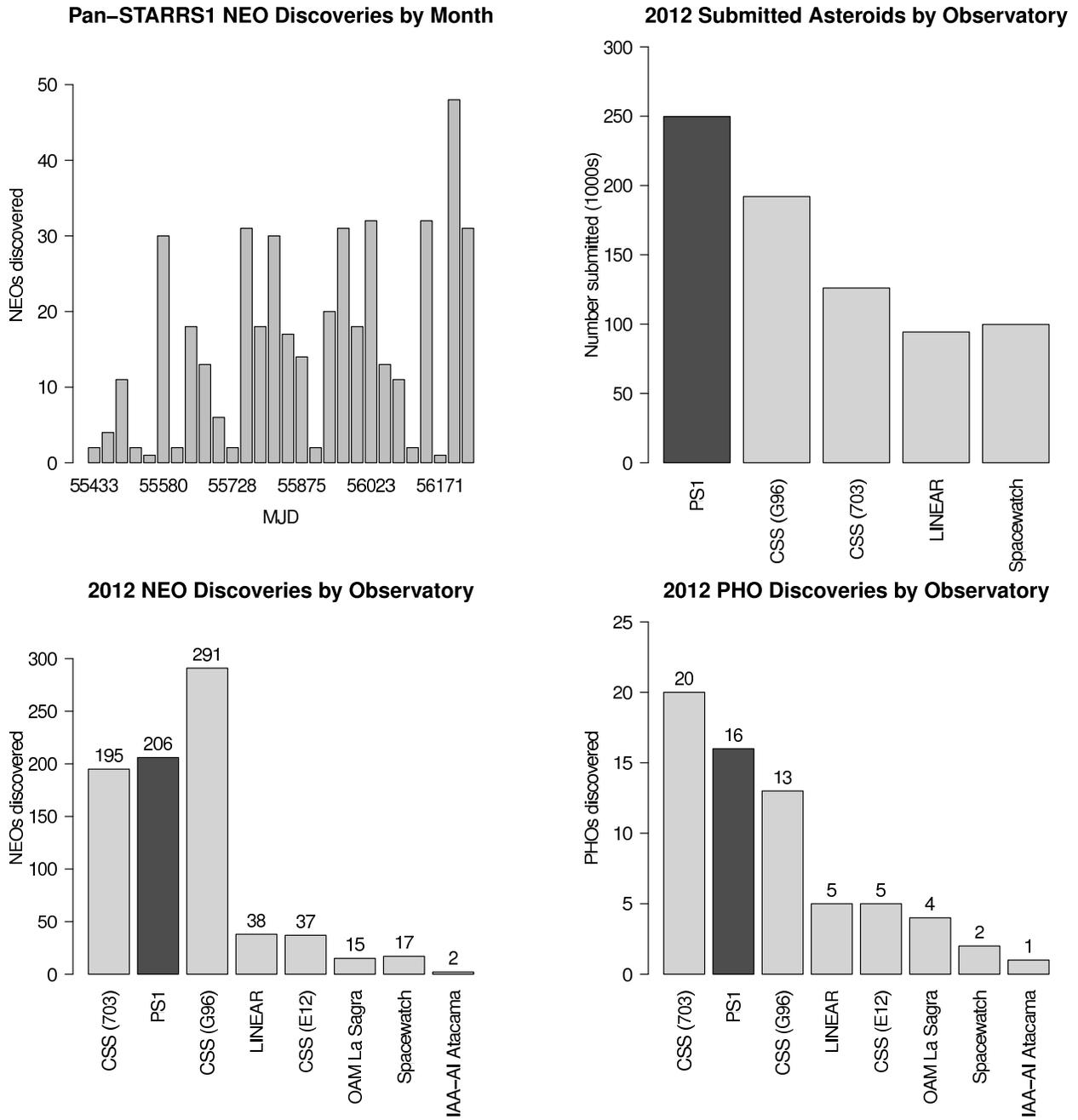}
\caption{\PSone\ discovery statistics as of November 2012 (from the
  Minor Planet Center).}
\label{f.ps1performance}
\end{figure}

Fig. \ref{f.ps1performance} shows NEO discovery and submitted object
totals for \PSone\ and other NEO surveys.  Although monthly totals can
vary with weather losses, since late 2010 the \PSone\ NEO discovery
rate has been increasing.  Using IPP's morphological characterization
and a MOPS comet candidate review procedure similar to that for NEOs,
\PSone\ has become a capable comet finder, discovering 30 to date and
8 in October 2012 alone.

\section{Availability \& Ongoing Development}

The MOPS software is available under the GNU General Public License Version 2, and can
be retrieved by sending a query to the PS1 Science Consortium ({\tt http://www.ps1sc.org}).
MOPS has become a large and somewhat unwieldly package, employing (too many) different
programming languages preferred by third-party software (not all available directly from the PS1SC), and containing
many installation dependencies for Perl and Python modules. Fortunately, documentation
and development notes are available via the PS1SC web site.

Not all MOPS subcomponents are freely available and some are not integrated into the MOPS software
distribution. In particular, the JPL Solar System Dynamics package that performs MOPS
differential correction must be obtained directly from JPL. Other packages such as OrbFit
and the SLALIB positional astronomy library must be downloaded separately.

Approximately 15-20 full-time equivalent (FTE) years have gone into the development of MOPS
and the S3M, and development continues in support of \PSone's ongoing NEO survey
and upcoming NEO surveys such as ATLAS \citep{Tonry2012}.  Prior to handling live \PSone\ data,
MOPS processed testing data from Spacewatch \citep[\eg][]{Gehrels1991,Gehrels1996}  
and raw data in support of the Thousand Asteroid Light Curve Survey \citep[TALCS;][]{Masiero2009}.
MOPS has additionally been instrumental in providing simulation results for \PS, LSST
and ATLAS, and as a research tool for many graduate students and postdocs.

\section{Future improvements}
\label{sss.FutureImprovements}

The future of MOPS likely falls in several areas:

\begin{itemize}

\item Optimizing execution speed for next-generation \PSfour-like data volumes through
reduction of database I/O and greater in-memory processing.

\item Improved MOPS installation and configuration for non-\PS\ surveys.

\item Modeling of photometric artifacts, e.g. partially masked PSFs and trails.

\item The linear processing model employed by MOPS is appropriate for a
mature survey but can be inadequate when evaluating production
parameters. It is
difficult to merge separate MOPS databases created under different
runtime configurations into a single coherent database.  Future versions of
MOPS should be more agile in its ability to perform
non-linear processing and merge datasets.

\item Increased fidelity of synthetics. For
example, although our simulations coarsely avoid the moon they do not
incorporate the effects of varying sky sensitivity or effects due to
the presence of bright objects in the field of view. In one sense,
however, these effects are really an aspect of detector sensitivity
and not MOPS itself, and they can be integrated via per-field
detection detection efficiency parameters if they are measured by IPP.

\item Fill-factor modeling.  There is experimental code to use the
mask of Fig.~\ref{f.fillfactor75} as a crude live-pixel server (LPS, \S \ref{ss.Simulations})
to simulate detections lost on dead areas of the detector.

\item Simplified processing for large-scale simulations, \eg\ two-body
  ephemerides instead of a perturbed dynamical model.

\end{itemize}

\section{Summary}

MOPS has proven to be a capable tool for detecting moving objects 
in the \PSone\ transient detection stream.  Despite the
differences between the targeted \PSfour\ capability and current \PSone\ 
performance, we have deployed MOPS effectively in our search for NEOs and 
comets and in characterizing the main belt.  Importantly,
we have estimates of our efficiency of object detection so that we can
provide a foundation for large-scale population studies.

Additionally, we have made progress toward our design goal of creating
a system that can detect objects and compute their orbits with $>99$\%
efficiency for most solar system populations when provided
next-generation survey astrometry and data quality.  Our experience
with \PSone\ shows that the desired performance is challenging to
achieve with real telescope data.

With nearly one year remaining in the \PSone\ survey, we expect further broad 
improvements in astrometry, photometry, difference imaging and sensitivity
in the final \PSone\ data release.  The \PStwo\ telescope \citep{Burgett2012} 
is under construction adjacent to the \PSone\ facility and will have a raw detector sensitivity and optical
performance exceeding that of \PSone{}.  Operating modes currently under consideration
for the two telescopes have them working in tandem, increasing performance even further.
Combined with optimization and tuning of the \PSone\ MOPS pipeline, we look forward to 
applying the full capability of MOPS to this data and to the subsequent riches 
that will lie in this catalog of moving objects.

\acknowledgements
\section*{Acknowledgments}

The \PSone\ Survey has been made possible through contributions of the
Institute for Astronomy, the University of Hawai`i, the Pan-STARRS
Project Office, the Max-Planck Society and its participating
institutes, the Max Planck Institute for Astronomy, Heidelberg and the
Max Planck Institute for Extraterrestrial Physics, Garching, The Johns
Hopkins University, Durham University, the University of Edinburgh,
Queen's University Belfast, the Harvard-Smithsonian Center for
Astrophysics, and the Las Cumbres Observatory Global Telescope
Network, Incorporated, the National Central University of Taiwan, and
the National Aeronautics and Space Administration under Grant
No. NNX08AR22G issued through the Planetary Science Division of the
NASA Science Mission Directorate. 

The design and construction of the 
Panoramic Survey Telescope and
Rapid Response System by the University of Hawaii Institute for
Astronomy was funded by the United States Air Force Research Laboratory
(AFRL, Albuquerque, NM) through grant number F29601-02-1-0268.

We thank the Jet Propulsion Laboratory, an operating division of the California Institute of Technology, for providing an executable version of their Rapid Comet and Asteroid Orbit Determination Process software (NTR-41180) to the Institute for Astronomy at the University of Hawai`i for use in MOPS.

We acknowledge the financial and technical contributions to
this work made by the Large Synoptic Survey Telescope (LSST)
Corporation team, in particular Tim Axelrod, Lynne Jones and Jeff
Kantor. In addition, we acknowledge the support by Pan-STARRS and LSST
management to enable and facilitate this productive collaborative
effort between the two projects.

Don Yeomans (JPL) and Ted Bowell (Lowell Observatory) provided expert feedback as
external reviewers for the MOPS system in the early years of its
development.

Many colleagues provided helpful feedback on the MOPS system operation
and usability as well as supporting MOPS development.  We would in
particular like to thank Wen-Ping Chen and Rex Chang
from the Institute of Astronomy, National Central University, Taiwan.

This research has made use of the SIMBAD database,
operated at CDS, Strasbourg, France.




\clearpage
\begin{deluxetable}{lccc}
\tabletypesize{\small}
\tablecaption{MOPS \PSone\ Filter Transformations\tablenotemark{*}}
\tablehead{
\colhead{Transform} & 
\colhead{Solar} & 
\colhead{Mean S+C}  &
\colhead{$t_{exp}$ (s) \tablenotemark{\dagger}}
} 
\startdata
$V$-\gps  & -0.217 & -0.28 & 43 \\
$V$-\rps  &  0.183 &  0.23 & 40 \\
$V$-\ips  &  0.292 &  0.39 & 45 \\
$V$-\zps  &  0.311 &  0.37 & 30 \\
$V$-\yps  &  0.311 &  0.36 & 30 \\
$V$-\wps  &  0.114 &  0.16 & 45\\
\enddata
\label{t.PS1FilterTransformations}
\tablecomments{Transformations from Johnson $V$ to the six
  \PS\ filters are provided for both a standard solar spectrum and
  asteroids with a mean S+C spectral type. These transformations have
  been implemented by the Minor Planet Center and AstDyS.}
\tablenotetext{*}{\citet{Tonry2012}.  }
\tablenotetext{\dagger}{Exposure times as of October 2012.}
\end{deluxetable}

\begin{deluxetable}{llcc}
\tabletypesize{\small}
\tablecaption{\PSone\ MOPS Hardware Components (see Fig.~\ref{f.cluster})}
\tablehead{
\colhead{Item} & 
\colhead{Purpose} & 
\colhead{Number} & 
\colhead{Total Capacity} 
} 
\startdata
Disk & Database Storage & 2 $\times$ 10 TB & 20 TB\tablenotemark{1} \\
Disk & Administrative & 2 $\times$ 2.7 TB & 5.4 TB \\
CPU & Cluster processing & 8 $\times$ 4 cores & 32 cores ($\sim$100 GFLOPS) \\
CPU & Administrative\tablenotemark{2} & 2 $\times$ 4 cores & 8 cores ($\sim$25 GFLOPS) \\
Network Switch & Network & 32 ports & 1 Gbps/port \\
\enddata
\label{t.MOPSHardwareComponents}
\tablenotetext{1}{All database storage employs RAID6 for data integrity.}

\tablenotetext{2}{Administrative functions include a user console,
  pipeline and Condor workflow management, and the web
  interface.}

\end{deluxetable}

\clearpage
\begin{deluxetable}{lccc}
\tabletypesize{\small}
\tablecaption{MOPS Database Storage Requirements}
\tablehead{
\colhead{Data Component} & 
\colhead{PS4 Estimated (GB)} & 
\colhead{PS1 Estimated (GB)} & 
\colhead{PS1 Actual\tablenotemark{1} (GB)} 
} 
\startdata
Fields & 0.2 & 0.06 & 0.02 \\
3$\sigma$ Detections & 500,000 & 150,000 & N/A\tablenotemark{2} \\
5$\sigma$ Detections & 500 & 150 & 122 \\
Derived Object Parameters & 1,000 & 300 & 0.152\tablenotemark{3} \\
Synthetic Object Parameters & 1,100 & 330 & 2.43\tablenotemark{4} \\
Tracklets & 1,200 & 360 & 2.19\tablenotemark{5} \\
Image Postage Stamps & N/A & N/A & 1,800\tablenotemark{6} \\
\enddata
\label{t.MOPSDatabaseStorageRequirements}
\tablenotetext{1}{As of October 2012, 2.5 years into the 3.5-year \PSone\ survey.}
\tablenotetext{2}{Processing of transients below 3$\sigma$ confidence is currently untenable for \PSone\ MOPS due to the systematic false tracklet rate.}
\tablenotetext{3}{MOPS produces derived object parameters for only a small subset of \PSone\ data until the pipeline can be tuned for \PSone\ performance.}
\tablenotetext{4}{\PSone\ operations uses a 1/10 sampled synthetic solar system model (S3M). A final \PSone\ processing of its transient catalog will include a full S3M.}
\tablenotetext{5}{\PSone\ single-exposure sensitivity reduces the actual tracklet count on top of the reduction from a shorter survey than \PSfour.}
\tablenotetext{6}{Image postage stamps (200$\times$200\,pixels) are stored on a network file system, not in the MOPS database.  They are locatable using database records.}
\end{deluxetable}

\clearpage
\begin{deluxetable}{cc}
\tabletypesize{\small}
\tablecaption{S3M Grid Model Semi-Major Axis Number Distribution (see Fig.~\ref{f.S3MGridTrackletBya}).}
\tablehead{
\colhead{Semi-major axis range (AU)} & 
\colhead{Number of objects} 
} 
\startdata
0.75-1.5 & 50,000 \\
1.5-6.0 & 200,000 \\
6.0-32 & 50,000 \\
32-50 & 200,000 \\
50-500 & 50,000 \\
500-5000 & 25,000 \\
\enddata
\label{t.gridSemiMajorAxis}

\tablecomments{The number of objects in each pseudo-logarithmic range
  and their absolute magnitudes were selected so that the sky-plane
  density of detected objects in the simulations will be similar.}

\end{deluxetable}

\clearpage

\clearpage
\begin{deluxetable}{cccccccccccc}
\tablecaption{Tracklet (intra-night) Efficiency\tablenotemark{*} \&
  Accuracy\tablenotemark{*} in the 2-year \PSfour\ MOPS full-density
  simulation.}
\tablehead{
\colhead{Avail.} & 
\colhead{Clean} & 
\colhead{\%} & 
\colhead{Unfound} & 
\colhead{\%} & 
\colhead{Mixed} & 
\colhead{\%} & 
\colhead{Bad} & 
\colhead{\%} & 
\colhead{Non-syn.} 
} 
\startdata
 24056199 & 24056191	& 100.0	& 8 & 0.0 & 417774 & 1.7	& 553475 & 2.3 & 1854648 \\
\enddata

\tablecomments{This simulation used a realistic solar system model
  (S3M) and ``next-generation'' ($0.01\arcsec$) astrometric
  uncertainty (about $10\times$ better than delivered by the best
  contemporary surveys like \PSone).  The tracklet terminology in the
  headings is described in detail in
  \S\ref{ss.MOPSEfficiencyConcepts}.}

\tablenotetext{*}{See \S\ref{s.Overview} and \S\ref{ss.MOPSEfficiencyConcepts} for definitions of efficiency and accuracy.}
\label{f.full_S1b_trk_eff}
\end{deluxetable}

\clearpage
\begin{deluxetable}{crrrrrrrrr}
\tabletypesize{\small}

\tablecaption{Derived Object Efficiency for
  eight\tablenotemark{\dagger} different classes of synthetic solar
  system objects in the 2-year \PSfour\ MOPS full-density simulation.}

\tablehead{
\colhead{Object} & 
\colhead{Avail.\tablenotemark{*}} & 
\colhead{Clean\tablenotemark{*}} & 
\colhead{\%} & 
\colhead{Pass} & 
\colhead{\%} & 
\colhead{Pass} & 
\colhead{\%} 
\\ 
\colhead{Class\tablenotemark{1}} &  
\colhead{}       & 
\colhead{(Linked)}       & 
\colhead{}   & 
\colhead{IOD\tablenotemark{2}} & 
\colhead{}    & 
\colhead{Diff.\tablenotemark{3}} & 
\colhead{} 
} 
\startdata
NEO	 & 5203  	& 4994	& 96.0	 & 4924 	& 94.6	 & 4924 	& 94.6 \\
MBO	 & 2043584  & 2032676 	& 99.5	 & 2029618 	& 99.3	 & 2029618	& 99.3 \\
TRO	 & 56214	& 56061	& 99.7	 & 55923	& 99.5	 & 55923	& 99.5 \\
CEN  & 557	& 556 	& 99.8 	 & 556 	& 99.8 	 & 556	& 99.8 \\
JFC	 & 551	 & 546 	& 99.1	 & 541 	& 98.2	 & 541	& 98.2 \\
LPC	 & 1714 	& 1713 	& 99.9	 & 1584 	& 92.4	 & 1584	 & 92.4 \\
SDO	 & 2289 	& 2286 	& 99.9	 & 2281 	& 99.7	 & 2281 	& 99.7 \\
TNO	 & 12719 	& 12710 	& 99.9	 & 12686 	& 99.7	 & 12686	& 99.7 \\
\hline
\enddata

\tablecomments{This simulation used a realistic solar system model
  (S3M) and ``next-generation'' ($0.01\arcsec$) astrometric
  uncertainty (about $10\times$ better than delivered by the best
  contemporary surveys like \PSone).  The tracklet terminology in the
  headings is described in detail in
  \S\ref{ss.MOPSEfficiencyConcepts}.}

\tablenotetext{*}{See \S\ref{s.Overview} and \S\ref{ss.MOPSEfficiencyConcepts} for definitions of efficiency and accuracy.}
\tablenotetext{\dagger}{The impactor and hyperbolic models are omitted because they did not exist in the S3M at the time of this simulation.}
\tablenotetext{1}{NEO - Near Earth Objects; MBO - Main Belt Objects; TRO - Trojans; JFC - Jupiter Family Comets; LPC - Long Period Comets; CEN - Centaurs; SDO - Scattered Disk Objects; TNO - Trans-Neptunian Objects}
\tablenotetext{2}{the number (and percentage) for which initial orbit determination was successful.}
\tablenotetext{3}{the number (and percentage) for which a differentially corrected orbit was successfully computed.}
\label{f.full_S1b_eff}
\end{deluxetable}

\clearpage
\begin{deluxetable}{crrrrrrrrr}
\tabletypesize{\small}
\tablecaption{Derived Object Efficiency for the 4-year \PSone\ MOPS NEO-only simulation.}
\tablehead{
\colhead{Object} & 
\colhead{Avail.\tablenotemark{*}} & 
\colhead{Clean\tablenotemark{*}} & 
\colhead{\%} & 
\colhead{Pass} & 
\colhead{\%} & 
\colhead{Pass} & 
\colhead{\%} 
\\ 
\colhead{Class\tablenotemark{1}} &  
\colhead{}       & 
\colhead{(Linked)}       & 
\colhead{}   & 
\colhead{IOD\tablenotemark{1}} & 
\colhead{}    & 
\colhead{Diff.\tablenotemark{2}} & 
\colhead{} 
} 
\startdata
NEO	 & 11515  	& 11025	& 95.7	 & 9404 	& 81.7	 & 9404 	& 81.7 \\
\hline
\enddata

\tablecomments{This simulation used a realistic population of $\sim 250,000$ NEOs with
  \PSone-like astrometric uncertainty ($0.1\arcsec$) simulation and
  a 100\% fill-factor detector.  This simulation was run during
  commissioning of the \PSone\ telescope before the actual
  \PSone\ survey was implemented.}

\tablenotetext{*}{See \S\ref{s.Overview} and \S\ref{ss.MOPSEfficiencyConcepts} for definitions of efficiency and accuracy.}
\tablenotetext{1}{the number (and percentage) for which initial orbit determination was successful.}
\tablenotetext{2}{the number (and percentage) for which a differentially corrected orbit was successfully computed.}
\label{f.neo_4yr_eff}
\end{deluxetable}

\clearpage
\begin{deluxetable}{cccccccccccc}
\tablecaption{
Tracklet Efficiency\tablenotemark{*} \& Accuracy\tablenotemark{*} in \PSone\ Observing Cycle 143 (2011-08-13 through 2011-09-10)\tablenotemark{\dagger}.
}
\tablehead{
\colhead{Avail.} & 
\colhead{Clean} & 
\colhead{\%} & 
\colhead{Unfound} & 
\colhead{\%} & 
\colhead{Mixed} & 
\colhead{\%} & 
\colhead{Bad} & 
\colhead{\%} & 
\colhead{Non-syn.} 
} 
\startdata
 246596 & 246558	& 100.0	& 38 & 0.0 & 1969 & 0.8	& 8884 & 3.6 & 1429073 \\
\enddata

\tablecomments{Measured performance for a single observing cycle of
  real \PSone\ telescope pointings using realistic astrometric
  uncertainty ($0.1\arcsec$) and a simplistic 75\% fill-factor model.}

\tablenotetext{*}{See \S\ref{s.Overview} and \S\ref{ss.MOPSEfficiencyConcepts} for definitions of efficiency and accuracy.}
\tablenotetext{\dagger}{Excluding fields within 15$\arcdeg$ of the Galactic Plane.}
\label{t.MOPSTrackletEfficiencyAndAccuracy}
\end{deluxetable}

\clearpage
\begin{deluxetable}{rrrrrrrrrrrr}
\tabletypesize{\small}
\tablecaption{Cumulative derived object efficiency for 10 different classes of synthetic solar system objects in \PSone\ Observing Cycle 143 (2011-08-13 through 2011-09-10).\tablenotemark{\dagger}}
\tablehead{
\colhead{Object} & 
\colhead{Avail.\tablenotemark{*}} & 
\colhead{Clean\tablenotemark{*}} & 
\colhead{\%} & 
\colhead{Pass} & 
\colhead{\%} & 
\colhead{Pass} & 
\colhead{\%} &
\colhead{Dup.\tablenotemark{4}} & 
\colhead{\%} 
\\ 
\colhead{Class\tablenotemark{1}} &  
\colhead{}       & 
\colhead{(Linked)}       & 
\colhead{}   & 
\colhead{IOD\tablenotemark{2}} & 
\colhead{}    & 
\colhead{Diff.\tablenotemark{3}} & 
\colhead{} &
\colhead{} &
\colhead{} 
} 
\startdata
IMP	 & 88		& 68	 	& 77.3	 & 67	 	& 76.1	 & 67		& 76.1	 & 0	& 0.0 \\
NEO	 & 24  	& 18	 	& 75.0	 & 18	 	& 75.0	 & 18	 	& 75.0	 & 0	& 0.0 \\
MBO	 & 18301  & 14608 	& 79.8	 & 14551 	& 79.5	 & 14461	& 79.0	 & 6	& 0.0 \\
TRO	 & 662	& 538	& 81.3	 & 538	& 81.3	 & 522	& 78.9	 & 0	& 0.0 \\
CEN\tablenotemark{5}  & 0 		& 0 		& 0.0 	 & 0 		& 0.0 	 & 0 		& 0.0 	 & 0	& 0.0 \\
JFC	 & 22	 	& 18	 	& 81.8	 & 18	 	& 81.8	 & 18	 	& 81.8	 & 0	& 0.0 \\
LPC	 & 50	 	& 39	 	& 78.0	 & 36	 	& 72.0	 & 36	 	& 72.0	 & 0	& 0.0 \\
SDO	 & 7	 	& 1	 	& 14.3	 & 1	 	& 14.3	 & 1	 	& 14.3	 & 0	& 0.0 \\
TNO	 & 17	 	& 15	 	& 88.2	 & 15	 	& 88.2	 & 15	 	& 88.2	 & 0	& 0.0 \\
HYP	 & 32	 	& 27	 	& 84.4	 & 19	 	& 59.4	 & 19	 	& 59.4	 & 0	& 0.0 \\
\hline
\enddata

\tablecomments{Measured performance for a single observing cycle of
  real \PSone\ telescope pointings using realistic astrometric
  uncertainty ($0.1\arcsec$) and a simplistic 75\% fill-factor model.}

\tablenotetext{*}{See \S\ref{s.Overview} and \S\ref{ss.MOPSEfficiencyConcepts} for definitions of efficiency and accuracy.}
\tablenotetext{\dagger}{Excluding fields within 15$\arcdeg$ of the galactic equator.}
\tablenotetext{1}{IMP - Earth Impactor; NEO - Near Earth Objects; MBO - Main Belt Objects; TRO - Trojans; JFC - Jupiter Family Comets; LPC - Long Period Comets; CEN - Centaurs; SDO - Scattered Disk Objects; TNO - Trans-Neptunian Objects; HYP - Hyperbolic (Interstellar) Objects}
\tablenotetext{2}{the number (and percentage) for which initial orbit determination was successful.}
\tablenotetext{3}{the number (and percentage) for which a differentially corrected orbit was successfully computed.}
\tablenotetext{4}{the number (and percentage) of duplicate derivations.  The duplicates contain non-identical but intersecting sets of detections of the same object.}
\tablenotetext{5}{Centaurs were inadvertently omitted from the S3M for this study.}
\label{t.MOPSDerivedObjectEfficiency}
\end{deluxetable}

\clearpage

\begin{deluxetable}{cccccccc}
\tabletypesize{\small}
\tablecaption{
Attribution Efficiency\tablenotemark{*} \& Accuracy\tablenotemark{*} in \PSone\ Observing Cycle 143 (2011-08-13 through 2011-09-10).\tablenotemark{\dagger}
}
\tablehead{
\colhead{Avail.\tablenotemark{*}} & 
\colhead{Clean\tablenotemark{*}} & 
\colhead{\%} & 
\colhead{Mixed\tablenotemark{*}} & 
\colhead{\%} & 
\colhead{Bad\tablenotemark{*}} & 
\colhead{\%} & 
\colhead{Non-syn.\tablenotemark{*}} 
} 
\startdata
9684	 & 8971	& 92.6	 & 5	& 0.1	 & 26 	& 0.3	 & 3625 \\
\enddata

\tablecomments{Measured performance for a single observing cycle of
  real \PSone\ telescope pointings using realistic astrometric
  uncertainty ($0.1\arcsec$) and a simplistic 75\% fill-factor model.}

\tablenotetext{*}{See \S\ref{s.Overview} and \S\ref{ss.MOPSEfficiencyConcepts} for definitions of efficiency and accuracy.}
\tablenotetext{\dagger}{Excluding fields within 15$\arcdeg$ of the Galactic Plane.}
\label{t.MOPSAttributionEfficiency}
\end{deluxetable}

\clearpage

\begin{deluxetable}{cccccccc}
\tabletypesize{\small}
\tablecaption{
MOPS Precovery Efficiency\tablenotemark{*} \& Accuracy\tablenotemark{*} in \PSone\ Observing Cycle 143 (2011-08-13 through 2011-09-10).\tablenotemark{\dagger}
}
\tablehead{
\colhead{Avail.\tablenotemark{*}} & 
\colhead{Clean\tablenotemark{*}} & 
\colhead{\%} & 
\colhead{Mixed\tablenotemark{*}} & 
\colhead{\%} & 
\colhead{Bad\tablenotemark{*}} & 
\colhead{\%} & 
\colhead{Non-syn\tablenotemark{*}} 
} 
\startdata
8557	& 6635	&  77.5	& 14	& 0.2	& 23	& 0.3	& 1966	 \\
\enddata

\tablecomments{Measured performance for a single observing cycle of
  real \PSone\ telescope pointings using realistic astrometric
  uncertainty ($0.1\arcsec$) and a simplistic 75\% fill-factor model.}

\tablenotetext{*}{See \S\ref{s.Overview} and \S\ref{ss.MOPSEfficiencyConcepts} for definitions of efficiency and accuracy.}
\tablenotetext{\dagger}{Excluding fields within 15$\arcdeg$ of the galactic equator.}
\label{t.MOPSPrecoveryEfficiency}
\end{deluxetable}

\clearpage

\begin{deluxetable}{rl}
\tabletypesize{\small}
\tablecaption{Orbit Identification Matching Tolerances}
\tablehead{
\colhead{Orbit} & 
\colhead{Tolerance} 
\\ 
\colhead{Element\tablenotemark{1}} & 
\colhead{} 
} 
\startdata
perihelion         & 0.1~AU \\
eccentricity       & 0.05 \\
inclination        & 0.1$\arcdeg$ \\
ascending node     & 1$\arcdeg$ \\
arg. of perihelion & 1$\arcdeg$ \\
time of perihelion\tablenotemark{2} & 0.1 \\
\enddata

\tablecomments{The maximum difference between two derived object's
  orbits before they are tested for consistency with being the same
  object.}

\tablenotetext{1}{Cometary form.}
\tablenotetext{2}{In terms of fractional orbit period.}
\label{t.OrbitIdentificationMatchingTolerances}
\end{deluxetable}

\clearpage
\begin{deluxetable}{cccccccc}
\tabletypesize{\small}
\tablecaption{
MOPS Orbit Identification Efficiency\tablenotemark{*} \& Accuracy\tablenotemark{*} in \PSone\ Observing Cycles 143 \& 144 (2011-08-13 through 2011-10-10).\tablenotemark{\dagger}
}
\tablehead{
\colhead{Avail.\tablenotemark{*}} & 
\colhead{Found\tablenotemark{*}} & 
\colhead{\%} & 
\colhead{Mixed\tablenotemark{*}} & 
\colhead{\%} & 
\colhead{Bad\tablenotemark{*}} & 
\colhead{\%} & 
\colhead{Non-syn.} 
} 
\startdata
  40  &  7  & 17.5  &  1  &  2.5  &  2  &  5.0  &  22  \\
\enddata

\tablecomments{Measured performance for a single observing cycle of
  real \PSone\ telescope pointings using realistic astrometric
  uncertainty ($0.1\arcsec$) and a simplistic 75\% fill-factor model.}

\tablenotetext{*}{See \S\ref{s.Overview} and \S\ref{ss.MOPSEfficiencyConcepts} for definitions of efficiency and accuracy.}
\tablenotetext{\dagger}{Excluding fields within 15$\arcdeg$ of the galactic equator.}
\label{t.MOPSOrbitIdentificationEfficiency}
\end{deluxetable}
\clearpage




\clearpage

\clearpage

\clearpage

\clearpage

\clearpage

\clearpage

\clearpage

\clearpage

\clearpage

\clearpage

\clearpage

\clearpage

\clearpage

\clearpage

\clearpage

\clearpage

\clearpage

\clearpage

\clearpage

\clearpage

\clearpage

\clearpage

\clearpage

\clearpage

\clearpage

\bibliography{msref}
\bibliographystyle{pasp}

\end{document}